\documentclass[10pt,a5paper]{scrbook}

\usepackage[passed={June 13, 2012}]{dissertation}

\newcommand\Title{Computational Aspects of Dependence Logic}
\newcommand\Author{Peter Lohmann}
\newcommand\Subject{Dissertation}
\newcommand\Keywords{dependence logic, computational complexity, modal logic, expressivity, satisfiability, model checking, two-variable logic, independence-friendly logic, intuitionistic logic}
\newcommand\Schlagworte{Dependence-Logik, Komplexität, Modallogik, Erfüllbarkeit, Model-Checking, Zwei-Variablen-Logik, Independence-friendly-Logik, Intuitionistische Logik}
\newcommand\SubjectClassifiers{F.2.2 Complexity of proof procedures; F.4.1 Modal logic, Computability theory, Model theory; F.1.3 Reducibility and completeness; D.2.4 Model checking}

\title\Title
\author\Author

\hypersetup{%
  pdftitle={\Title},
  pdfsubject={\Subject},
  pdfauthor={\Author},
  pdfkeywords={\Keywords},
  colorlinks=false,bookmarksopen,pdfdisplaydoctitle=false,pdfborder={0 0 0}
}

\begin{document}
  \selectlanguage{ngerman}
  \frontmatter
  \pagestyle{plain}

  \newlength{\titlepageskip}
  \setlength{\titlepageskip}{20pt}
  \begin{titlepage}

    \begin{center}
      \leavevmode
      \vskip \titlepageskip
      
      {\usekomafont{disposition}\huge 
        Computational Aspects \\[-4pt]
        of \\[-4pt]
        Dependence Logic\\
      }
      \vskip 3\titlepageskip
      
      \ifpassed
        Von der
      \else 
        Der
      \fi
      Fakultät für Elektrotechnik und Informatik\\
      der Gottfried Wilhelm Leibniz Universität Hannover\\
      zur Erlangung des Grades
      \vskip \titlepageskip
      
      Doktor der Naturwissenschaften \\ Dr.\ rer.\ nat.
      \vskip \titlepageskip
      
      \ifpassed
        genehmigte 
      \else 
        vorgelegte
      \fi
      Dissertation\\
      von
      \vskip \titlepageskip
      
      Dipl.-Math.\ Peter Lohmann
      \vskip \titlepageskip
      
      geboren am 23.~September 1985 in Hannover
      \vskip 3\titlepageskip
        
      \ifpassed
        2012
      \fi
    \end{center}
  \end{titlepage}
  
  \leavevmode
  \vfill
  
    \begin{tabular}{ll}
      Referent:          & Heribert Vollmer, Leibniz Universität Hannover\\
      Korreferent:       & Lauri Hella, Tampereen Yliopisto\\
      Tag der Promotion: & 13.~Juni 2012 \\
    \end{tabular}
    
    \vskip 10ex
    
{\small
\begin{tabular}[c]{@{}ll@{}}
\includegraphics[height=5ex]{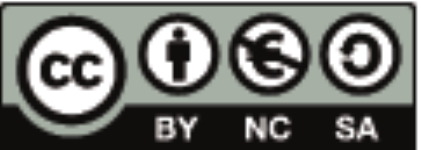}&
\begin{tabular}[b]{@{}l@{}}
\textcopyright\,Peter Lohmann\\
{\scriptsize licensed under a}\\[-0.5ex]
{\scriptsize \href{http://creativecommons.org/licenses/by-nc-sa/3.0/}{Creative Commons Attribution-NonCommercial-ShareAlike 3.0 Unported License}}
\end{tabular}
\end{tabular}}

  \cleardoublepage

  \begin{titlepage}
    \begin{center}
      \leavevmode
      \vskip \titlepageskip
      
      {\usekomafont{disposition}\huge 
        Computational Aspects \\[-4pt]
        of \\[-4pt]
        Dependence Logic\\
      }
      \vskip 3\titlepageskip
      
      Dissertation\\
      to obtain\\
      the academic degree of
      \vskip \titlepageskip
      
      Dr.\ rer.\ nat.\\
      PhD
      \vskip \titlepageskip
      
      authorized by the\\
      Faculty of Electrical Engineering and Computer Science\\
      of the Gottfried Wilhelm Leibniz Universität Hannover\\
      \vskip 1.5\titlepageskip
      
      written by\\
      Dipl.-Math.\ Peter Lohmann
      \vskip \titlepageskip
      
      born September 23, 1985\\
      in Hannover, Germany
      \vskip 2\titlepageskip
        
      2012
    \end{center}
  \end{titlepage}
  
  \leavevmode
  \vfill

    \begin{tabular}{ll}
      First Reviewer:          & Heribert Vollmer, Leibniz Universität Hannover\\
      Second Reviewer:       & Lauri Hella, Tampereen Yliopisto\\
      Graduation Date: & \passeddate \\
    \end{tabular}

    \vskip 10ex
    
{\small
\begin{tabular}[c]{@{}ll@{}}
\includegraphics[height=5ex]{cc-by-nc-sa}&
\begin{tabular}[b]{@{}l@{}}
\textcopyright\,Peter Lohmann\\
{\scriptsize licensed under a}\\[-0.5ex]
{\scriptsize \href{http://creativecommons.org/licenses/by-nc-sa/3.0/}{Creative Commons Attribution-NonCommercial-ShareAlike 3.0 Unported License}}
\end{tabular}
\end{tabular}}

  \cleardoublepage
  
  \selectlanguage{english}
  \leavevmode
  \vskip 0pt plus 1 fill
  \begin{center}
    \itshape%
    I am interested in mathematics only as a creative art.
  \end{center}
  \hfill G.~H.~Hardy
  \vskip 0pt plus 1 fill

  \cleardoublepage
  
  \leavevmode
  \vskip 0pt plus 1 fill
  \selectlanguage{ngerman}
  \section*{Danksagung}
  Zuerst möchte ich ganz herzlich meinem Doktorvater Heribert Vollmer für die Betreuung, Hilfestellung und gemeinsame Forschung während meiner Promotionszeit danken -- und dafür, dass er zu der besten Sorte Chef gehört, die ich mir vorstellen kann.

Weiter danke ich meinem Kollegen, Koautor und Büronachbarn Johannes Ebbing für zahlreiche gewinnbringende Diskussionen und gemeinsame Forschungsergebnisse.
Des Weiteren möchte ich mich bei meinen weiteren Koautoren für die gemeinsame Forschung bedanken.
Auch danke ich Juha Kontinen und Lauri Hella für meine erfolgreichen Besuche in Helsinki und Tampere.
Ganz besonders bedanke ich mich bei meinem Kollegen Arne Meier für viele detaillierte Anregungen zur Verbesserung dieser Arbeit.

Außerdem danke ich meinen Eltern, dass sie mich immer in meinen Fähig\-keit\-en bestärkt haben.
Und am meisten danke ich meiner Frau Anna für ihre Unterstützung und unendliche Geduld auch während schwieriger Phasen.

  \selectlanguage{english}
  \section*{Acknowledgements}
  First I want to thank my advisor Heribert Vollmer for his supervision, support and joint research during my time as a PhD student -- and for being the best kind of boss I can imagine.

Further I thank my colleague, co-author and office neighbour Johannes Ebbing for numerous fruitful discussions and joint results.
In addition, I want to thank my other co-authors for our joint research.
Also, I thank Juha Kontinen and Lauri Hella for my successful visits to Helsinki and Tampere.
Especially, I thank my colleague Arne Meier for many detailed suggestions for improving this thesis.

Furthermore I thank my parents for always having encouraged me in my abilities.
And above all I thank my wife Anna for her support and unlimited patience even during difficult phases.

  \vskip 0pt plus 2 fill
  
  \cleardoublepage
  
  \pagestyle{headings}
  \renewcommand{\chaptermark}[1]{\markboth{\chaptername\ \thechapter\ \ #1}{}}
  \renewcommand{\sectionmark}[1]{\markright{\thesection\ \ #1}}

  \selectlanguage{ngerman}
  \chapter*{Zusammenfassung}
  In dieser Arbeit wird (modale) Dependence-Logik untersucht. Diese wurde 2007 von Jouko V\"a\"an\"anen eingeführt und ist eine Erweiterung der Logik erster Stufe (bzw.~der Modallogik) um den Dependence-Operator $\dep$. Für Variablen erster Stufe (bzw.\allowbreak~propositionale Variablen) $x_{1},\dots,x_{n}$ bedeutet $\dep[x_{1},\dots,x_{n-1},{x_{n}}]$, dass der Wert von $x_n$ durch die Werte von $x_1,\dots,x_{n-1}$ bestimmt wird.

Wir betrachten Fragmente modaler Dependence-Logik, die durch Be\-schr{\"a}n\-ken der Menge erlaubter modaler und aussagenlogischer Operatoren definiert sind. Wir klassifizieren diese Fragmente in Bezug auf die Komplexität ihres Erfüllbarkeits- und Model-Checking-Problems. Für Erfüllbarkeit erhalten wir Komplexitätsgrade von $\P$ über $\NP$, $\SigmaP{2}$ und $\P\SPACE$ bis hin zu $\NEXP$, während wir die Fragmente für Model-Checking nur in Bezug auf ihre Praktikabilität klassifizieren, d.h.~wir zeigen entweder $\NP$-Vollständigkeit oder Enthaltensein in $\P$.

Anschließend untersuchen wir die Erweiterung modaler Dependence-Logik um die sogenannte intuitionistische Implikation. Für diese Erweiterung klassifizieren wir wiederum die Fragmente in Bezug auf die Komplexität ihres Model-Checking-Problems. Hierbei erhalten wir Komplexitätsgrade von $\P$ über $\NP$ und $\co\NP$ bis hin zu $\P\SPACE$.

Zuletzt analysieren wir noch erststufige Dependence-Logik, Inde\-pen\-dence-friendly-Logik und deren Zwei-Variablen-Fragmente. Wir beweisen, dass das Erfüll\-bar\-keitsproblem für Zwei-Variablen-Dependence-Logik $\NEXP$-voll\-stän\-dig ist, während es für Zwei-Variablen-Independence-friendly-Logik unentscheidbar ist; und benutzen dieses Resultat, um zu beweisen, dass letztere Logik zudem ausdrucksstärker als die vorherige ist.

\bigskip
\bigskip
\noindent{\bfseries Schlagworte:} \Schlagworte

  \cleardoublepage
  
  \selectlanguage{english}
  \chapter*{Abstract}
  In this thesis (modal) dependence logic is investigated. It was introduced in 2007 by Jouko V\"a\"an\"anen as an extension of first-order (resp.~modal) logic by the dependence operator $\dep$. For first-order (resp.~propositional) variables $x_{1},\dots,x_{n}$, $\dep[x_{1},\dots,x_{n-1},{x_{n}}]$ intuitively states that the value of $x_{n}$ is determined by those of $x_{1},\dots,x_{n-1}$.

We consider fragments of modal dependence logic obtained by restricting the set of allowed modal and propositional connectives.
We classify these fragments with respect to the complexity of their satisfiability and model-checking problems. For satisfiability we obtain complexity degrees from $\P$ over $\NP$, $\SigmaP{2}$ and $\P\SPACE$ up to $\NEXP$, while for model-checking we only classify the fragments with respect to their tractability, \ie we either show $\NP$-completeness or containment in $\P$.

We then study the extension of modal dependence logic by intuitionistic implication. For this extension we again classify the complexity of the model-checking problem for its fragments. Here we obtain complexity degrees from $\P$ over $\NP$ and $\co\NP$ up to $\P\SPACE$.

Finally, we analyze first-order dependence logic, independence-friendly logic and their two-variable fragments. We prove that satisfiability for two-variable dependence logic is $\NEXP$-complete, whereas for two-variable inde\-pen\-dence-friendly logic it is undecidable; and use this to prove that the latter is also more expressive than the former.

\bigskip
\bigskip
\noindent{\bfseries Keywords}: \Keywords

\bigskip
\bigskip
\noindent{\bfseries ACM Subject Classifiers}: \SubjectClassifiers

  \cleardoublepage
  
  \tableofcontents
  
  \listoffigures
  
  \listoftables
  
  \mainmatter
  
  \chapter{Introduction}\label{chap:intro}
\section{Why do we need dependence?}
One of the purposes of formal logic is to aid in the translation of facts from natural language into mathematical language. Consider, for example, the following sentence:
\begin{center}
\qte{\emph{There are evil hackers.}}
\end{center}
This could be translated to the first-order formula
\[\exists x\, \big(\mathrm{evil}(x)\wedge \mathrm{hacker}(x)\big).\]
Now one could make the claim
\begin{center}
\qte{\emph{Every hacker named Eve or Mallory is evil.}}
\end{center}
And again we can translate to first-order logic:
\[\forall x\,\Big( \big(\mathrm{hacker}(x) \wedge \big(\mathrm{Eve}(x)\sor\mathrm{Mallory}(x)\big)\big) \imp \mathrm{evil}(x) \Big)\]
A more general claim of this sort is the following:
\begin{equation}\label{evil degree}
\textnormal{\qte{\emph{The degree of evilness of someone is determined by his/her name.}}}
\end{equation}
But now we have a problem when translating to first-order logic since the concept of some arbitrary \cim{dependence} or \cim{determination} is not expressible in common first-order logic. We can state a fully specified dependence as above where we know that \emph{hacker} together with \emph{Eve or Mallory} implies \emph{evil}. But when we only know that there is a dependence between variables or terms but do not know its details, \ie in this case which names belong to which grades of evilness, then first-order logic is not expressive enough.

\bigskip
In first-order logic, dependence between variables is determined by the order in which the variables are quantified -- and therefore is always linear. For example, when using game 
theoretic semantics to evaluate the formula
\[ 
\forall x_0\exists x_1\forall x_2\exists x_3\, \phi,
\] 
the choice for $x_1$ depends on the value of $x_0$ and the choice for $x_3$ depends on the value of both universally quantified variables $x_0$  and  $x_2$.
What we need is the possibility to express non-linear dependencies between variables -- which cannot be expressed in first-order logic (\fo)\index{FO@$\fo$}.

The first step in this direction was taken by Henkin \cite{he61} with his \cim[partially-ordered quantifier]{partially-ordered quantifiers}
\begin{equation}\label{poc}
\left(\begin{array}{cc}\forall x_0& \exists x_1\\ \forall
x_2&\exists x_3\end{array}\right)\phi,
\end{equation}
where $x_1$ depends only on $x_0$ and  $x_3$ depends only on $x_2$.

The second step was taken by Hintikka and Sandu \cite{hisa89,hi96} who introduced \cim{independence-friendly logic} (\ifl)\index{IF@$\ifl$} which extends \fo in terms of \emph{slashed quantifiers}\index{slashed quantifier}.
For example, in 
\[\forall x_0\exists x_1\forall x_2\exists x_3/\forall x_0\, \phi\]
the quantifier $\exists x_3/\forall x_0$ means that
$x_3$ is \emph{independent}\index{independence} of $x_0$ in the sense that a choice for
the value of $x_3$ should not depend on what the value of $x_0$
is. The semantics of \ifl was first formulated in game theoretic terms and
therefore \ifl can be regarded as a game theoretically motivated generalization of \fo.
Whereas the semantic game of \fo is a game of perfect information, the game of \ifl is a game of imperfect information.
Later, the team semantics of \ifl, also used in this thesis, was introduced by Hodges \cite{ho97,ho97b}.

\emph{Dependence logic}\index{dependence logic} (\df)\index{D@$\df$}, introduced by V\"a\"an\"anen \cite{va07},
is inspired by \ifl-logic, but here the 
dependencies between variables are written in terms of
atomic dependence formulas. For example, the partially-ordered quantifier \eqref{poc} can be expressed in dependence logic as
\[ 
\forall x_0\exists x_1\forall x_2\exists
x_3\,(\dep[x_2,x_3]\wedge\phi).
\]\index{$\dep$}
The  atomic formula $\dep[x_2,x_3]$ has
the explicit meaning that $x_3$ is completely determined by $x_2$ and nothing else. For $x_1$ on the other hand we do not need a dependence atom since it is quantified before $x_2$ and therefore automatically completely determined by $x_0$ alone.
In general, functional dependence of the value of a term $t_{n}$ from the values of the terms $t_{1},\dots,t_{n-1}$, denoted by $\dep[t_1,\dots,t_{n-1},t_n]$, states that there is a function, say $f$, such that $t_{n}=f(t_{1},\dots,t_{n-1})$ holds, \ie the value of $t_{n}$ is completely determined by those of $t_{1},\dots,t_{n-1}$.

\medskip
First-order logic with partially-ordered quantifiers, independence-friendly logic and dependence logic are all \emph{conservative extensions}\index{conservative extension} of first-order logic, \ie they agree with $\fo$ on sentences which syntactically are $\fo$-sentences, and are expressively equivalent to \emph{existential second-order logic} (\eso)\index{second-order logic!existential}\index{ESO@$\eso$} \cite{en70,wa70,ho97b,va07}.

Hence, reconsidering our hacker example \eqref{evil degree}, we can formalize it in any of the above extensions of first-order logic.
In dependence logic we can write it as
\[\forall x \exists y \exists z\,\big(y=\mathrm{evilness}(x) \wedge z=\mathrm{name}(x) \wedge \dep[z,y]\big),\]
in \ifl as
\[\forall x \exists z \exists y/\forall x\,\big(y=\mathrm{evilness}(x) \wedge z=\mathrm{name}(x)\big)\]
and with a partially-ordered quantifier as
\[\begin{array}{rcl}
\left(\begin{array}{ccc}\forall x& \exists y&\exists z\\
\forall x'&\exists y'&\exists z' \end{array}\right)&&
\big(y=\mathrm{evilness}(x) \wedge z=\mathrm{name}(x)\\
&\wedge& y'=\mathrm{evilness}(x') \wedge z'=\mathrm{name}(x')\\
&\wedge&  z=z'\; \imp\; y=y'\big).
\end{array}\]

Note that although the statement can be expressed in all three logics its formulation is not equally intuitive in each one of them. With partially-ordered quantifiers we have to introduce additional variables (which are later identified with one another) and in \ifl we have to arrange the quantifiers in a specific order whereas in \df we do not have any of such restrictions.

\section{The nature of dependence}
Of course, dependence does not manifest itself in a single world, play, event or observation. Important for a dependence to make sense is a collection of such worlds,  plays, events or observations. These collections are called \emph{teams}\index{team}.
They are the basic objects in the definition of the semantics of dependence logic.
A team can be a set of rounds in a game. Then $\dep[x_{1},\dots,x_{n-1},x_{n}]$ intuitively states that in each round move $x_{n}$ is determined by moves $x_{1},\dots,x_{n-1}$.
A team can be a database. Then $\dep[x_{1},\dots,x_{n-1},x_{n}]$ intuitively states that in each line the value of attribute $x_{n}$ is determined by the values of attributes $x_{1},\dots,x_{n-1}$, \ie $x_{n}$ is functionally determined by $x_{1},\dots,x_{n-1}$.

More formally, in first-order logic a team $X$ is a set of assignments and\linebreak[1] $\dep[x_{1},\dots,\allowbreak x_{n-1},x_{n}]$ states that in each assignment the value of $x_{n}$ is determined by the values of $x_{1},\dots,x_{n-1}$, \ie there is a function $f\colon A^{n-1}\to A$ (where $A$ is the universe of a first-order structure) such that for all $s\in X$ it holds that $s(x_n)=f(s(x_1),\dots,s(x_{n-1}))$. Dependence logic (\df) is defined by simply adding these dependence atoms to usual first-order logic \cite{va07} and is a conservative extension of the latter.

In modal logic a team is a set of worlds in a Kripke structure and $\dep[p_{1},\dots,\allowbreak p_{n-1},p_{n}]$ states that in each of these worlds the value of the propositional variable $p_{n}$ is determined by the values of $p_{1},\dots,p_{n-1}$, \ie there is a Boolean function $f\colon \{\true,\false\}^{n-1}\to\{\true,\false\}$ that determines the value of $p_n$ from the values of $p_1,\dots,p_{n-1}$ for all worlds in the team.
Modal dependence logic (\MDL) is the conservative extension of modal logic defined by introducing dependence atoms to modal logic \cite{va08}.

\section{Computational complexity}
In this thesis we will investigate fragments of first-order as well as modal dependence logic from a computational point of view, \ie we will determine the difficulty of computational problems involving logical formulas. Naturally, the question that arises is: What, precisely, do we mean by \emph{difficulty}?
To answer this question we use methods from the field of computational complexity which will be introduced in the following.

\smallskip
When designing an algorithm to solve a \ci{problem}, it is, of course, important whether your algorithm needs to run a minute or a year to solve the problem. If you have only one problem to solve you can simply specify the ressources needed by the algorithm by noting its runtime\index{time} and the memory space\index{space} it uses.
But now imagine that you are designing an algorithm to solve a type of problems, \eg an algorithm to compute the fastest car route from city A to city B. Now the time and space needed by the algorithm depend on the distance between the two cities, the number of citites in between them and many other things.

One way to specify the time and space of a generic algorithm is to express the worst-case runtime and memory space needed by the algorithm as a function of the size of the given problem. Here, \emph{size} can, for example, mean the number of nodes in a graph, the number of rows in a database or the number of streets on a map that includes both city A and city B.
The \cim{computational complexity} of a problem type can then be specified by finding an optimal algorithm to solve it and examining the algorithm's worst-case time and space needs as a function of the input size.

The probably most important complexity class definable in this way is the class \P of all problems for which there is an algorithm solving them whose runtime is bounded by a polynomial in the input size. Another important class is \EXP which is defined analogous to \P but instead of a polynomial function there only has to be an exponential function for the runtime bound.

For these two classes (as for many other pairs of classes) there is a fundamental difference between the complexity of a problem A in class \P and a problem B in class \EXP.
Assume, for example, that for a given input size the optimal algorithms for both problems can compute a solution in one minute. Now, if we increase the input size by one unit, the algorithms for the problems will, for example, both take two minutes. But if we further increase the input size by one more unit then problem A will typically be solved in three minutes while problem B will take four minutes. And by each further incrementation of the input size, problem A will take one more minute while the time needed to solve problem B will double. So the latter has a fundamentally worse asymptotic runtime than the former.

This example demonstrates that a single complexity class typically covers a wide range of concrete runtime or space usage functions and that the differences between two classes are usually of a fundamental nature which cannot be eliminated by simply using faster computers or optimized algorithms.

\section{Modal dependence logic}\index{dependence logic!modal}
%
V\"a\"an\"anen \cite{va08} introduced both an inductive semantics and an equivalent game-theoretic semantics for modal dependence logic. Sevenster \cite{se09} proved that on singleton teams, \ie teams which contain only one world, there is a translation from $\MDL$\index{MDL@$\MDL$} to usual modal logic while on arbitrary teams there is no such translation. Sevenster also initiated the complexity-theoretic study of modal dependence logic by proving that the satisfiability problem of \MDL is complete for the class \NEXP of all problems decidable by a nondeterministic Turing machine in exponential time.

In this thesis, we continue the work of Sevenster by presenting a more thorough study on complexity questions related to \MDL. We will systematically investigate fragments of \MDL by restricting the propositional and modal operators allowed in the language.
This method of classifying the complexity of logic related problems goes back to Lewis who used this method for the satisfiability problem of propositional logic \cite{le79}. Recently it was, for example, used by Hemaspaandra et al.~for the satisfiability problem of modal logic \cite{he05,hescsc10} and by Meier et al.~for the satisfiability problem of temporal logic \cite{memuthvo09}.
The motivation for this approach is that by a systematic study of all fragments of a logic one might find a fragment with efficient algorithms but still high enough expressivity to be useful in practice.
On the other hand, this systematic approach usually leads to insights into the sources of hardness, \ie the exact components of the logic that make satisfiability, model checking, etc.~hard.

We follow the same approach here. We consider all subsets of modal operators $\Box,\Diamond$ and propositional operators $\wedge$, $\vee$, $\aneg$ (atomic negation), $\true, \false$ (the Boolean constants true and false), \ie we study exactly those operators considered by V\"a\"an\"anen \cite{va08}, and each time examine the satisfiability problem and the model checking problem of \MDL restricted to formulas built only by operators from the chosen subset. Here, by \cim{satisfiability} we mean the problem where a formula is given and the task is to decide whether there is a Kripke structure satisfying the formula. By \cim{model checking} we mean the problem where a formula, a Kripke structure and a team are given and the task is to decide whether the Kripke structure and the team satisfy the formula.

We also extend the logical language of \cite{va08} by adding classical disjunction\index{$\nor$}\index{disjunction!classical}\index{classical disjunction} (denoted here by $\nor$) besides the usual dependence disjunction\index{disjunction!dependence}\index{dependence!disjunction}. Connective $\nor$ was already considered by Sevenster (he denoted it by $\bullet$),
but not from a computational point of view.
Furthermore it seems natural to not only restrict modal and propositional operators but to also impose restrictions on the dependence atoms.
We consider one such restriction, where we limit the arity of the dependence atoms, \ie the number $n$ of variables $p_1,\dots,p_n$ by which $q$ has to be determined to satisfy the formula $\dep[p_1,\dots,p_n,q]$, to a fixed upper bound $k\geq 0$ (the logic is then denoted by \MDLk\index{MDLk@$\MDLk$}\index{dependence logic!modal!bounded arity}\index{arity!bounded}).

We include both, the classical disjunction and the arity restricted dependence atoms, into the set of operators which we systematically restrict.
Then we analyze the complexity of the satisfiability problem and the model checking problem of \MDL and \MDLk for all subsets of operators studied by V\"a\"an\"anen and Sevenster.

\section{Intuitionistic dependence logic}\label{sec:idl intro}
In classical logic (modal as well as first-order and propositional) there is a well-defined canonical notion of \cim{implication} which is defined by interpreting $\phi \imp \psi$ as a shortcut for $\neg \phi \sor \psi$. In dependence logic, however, there is no such canonical notion. The reason for this is that $\neg \phi$ is not well-defined for an arbitrary formula $\phi$ since negation is only atomic in dependence logic. It is, of course, possible to define arbitrary negation $\neg \phi$ to mean the dual formula $\dual{\phi}$ obtained from $\neg \phi$ by pushing the negation down to the atomic level by use of de Morgan's law.

However, when using this syntactically-defined non-atomic negation the obtained notion of implication misses some of the usually desired properties. For example, it is possible that in a team neither $\phi$ nor $\psi$ holds but neither does the formula $\neg \phi \sor \psi$.
Therefore Abramsky and V\"a\"an\"anen \cite{abva08} have investigated two other notions of implication, namely \emph{linear implication}\index{implication!linear} $\limp$ and \emph{intuitionistic implication}\index{implication!intuitionistic} $\imp$. The latter satisfies axioms of intuitionistic logic and is defined by $\phi \imp \psi$ being satisfied in a team iff in all subteams where $\phi$ is satisfied it also holds that $\psi$ is satisfied.

Yang \cite{ya11} proved that first-order dependence logic extended by intuitionistic implication has the same expressive power as full second-order logic on the level of sentences. Additionally, this still holds for the fragment called \cim{intuitionistic dependence logic} (\IDL)\index{IDL@$\IDL$}\index{dependence logic!intuitionistic|see{intuitionistic dependence logic}} which only uses quantifiers, $\wedge$, $\nor$, $\neg$, $\imp$ and dependence atoms, \ie it misses $\sor$.



In this thesis we consider modal dependence logic extended with intuitionistic implication, called \emph{modal intuitionistic dependence logic}\index{intuitionistic dependence logic!modal} (\MIDL)\index{MIDL@$\MIDL$}. We follow the same approach as for \MDL and investigate the computational complexity of the model checking problem of fragments of \MIDL obtained by using only subsets of the operators.
%
Amongst others we discuss a natural variant of first-order \IDL, namely \emph{propositional intuitionistic dependence logic}\index{intuitionistic dependence logic!propositional}\index{PIDL@$\PIDL$} (\PIDL) which in our setting can be identified with  \MIDL without modalities and without disjunction $\sor$. 


\section{Two-variable logic}

The satisfiability problem of first-order logic \fo was shown to be undecidable in \cite{ch36,tu36} and, ever since, logicians have been searching for decidable fragments of \fo.
Henkin \cite{he67} was the first to consider the logics $\fo^k$, \ie the fragments of first-order logic with only $k$ variables.\index{two-variable logic}\index{first-order logic!$k$-variable}\index{FOk@$\fo^k$}
The fragments $\fo^k$ for $k\ge 3$ were easily seen to be undecidable but the case for $k=2$ remained open.
Earlier, Scott \cite{sc62} had shown that \fotwo without equality is decidable and then Mortimer \cite{mo75} extended the result to \fotwo with equality and showed that every satisfiable \fotwo formula has a model whose size is at most doubly exponential in the length of the formula. His result established that the satisfiability and finite satisfiability problems of \fotwo are contained in \NEEXP.
Finally, Gr\"adel, Kolaitis and Vardi \cite{grkova97} improved the result of Mortimer by establishing that every satisfiable \fotwo formula  has a model of  exponential size. Furthermore, they showed that the satisfiability problem of \fotwo is \NEXP-complete.

Since then, the decidability of the satisfiability problem of various extensions of \fotwo has been studied (\eg \cite{grotro97, grot99, etvawi02, kiot05}). One such extension, \foctwo,\index{FOCk@$\foc^k$}\index{first-order logic!$k$-variable!with counting} is acquired by extending \fotwo with counting quantifiers $\existk$\index{counting quantifier}\index{$\existk$}. The meaning of a formula of the form $\existk x\, \phi(x)$ is that $\phi(x)$ is satisfied by  at least $k$ distinct elements.
The satisfiability problem of the logic \foctwo
was shown to be decidable by Gr\"adel et al.~\cite{grotro97a} and shown to be in \NEEXP by Pacholski et al.~\cite{paszte97}. 
Finally, Pratt-Hartmann \cite{pr05} established that the problem is \NEXP-complete.

In this thesis we study the decidability and complexity of the satisfiability problems for the two-variable fragments of in\-de\-pendence-friendly logic and dependence logic. For this purpose we will use the result of Pratt-Hartmann to determine the complexity of satisfiability for two-variable dependence logic.


%
%
%
%
%
%

\section{Results}

In \cref{sec:mdl-sat} we completely classify the complexity of the satisfiability problem of all fragments of \MDL defined by taking only some of the operators and constants $\AX$, $\EX$, $\wedge$, $\sor$, $\nor$, $\neg$, $\true$, $\false$ and $\dep$.

One of our main and technically most involved contributions addresses a fragment that has been called \cim{Poor Man's Logic} in the literature on modal logic \cite{he01}, \ie the language without disjunction $\vee$. We show that for unbounded arity dependence logic we still have full complexity (\Cref{poor man dep complexity}%
),\linebreak[1] \ie we show that Poor Man's Dependence Logic is \NEXP-complete. If we also forbid negation then the complexity drops down to $\SigmaP{2}(=\NP^\NP)$, \ie Monotone Poor Man's Dependence Logic is $\SigmaP{2}$-complete (\Cref{poor man bullet complexity}, but note that we need $\nor$ here). And if we, instead of forbidding negation, restrict the logic to only contain dependence atoms of arity less or equal $k$ for a fixed $k\geq 3$ then the complexity drops to $\SigmaP{3}(=\NP^{\SigmaP{2}})$, \ie bounded arity Poor Man's Dependence Logic is $\SigmaP{3}$-complete (\Cref{bounded dep concrete}b).

All of our complexity results are summarized in \cref{mdl sat results} for dependence atoms of unbounded arity and in \cref{mdl sat results bounded} for dependence atoms whose arity is bounded by a fixed $k\geq 3$.

\medskip
In \cref{sec:mdl-mc} we classify the complexity of the model checking problem of fragments of \MDL with unbounded as well as bounded arity dependence atoms.
For plain modal logic this problem is solvable in \P as shown by Clarke et al.~\cite{clemsi86}. A detailed complexity classification for the model checking problem over fragments of modal logic was given by Beyersdorff et al.~\cite{bememuscthvo11} (who investigate the temporal logic \CTL which contains plain modal logic as a special case).

In the case of \MDL it turns out that model checking is \NP-complete in general and that this still holds for several seemingly quite weak fragments of \MDL, \eg the one without modalities or the one where nothing except dependence atoms and $\EX$ is allowed%
.
Interestingly, this also holds for the case where only both disjunctions $\sor$ and $\nor$ are allowed and not even dependence atoms occur%
.

We are able to determine the tractability of each fragment except the one where formulas are built from atomic propositions and unbounded dependence atoms only by dependence disjunction and negation%
.
In each of the other cases we either show \NP-completeness or show that the model checking problem admits an efficient (deterministic polynomial time) solution.

For the restriction to bounded arity dependence atoms, model checking remains \NP-complete in general but for the fragment with only the $\EX$ operator allowed this does not hold any more%
. In this case either $\wedge$
 or $\vee$
 is needed to still get \NP-hardness.

In \cref{mdl mc results} we list all of our complexity results 
for the cases with unbounded arity dependence atoms and in \cref{mdl mc results bounded} for the cases with bounded arity.

\medskip
In \cref{chap:midl} we extend our classification from \cref{sec:mdl-mc} to include intuitionistic implication into the set of operators, \ie we study the complexity of model checking for \MIDL and its fragments. We prove that model checking for \MIDL in general is $\P\SPACE$-complete but for the \MIDL fragment where formulas are not allowed to contain $\EX$ or $\sor$ the problem turns out to be \co\NP-complete. In particular, model checking for \PIDL is \co\NP-complete.
Our detailed results are listed in \cref{midl mc results}.

\medskip
In \cref{chap:dtwo} we examine the expressiveness, satisfiability and finite satisfiability for bounded-variable \ifl-logic and dependence logic.
We show that there is an effective translation from \dtwo to \iftwo (\Cref{dtoif}) and from \iftwo to $\df^3$ (\Cref{iftod}).
We also show that \iftwo is strictly more expressive than \dtwo (\Cref{d less than if}).
This result is a by-product of our proof 
that the satisfiability problem of \iftwo is undecidable (\Cref{iftwo sat complexity} shows \pizeroone-completeness). The proof can be adapted to the context of finite satisfiability, \ie the problem of determining for a given formula $\phi$ whether there is a finite structure $\mA$ such that $\mA\models \phi$ holds (\Cref{iftwo finsat complexity} shows \sigmazeroone-completeness). The undecidability proofs are based on tiling arguments.
Finally, 
we study the decidability of the satisfiability and finite satisfiability problems of \dtwo. 
For this purpose we reduce the problems to the (finite) satisfiability problem of \foctwo (\Cref{DtoESO}) and thereby show 
that they are \NEXP-complete (\Cref{dtwo nexptime}).

\section{Publications}
\Cref{sec:mdl-sat} is based on \cite{lovo10} -- although \cref{sec:bounded sat} about the bounded arity cases is new.
\Cref{sec:mdl-mc} is based on \cite{eblo12} -- while the investigation of the $\nor$ operator is new (\Cref{sec:mdl-mc-nor} and new version of \Cref{diamond-wedge-bounded}).
\Cref{chap:midl} contains unpublished work with Fan Yang and Johannes Ebbing.
Finally, \cref{chap:dtwo} is based on \cite{kokulovi11} -- while \cref{bordered tiling sigmazeroone} and the proofs in \cref{iffinsatsection} are new.

\chapter{Preliminaries}\label{chap:prelim}
\section{Predicate logic}\label{sec:predicate logic}
We assume familiarity with the basic notions of first- and second-order logic and thus will not explain what a vocabulary or a structure is. The interested reader may refer to any introductory logic textbook, \eg \cite{ebflth94}. We will, however, restate the semantics of first-order logic in terms of team semantics since this will be used in the next \lcnamecref{chap:dep} to define the semantics of (first-order) dependence logic.

First, let us quickly recall the syntax of first-order logic.
\begin{definition}[Syntax of first-order logic]\label{def:fo syntax}\index{first-order logic}\index{FO@$\fo$|textbf}
Let $\tau$ be a first-order vocabulary, $t_1,\allowbreak{}\dots,\allowbreak{}t_n$ arbitrary $\tau$-terms, $R$ an arbitrary relation symbol in $\tau$ (of arity $n$) and $x$ an arbitrary first-order variable.
Then \fo is the set of all formulas built by the rules
\[\begin{array}{rcl}
\phi & \ddfn & \true \;\mid\; \false \;\mid\; R(t_1,\dots,t_n) \;\mid\; \neg R(t_1,\dots,t_n) \;\mid\; t_1=t_2 \;\mid\; \neg t_1=t_2 \;\mid\\
&& \phi \wedge \phi \;\mid\; \phi \vee \phi \;\mid\; \forall x \phi \;\mid\; \exists x \phi.
\end{array}\]
\end{definition}

Note that \wlg we only define formulas in \ci{negation normal form}. The rationale for this will become clear in the next \lcnamecref{chap:dep} when we see that for dependence logic we have to use negation normal form since negation is only properly defined for atomic formulas.

%

To define team semantics we first introduce the concept of a \emph{team}.
Let $\mA$ be a \cim{model} (or \cim[structure!first-order]{structure}) with domain $A$ (in the future, we will assume that if $\mA$ (resp.~$\mB$, $\mC$,\dots) is a model then $A$ (resp.~$B$, $C$,\dots) is its domain). \cim[assignment]{Assignments} over $\mA$ are finite functions that map variables to elements of $A$.
The value of a term $t$ in an assignment $s$ is denoted by $t^{\mA}\langle s\rangle$\index{$t^{\mA}\langle s\rangle$}.
If $s$ is an assignment, $x$ is a variable and $a\in A$ then $s(a/x)$ denotes the assignment (with the domain $\dom{s}\cup \{x\}$)  which agrees with $s$ everywhere except that it maps $x$ to $a$.

Let $\{x_1,\ldots,x_k\}$ be a finite (possibly empty) set  of
variables. 
A \emph{team}\index{team|textbf} $X$ of $A$ with domain
$\dom{X}=\{x_1,\ldots,x_k\}$\index{dom@$\dom{\cdot}$} is a set of assignments from the variables
$\{x_1,\ldots,x_k\}$ into $A$.
We denote by $\rel{X}$\index{rel@$\rel{\cdot}$} the $k$-ary relation of $A$ corresponding to $X$:
\[\rel{X} \dfn \{(s(x_1),\ldots,s(x_k)) \mid s\in X \}.\]
If $X$ is a team of $A$ and $F\colon X\to A$, we use \ci[$X(F/x)$] to denote the team
\[\set{s(F(s)/x)}{$s\in X$}\]
and \ci[$X(A/x)$] for the team
\[\set{s (a/x)}{$s\in X$ and $a\in A$}.\]

If $\phi$ and $\theta$ are formulas and $\psi$ is a subformula of $\phi$ then we define $\phi(\theta / \psi)$ to be the formula generated from $\phi$ by \emph{substituting} all occurrences of $\psi$ with $\theta$.\index{substitution}\index{$\phi(\theta / \psi)$}

\begin{definition}[Team semantics of \fo]\label{def:fo semantics}\index{semantics!team}\index{team!semantics}
Let $\mA$ be a model and let $X$ be a team of $A$. The \cim{satisfaction relation}
$\mA\models_X \phi$ is defined as follows:
\begin{itemize}
\item If $\phi$ is an atomic formula then $\mA\models_X \phi$ iff for all $s\in X$: $\mA,s\models\phi$ (in the usual single-assignment based semantics).
\item If $\psi$ is atomic then $\mA\models_X \neg \psi$ iff for all $s\in X$: $\mA,s\not\models\psi$.
\item $\mA\models_X \psi \wedge \phi$ iff $\mA\models _X \psi$ and $\mA\models _X \phi$.
\item $\mA\models_X \psi \vee \phi$ iff there exist teams $Y$ and $Z$ such that $X=Y\dcup Z$, $\mA\models_Y \psi$ and $\mA\models _Z \phi$.

\item   $\mA \models_X \exists x \psi$ iff $\mA \models _{X(F/x)} \psi$ for some $F\colon X\to A$.

\item $\mA \models_X \forall x\psi$ iff $\mA \models _{X(A/x)} \psi$.
\end{itemize}

Above, we assume that the domain of $X$ contains $\frr[\phi]$\index{free@$\fr{\cdot}$}, \ie it contains all variables occurring in $\phi$ outside the scope of a quantifier. Finally, a sentence $\phi$ is true in a model $\mA$  ($\mA\models \phi$)  iff  $\mA\models _{\{\emptyset\}} \phi$ holds.
\end{definition}

\begin{remark}\label{fo single equals team}
Note that this team semantics coincides with the usual single-assign\-ment based semantics for first-order logic, \ie for all structures $\mA$, formulas $\phi\in\fo$ and assignments $s\colon \fr{\phi}\to A$ it holds that
\[\mA\models_{\{s\}} \phi  \quad\text{iff}\quad  \mA,s\models_{\fo} \phi.\]
 \end{remark}

For second-order logic we assume the usual single-assignment based semantics and only recall the syntax here.

\begin{definition}[Syntax of second-order logic]\label{def:so syntax}\index{second-order logic}\index{SO@$\so$}
The syntax of $\so$ extends the syntax of $\fo$ by the rules
\[\phi \ \ddfn\ \exists S \phi \;\mid\; \forall S\phi, \]
where $S$ is a second-order variable, \ie it is interpreted by a relation (with an a priori given arity) and used in the inner formula as if it belonged to the vocabulary.
\end{definition}



The following fragment of \so will be of special interest.
\begin{definition}[$\eso$]
\label{def:so fragments}
 \cim[second-order logic!existential|textbf]{Existential second-order logic (\eso)}\index{ESO@$\eso$|textbf} consists of all \so-for\-mu\-las that are of the form $\exists R_1\dots \exists R_n\phi$ for a \fo-formula $\phi$.
\end{definition}

\subsection{Fragments of first-order logic}

For first-order logic we also investigate fragments where only a fixed number of names for variables are allowed inside a formula.

\begin{definition}[$\fo^k$]\label{def:fok syntax}
Let $k>0$ and $\tau$ a relational vocabulary, \ie $\tau$ does not contain function or constant symbols.
Then \emph{$k$-variable first-order logic}\index{two-variable logic|textbf}\index{first-order logic!$k$-variable|textbf}\index{FOk@$\fo^k$|textbf} ($\fo^k$) is the set of all \fo-formulas that contain no more than $k$ distinct variables (we will usually use $x_1,\dots,x_k$ and in the case $k=2$ we use $x$ and $y$).
\end{definition}
Note that the above definition does not forbid requantification of variables, \eg $\exists x \exists y (Pxy \wedge \exists x Qxy)$ is in \fotwo.

In common first-order logic there are only two quantifiers, one asking about the existence of \emph{at least one} element with a certain property and the other one, the dual, stating that the property holds for all elements, \ie its complement holds for \emph{less than one} element.

A natural generalization of this is the introduction of \cim[counting quantifier|textbf]{counting quantifiers} $\existk$\index{$\existk$|textbf} asking about the existence of \emph{at least $k$} pairwise-different elements. It is possible to define $\existk$ for any $k\in\N$ with the ordinary $\exists$ quantifier by the translation
\[\existk x \,\phi(x) \quad\mapsto\quad \exists x_1\dots\exists x_k\big(\bigwedge_{\substack{i,j\in\{1,\dots,k\},\\i\neq j}} x_i\neq x_j\,\wedge \bigwedge_{i=1}^k \phi(x_i)\big).\]

In the case of $\fo^k$ it is, however, no longer possible to define $\existk[k+1]$. Therefore we will investigate the following extension of $\fo^k$.
\begin{definition}[$\foc^k$]\label{def:fock}
$k$-variable first-order logic with counting\index{FOCk@$\foc^k$|textbf}\index{first-order logic!$k$-variable!with counting|textbf} ($\foc^k$) is the extension of $\fo^k$ obtained by allowing the counting quantifiers $\existk[\ell]$ (for all $\ell\in\N$) as well as their negations $\exists^{< \ell}$.
\end{definition}

For an introduction into bounded-variable first-order logic with counting we refer to \cite{ot97}.

\section{Modal logic}
We will only briefly restate the definition of modal logic here. For a more in-depth introduction the reader may refer to \cite{blrive02}.

\begin{definition}[Syntax of modal logic]\label{modal logic}\index{modal logic}\index{ML@$\ML$}
Let $\ap$\index{atomic proposition}\index{prop@$\ap$} be an arbitrary countable set (of atomic propositions). Then \ML is the set of all formulas $\phi$ built by the rules
\[
\phi \ddfn \true \;\mid\; \false \;\mid\; p \;\mid\; \neg p  
\;\mid\; \phi\wedge\phi \;\mid\;  \phi\vee \phi \;\mid\;  \AX\phi\; \;\mid\; \EX\phi,
\]
where $p \in \ap$.

\CL (\cim{classical logic})\index{CL@$\CL$} is the set of all formulas of \ML that neither contain $\AX$ nor $\EX$ operators, \ie \CL is the set of all purely-propositional Boolean formulas.
\end{definition}
We sometimes write $\AX^k$ (resp.~$\EX^k$) for $\underbrace{\AX\,\AX\dots\AX}_{k \text{ times}}$ (resp.~$\underbrace{\EX\EX\dots\EX}_{k \text{ times}}$).
Note that we have defined $\true$ and $\false$ as parts of the language although we could equivalently have defined them as abbreviations for $p \sor \neg p$ and $p \wedge \neg p$, respectively. The rationale for this is that we will forbid negation and/or disjunction in parts of \cref{sec:mdl-sat} but still want to allow the constants.

Note that we define negation\index{negation normal form} 
to be only atomic, \ie it is only defined for atomic propositions.
Although in the literature negation is usually defined for arbitrary formulas, our definition is in fact not a restriction of this general case. This is due to the fact that arbitrary negation can be defined in terms of dual formulas by
\[\neg \phi \dfn \dual{\phi},\]
where dual formulas are defined as follows.
\begin{definition}[Dual formulas]\label{dual modal logic formulas}
Let $\phi \in \ML$. Then the \cim[dual formula]{dual} of a formula  is defined by the following translation $\phi \mapsto \dual{\phi}$:\index{$\dual{\cdot}$}
\[\begin{array}{rcl}
\true                   &\mapsto&       \false\\
\false                  &\mapsto&       \true\\
p                       &\mapsto&       \neg p\\
\neg p                  &\mapsto&       p\\
\psi\wedge\theta        &\mapsto&       \dual{\psi}\vee\dual{\theta}\\
\psi\vee\theta          &\mapsto&       \dual{\psi}\wedge\dual{\theta}\\
\AX\psi                 &\mapsto&       \EX\dual\psi\\
\EX\psi                 &\mapsto&       \AX\dual\psi.
\end{array}\]

Note that an analogous defintion of dual formulas can be given for first-order formulas.
\end{definition}

We will define the semantics of \ML via a translation to \fotwo.
\begin{definition}\label{def:ml to fotwo}
We define a translation $\phi \mapsto \fom{\phi}(x)$ from \ML to \fotwo as follows:
\[\begin{array}{rcl}
\true                   &\mapsto&       \true\\
\false                  &\mapsto&       \false\\
p                       &\mapsto&       P(x)\\
\neg p                  &\mapsto&       \neg P(x)\\
\psi\wedge\theta        &\mapsto&       \fom{\psi}(x) \wedge \fom{\theta}(x)\\
\psi\vee\theta          &\mapsto&       \fom{\psi}(x) \vee \fom{\theta}(x)\\
\AX\psi                 &\mapsto&       \forall y\,\big(R(x,y)\imp \fom{\psi}(y)\big)\\
\EX\psi                 &\mapsto&       \exists y\,\big(R(x,y)\wedge \fom{\psi}(y)\big).
\end{array}\]
\end{definition}
Note that if $\phi$ is a \ML-formula over the atomic propositions $\ap = \{p_1,\allowbreak p_2,\allowbreak \dots\}$ then $\fom\phi$ is a \fotwo-formula over the vocabulary $\tau_{\ap} \dfn \{R,P_1,P_2,\dots\}$.

\begin{definition}[Team semantics of modal logic]\label{modal logic semantics}
Let $\phi\in\ML$, $\mA$ a $\tau_{\ap}$-struc\-ture and $T\subseteq A$ a \cim{team}.
Then we say that $\mA$ in $T$ \emph{satisfies} $\phi$ ($\mA,T \models \phi$) if and only if $\mA \models_X \fom\phi(x)$ where $X \dfn \set{\{x\mapsto a\}}{$a\in T$}$, \ie $X$ is the first-order team that has one unary assignment $\{x\mapsto a\}$ for each element $a$ in $T$.
\end{definition}

Note that in the above definition we still call $T$ a \emph{team} of $\mA$ although it is a set of elements rather than a set of assignments.

The structure $\mA$ is often expressed in the form of a \cim{Kripke structure}\index{structure!Kripke} (or \cim{frame}) -- which is nothing else then a first-order $\tau_{\ap}$-structure with a special notation for representing the unary relations.

\begin{definition}[Kripke structure]\label{def:kripke}
Let \ap be a countable set of \emph{atomic propositions}\index{atomic proposition}\index{prop@$\ap$}. Then a \emph{$\ap$-Kripke structure} is a tuple $K=(S,R,\pi)$ where $S$ is an arbitrary non-empty set of \cim[world]{worlds} or \cim[state]{states}, $R\subseteq S\times S$ is the \cim{accessibility relation} and $\pi\colon S \to \powerset{\ap}$ is the \cim{labeling function}.
\end{definition}
The corresponding first-order $\tau_{\ap}$-structure $\fom{K}$ has universe $S$ and contains the binary relation $R$ and the unary relations $P_1,P_2,\dots$ defined by $P_i \dfn \set{s}{$p_i\in\pi(s)$}$.

Note that this team semantics via translation to first-order logic coincides with the usual single-assignment non-translational semantics for modal logic, \ie for all Kripke structures $K=(S,R,\pi)$, formulas $\phi\in\MDL$ and worlds $s\in S$ it holds that
\[K,\{s\}\models \phi  \quad\text{iff}\quad  K,s\models_{\ML} \phi.\]

 For any team $T$ of a Kripke model $K$, we define
\[R(T)=\{s\in K \mid \exists s'\in T\text{ such that }(s',s)\in R\}.\]\index{$R(T)$}\index{R@$R(T)$}
In the case that $T=\{s\}$, we write $R(s)$ instead of $R(\{s\})$. Elements in $R(s)$ are called \cim[world!successor]{successors}\index{successor} of $s$. 
Furthermore we define
\[\sucteams[R]{T}=\{T'\mid T'\subseteq R(T)\text{ and for all $s\in T$ with $R(s)\neq \emptyset$: $R(s)\cap T'\neq \emptyset$}\}.\]\index{$\sucteams[R]{T}$}
If $R$ is clear from the context we sometimes just write $\sucteams{T}$. Elements in $\langle T\rangle$ are called successor teams\index{successor!team}.
Clearly, $R(T)\in \sucteams{T}$.


\section{Complexity theory}
We assume familiarity with the basic notions from computational complexity theory. An introduction to this can, for example, be found in \cite{arba09} or in \cite{pa94}.
We will just briefly recall the relevant complexity classes and results.

\subsection{Important classes}
\begin{definition}[$\P$, $\NP$, $\P\SPACE$, $\EXP$, $\NEXP$]\label{def:basic complexities}
\ci[P@\P]{\P} (resp.~\NP)\index{NP@\NP} is the class of all (decision) problems\index{problem!decision}, \ie \qte{yes-or-no-questions}, solvable by a deterministic (resp.~non-deterministic) Turing machine in polynomial time. Analogously, \ci[EXP@\EXP]{\EXP} (resp.~{\NEXP})\index{NEXP@\NEXP} contains all problems solvable (non-)deterministically in exponential time and \ci[PSPACE@$\P\SPACE$]{$\P\SPACE$} contains all problems solvable in polynomial space.
\end{definition}

\begin{definition}[$\PH$, $\SigmaP{k}$, $\PiP{k}$, $\AP$]\label{def:ph}
\index{AP@\AP}
The \cim{polynomial hierarchy} is defined as
\[\ci[PH@\PH]{\PH} \dfn \bigcup_{k=1}^\infty\,\SigmaP{k}\]\index{$\SigmaP{k}$}\index{$\PiP{k}$}
with
\[\begin{array}{l}
 \SigmaP{0} \dfn \PiP{0} \dfn \P,\\
 \SigmaP{k+1} \dfn \NP^{\SigmaP{k}}\text{ and }\\
 \PiP{k+1} \dfn \co\NP^{\SigmaP{k}},
\end{array}\]
where $\NP^\calC$ describes the class of all problems solvable by a non-deterministic \cim{oracle Turing machine}, using a problem from class $\calC$ as oracle, in polynomial time and
\[A\in\co\calC  \quad\text{iff}\quad  \widebar{A}\dfn\set{x}{$x\notin A$} \in \calC.\]\index{coC@$\co\calC$}\index{coNP@$\co\NP$}

Equivalently, $\PH$ can be defined as the class of all problems for which there is an \cim{alternating Turing machine} which decides the problem in polynomial running time and for which there is a $k\in \N$ such that for all inputs $x$ the machine does not alternate between guessing modes more than $k$ times when started with input $x$. Note that the maximum number of alternations, $k\in \N$, is arbitrary but fixed for the problem and thus must not depend on the input given to the machine. If this restriction is dropped we get the class $\AP$ of all problems for which there is an alternating Turing machine deciding the problem in polynomial running time. In \cite{chkost81} it was shown that $\AP=\P\SPACE$.
\end{definition}

\begin{definition}[$\sigmazeroone$, $\pizeroone$, decidable]\label{def:decidable}
$\sigmazeroone$\index{$\sigmazeroone$} is the class of all decision problems $A$ for which there is a Turing machine that, given input $x$, holds in an accepting state if $x\in A$ and runs infinitely long if $x\notin A$. $\pizeroone$\index{$\pizeroone$} is defined as $\co\sigmazeroone$.

A problem is said to be \cim{decidable} iff there is a Turing machine that solves it, \ie for all inputs it holds with the correct answer after a finite number of computation steps.
\end{definition}

\Cref{fig:complexity inclusions} shows an inclusion diagram of all of the above complexity classes.

\begin{figure}[ht]
\begin{center}
\hspace*{2cm}\begin{tikzpicture}[auto, x=4em, y=4em]
\node                (p)      at    (5,0.2)                   {$\P$};
\node               (np)      at    (4,1)                   {$\NP=\SigmaP{1}$};
\node              (cnp)      at    (6,1)                   {$\co\NP=\PiP{1}$};
\node               (sp)      at    (4,2)                   {$\SigmaP{2}$};
\node               (pp)      at    (6,2)                   {$\PiP{2}$};
\node             (dots)      at    (5,2.5)                   {\Large $\vdots$};
\node               (ph)      at    (5,3.2)                   {$\PH$};
\node               (ap)      at    (5,4)                   {$\AP=\P\SPACE$};
\node              (exp)      at    (5,4.8)                   {$\EXP$};
\node             (nexp)      at    (5,5.6)                   {$\NEXP$};
\node              (dec)      at    (5,6.4)                   {decidable};
\node               (si)      at    (4,7.2)                   {$\sigmazeroone$};
\node               (pi)      at    (6,7.2)                   {$\pizeroone$};

\path     (p)      edge[dashed]                (np);
\path     (p)      edge[dashed]                (cnp);
\path    (np)      edge[dashed]                (sp);
\path    (np)      edge[dashed]                (pp);
\path   (cnp)      edge[dashed]                (sp);
\path   (cnp)      edge[dashed]                (pp);
\path    (sp)      edge[dashed]                (dots);
\path    (pp)      edge[dashed]                (dots);
\path  (dots)      edge[dashed]                (ph);
\path    (ph)      edge[dashed]                (ap);
\path    (ap)      edge[dashed]                (exp);
\path   (exp)      edge[dashed]                (nexp);
\path  (nexp)      edge                        (dec);
\path   (dec)      edge                        (si);
\path   (dec)      edge                        (pi);
\draw     (p)      .. controls (9,1) and (7,3.5) ..           (exp);
\path    (np)      edge[bend left=40]              (nexp);
\end{tikzpicture}
\caption{Diagram of complexity class inclusions}
{\small Dashed lines mark \qte{$\subseteq$} inclusions and solid lines mark \qte{$\subsetneqq$} inclusions.}
\end{center}
\label{fig:complexity inclusions}
\end{figure}

We will later need the \cim{complexity operator} $\exist$\index{$\exist$}.
If $\calC$ is an arbitrary complexity class then $\exist\calC$ denotes the class of all sets $A$ for which there is a set $B\in\calC$ and a polynomial $p$ such that for all $x$,
\[x\in A \quad \text{iff\quad there is a }y\text{ with }|y|\leq p(|x|)\text{ and }\enc{x,y}\in B.\]
Note that for every class $\calC$, $\exist\calC\subseteq \NP^\calC$. However, the converse does not hold in general.
We will only need the following facts: $\exist \co\NP=\SigmaP{2}$, $\exist \PiP{2} = \SigmaP{3}$, $\exist\P\SPACE=\P\SPACE$ and $\exist \NEXP = \NEXP$.

\subsection{Reducibility and completeness}
\begin{definition}\label{def:leqpm}
Let $C$ be a countable set and $A,B\subseteq C$. Then $A$ is \emph{polynomial-time many-one reducible}\index{reduction!polynomial-time}\index{$\leqpm$} to $B$, in symbols $A\leqpm B$, iff there is a \cim[reduction!function]{reduction} function $f\colon C \to C$ such that $f$ is computable in polynomial time and for all $x\in C$ it holds that $x\in A$ iff $f(x)\in B$.

We write $A \equivpm B$ iff both $A\leqpm B$ and $B\leqpm A$ hold.\index{equivalence!polynomial-time}\index{$\equivpm$}
\end{definition}

Note that most complexity classes $\calC$ with $\P \subseteq \calC$ are closed under $\leqpm$, \ie if $A\leqpm B$ and $B\in \calC$ then also $A\in \calC$. This holds for all classes relevant to us.

\begin{definition}\label{def:completeness}
Let $A$ be a computational problem and let $\calC$ be a complexity class with $\P \subseteq \calC$. Then $A$ is called \emph{$\calC$-hard}\index{hardness} (\wrt $\leqpm$-reductions) iff for all $B\in\calC$ it holds that $B\leqpm A$. $A$ is called \emph{$\calC$-complete}\index{completeness} (\wrt $\leqpm$-reductions) iff $A\in\calC$ and $A$ is $\calC$-hard.
\end{definition}

\section{Logical problems}
Whenever one is investigating a logic there are several basic problems which are naturally interesting. We will now introduce three of them -- from the complexity theory point of view.

\subsection{Satisfiability}
The first of the problems is the satisfiability of a given formula, \ie the question whether there exists a structure that satisfies the formula. This question for example arises in the context of system design -- when a formula is used to specify certain properties and one wants to know whether there is a system with such properties. Also, it is often the case that many other problems related to a logic can be reduced to its satisfiability problem.

\begin{definition}\label{def:sat}
If $\LL$ is a logic and its semantics, \ie the satisfaction of formulas, is defined over a class of structures $\calC$ then its \cim[satisfiability|textbf]{satisfiability problem}\index{sat@$\sat$} is defined as
\problemdef{
$\LL\text{-}\calC\hsat$
}{
A formula $\phi\in\LL$.
}{
Is there a structure $\mA\in\calC$ such that $\mA\models \phi$ holds?
}

If $\calC$ is clear from the context we just write $\LL\hsat$ instead of $\LL\text{-}\calC\hsat$.
The finite satisfiability problem $\LL\hfinsat$\index{satisfiability!finite}\index{finsat@$\finsatt$} is the analogue of $\LL\hsat$ in which we require the structure $\mA$ to be finite.

Sometimes the formal definition of the satisfaction relation entails the necessity to not only have a structure but also an element or assignment (or a set of elements or assignments) from the structure to satisfy the formula. The question then becomes
\begin{center}
\begin{tabular}{p{0.8\linewidth}}
Is there a structure $\mA\in\calC$ and an element $a\in A$ / an assignment $s\colon \dom{\phi}\to A$ (resp.~a set $T\subseteq A$ of elements / $X$ of assignments) such that $M,a\models \phi$ / $M,s\models \phi$ (resp.~$M,T\models \phi$ / $M\models_X \phi$)?
\end{tabular}
\end{center}
In the case of a set of elements / assignments it is also required that $T\neq \emptyset$ (resp.~$X\neq\emptyset$) since for all our logics it is the case that the empty set satisfies all formulas and hence the question would become trivial if we allowed it here.
\end{definition}

The following observation will be useful in \cref{chap:dtwo}.

\begin{remark}\label{eso sat}
If $\phi$ is a formula over the vocabulary $\tau$ and
\[\psi \dfn \exists R_1 \dots \exists R_n \exists f_1 \dots \exists f_m \phi\]
with $R_1,\dots,R_n,f_1,\dots,f_m\in \tau$, then $\phi$ is satisfiable iff the second-order formula $\psi$ is satisfiable.
\end{remark}

\subsection{Model checking}
The next problem is the model checking of a given structure for a given formula, \ie deciding whether a given structure satisfies a given formula. This question for example arises in system verification -- where it shall be checked if an already designed system does or does not have some specific properties.

\begin{definition}\label{def:mc}
Let $\LL$ be a logic and its semantics be defined over a class of structures $\calC$. Then its \cim[model checking|textbf]{model checking problem}\index{mc@$\problem{MC}$} is defined as
\problemdef{
$\LL\text{-}\calC\hmc$
}{
A formula $\phi\in\LL$ and a structure $\mA\in\calC$.
}{
Does $\mA$ satisfy $\phi$, \ie $\mA\models \phi$?
}

Again, if $\calC$ is clear from the context we leave it out. Also, as above, it may sometimes be necessary to not only have a structure but also one or more elements from the structure. In this case these elements are part of the input, \ie they have to be given together with the structure when the question is asked.
\end{definition}

\subsection{Expressiveness}\label{sec:expressiveness}
The last of our problems is \cim{expressiveness}. This makes most sense in comparing two logics $\LL$ and $\LL'$. In this case the question is whether every property definable in one logic is also definable in the other. More formally we have the following definitions.

\begin{definition}\label{def:feqv}
Let $\phi$ and $\psi$ be two formulas with their semantics defined over the class of structures $\calC$. We say that $\psi$ is a \emph{logical consequence} of $\phi$, in symbols $\phi\fimp\psi$, iff for all structures $\mA\in\calC$ it holds that\index{$\fimp$}
\[\mA\models \phi  \quad\text{implies}\quad  \mA\models \psi.\]
We say that $\phi$ and $\psi$ are \emph{logically equivalent} (in $\calC$), in symbols $\phi\feqv\psi$, iff $\phi\fimp\psi$ and $\phi\frimp\psi$.\index{equivalence!formulas}\index{$\feqv$}
\end{definition}

\begin{definition}
Let $\LL$ and $\LL'$ be two logics with their semantics defined over the classes of structures $\calC$ and $\calC'$ with $\calC \subseteq \calC'$. Then we say that $\LL$ is \emph{reducible} to $\LL'$ ($\LL \lleq \LL'$) iff for all formulas $\phi\in\LL$ there is a formula $\psi\in\LL'$ such that $\phi \feqv \psi$ (in $\calC$) holds.\index{$\lleq$}\index{reduction!logics}

We say that $\LL$ and $\LL'$ are \emph{equivalent} ($\LL \leqv \LL'$) iff $\LL\lleq \LL'$ and $\LL'\lleq \LL$ hold. And we write $\LL \llneq \LL'$ iff both $\LL\lleq\LL'$ and $\LL'\not\lleq\LL$ hold.\index{$\leqv$}\index{equivalence!logics}
\end{definition}

For extensions of first-order logic we usually restrict our attention to sentences when dealing with expressiveness, \eg $\fo$ only means the set of first-order sentences in this context and not the set of all first-order formulas.


One could think that showing equivalence of two logics $\LL$ and $\LL'$ would also show that their model-checking and satisfiability problems are of the same complexity. Unfortunately, this is not always the case, \ie it is possible that $\LL\leqv \LL'$ but $\LL\hsat\not\equivpm\LL'\hsat$ and/or $\LL\hmc\not\equivpm\LL'\hmc$. This is caused by a possible blow-up in translating from one logic to the other.

\chapter{Dependence Logic}\label{chap:dep}
\section{First-order dependence logic}\label{sec:df logic}
In this thesis, the semantics of \df and \ifl are both  based on the concept of teams. For \ifl, we follow the exposition of \cite{cadeja09} and the recent monograph \cite{masase11}. 

\begin{definition}[Syntax of \ifl]\label{def:ifl syntax}\index{independence-friendly logic|textbf}\index{IF@$\ifl$|textbf}
The syntax of $\ifl$ extends the syntax of $\fo$
by the rules\index{$\exists x/W$}
\[\phi \ \ddfn\  \exists x/W\,\phi \;\mid\; \forall x/W\,\phi,\]
where $x$ is a first-order variable and $W$ a finite set of first-order variables.
The quantifiers are called \emph{slashed quantifiers}\index{slashed quantifier|textbf} in this case.
\end{definition}

\begin{definition}[Syntax of \df, \cite{va07}]\label{def:df syntax}\index{dependence logic|textbf}\index{dependence logic!first-order}\index{D@$\df$|textbf}
The syntax of $\df$ extends the syntax of $\fo$
by the rules\index{$\dep$|textbf}\index{$\nor$|textbf}
\[\phi\;\; \ddfn\;\; \dep[t_1,\ldots,t_n] \;\mid\; \neg\dep[t_1,\ldots,t_n] \;\mid\; \phi \nor \phi,\]
where $t_1,\ldots,t_n$ are terms.
An atomic formula of the form $\dep[t_1,\ldots,t_n]$ is called \emph{dependence atom}\index{dependence!atom} and in this thesis the terms $t_i$ will mostly be plain variables.

The \emph{arity}\index{arity!dependence atom} of a dependence atom $\dep[x_1,\dots,x_{n-1},x_n]$ is defined as $n-1$, \ie the arity of a dependence atom is the arity of the determinating function whose existence it asserts.

In the following chapters we usually denote by $\df$ the above logic but without the $\nor$ operator. The version that includes $\nor$ is primarily used for the translations in \cref{def:mdl semantics}.
\end{definition}

Note that we could have defined negation to be arbitrary and not only atomic. A sentence that is not in negation normal form could then be evaluated by first pushing the negations down to the atomic level by using de Morgan's law and other well-known rules.
However, as we will see in \cref{sec:d props}, non-atomic negation would be purely syntactical and not semantical (\emph{semantical} meaning that $\neg \phi$ is satisfied iff $\phi$ is not satisfied) -- whereas for plain first-order logic the syntactical and semantical notions coincide.


For a set $W\subseteq \dom{X}$ we call $F$ \emph{$W$-independent} (resp.~\emph{$W$-determined}) if for all $s,s' \in X$ with $s(x)=s'(x)$ for all $x\in \dom{X}\setminus W$ (resp.~for all $x\in W$) we have that $F(s)=F(s')$.

\begin{definition}[Semantics of \df and \ifl, \cite{ho97,va07}]\label{def:ifl semantics}
Let $\mA$ be a mo\-del and $X$ a team of $A$. For \ifl the satisfaction relation
$\mA\models _X \phi$ from \cref{def:fo semantics} is extended by the rules
\begin{itemize}
  \item $\mA\models_X \exists x/W\phi$ iff $\mA\models_{X(F/x)}\phi$ for some $W$-independent function\\ $F\colon X \to A$,
  \item $\mA\models_X \forall x/W\phi$ iff $\mA\models_{X(A/x)}\phi$
\end{itemize}
and for \df by the rules
\begin{itemize}
\item $\mA\models_X \dep[t_{1},\ldots,t_{n}]$ iff for all $s,s'\in
X$ such that $t_1^{\mA}\langle s\rangle  =t_1^{\mA}\langle
s'\rangle  ,$\\$\ldots, t_{n-1}^{\mA}\langle s\rangle
=t_{n-1}^{\mA}\langle s'\rangle $ it holds that $t_n^{\mA}\langle
s\rangle  =t_n^{\mA}\langle s'\rangle  $,
\item $\mA \models_X \neg \dep[t_{1},\ldots,t_{n}]$ iff $X=\emptyset$,
\item $\mA \models_X \phi \nor \psi$ iff $\mA\models _X \phi$ or $\mA\models _X \psi$.
\end{itemize}
Above, we assume that the domain of $X$ contains $\frr[\phi]$. Finally, a sentence $\phi$ is true in a model $\mA$  ($\mA\models \phi$)  if $\mA\models _{\{\emptyset\}} \phi$.
\end{definition}

Note the seemingly rather strange semantics of negated dependence atoms. The rationale for this, given by V\"a\"an\"anen \cite[p.~24]{va07}, is the fact that if we negate dependence atoms and maintain the same duality as between positive and negative first-order literals (cf.~\cref{def:fo semantics}) we get the condition
\[\begin{array}{l@{\ }l}
\text{$\forall s,s'\in X$:}&t_1^{\mA}\langle s\rangle  =t_1^{\mA}\langle s'\rangle  ,\ldots, t_{n-1}^{\mA}\langle s\rangle = t_{n-1}^{\mA}\langle s'\rangle\\
&\text{and} \quad   t_n^{\mA}\langle s\rangle  \neq  t_n^{\mA}\langle s'\rangle
\end{array}\]
and this is only true if $X=\emptyset$.

Above, we denote \emph{dependence disjunction}\index{disjunction!dependence|textbf}\index{dependence!disjunction|textbf} instead of \emph{classical disjunction}\index{disjunction!classical|textbf}\index{classical disjunction|textbf} with $\vee$ because the semantics of dependence disjunction is an extension of the semantics of usual first-order disjunction and thus we preserve downward compatibility of our notation in this way. However, we still call the $\nor$ operator ``classical'' because in a higher level context -- where our sets of states are viewed as single objects themselves -- it is indeed the usual disjunction, cf.~\cite{abva08}. Note that
\[\phi \nor \psi \quad\fimp\quad  \phi \vee \psi\]
holds but not the other way around.

\subsection{\texorpdfstring{Basic properties of \df and \ifl}{Basic properties of D and IF}}\label{sec:d props}
Now we will state some basic properties of \df and \ifl.

First we will see that the game theoretically motivated negation $\neg$ of \df and \ifl does not satisfy the \emph{law of excluded middle} and is therefore not the classical Boolean negation.
\begin{proposition}[Law of excluded middle failure]\label{no excluded middle}\index{law of excluded middle}
Let $\phi\dfn \dep[x]$, $\mA$ a structure with universe $A\dfn\{a,b\}$ and $X\dfn \{ \{x\mapsto a\}, \{x\mapsto b\}\}$.
Then $\mA\not\models\phi$ and $\mA\not\models\neg\phi$.
\end{proposition}

Also, many familiar propositional equivalences of connectives do not hold in \df and \ifl. For example, the idempotence of disjunction fails, which can be used to show that the  distributivity laws of disjunction and conjunction do not hold either.
We refer the interested reader to \cite[Section~3.3]{va07} for a detailed exposition on propositional equivalences of connectives in \df (and also \ifl).

The following fact is a fundamental property of all formulas of \df and \ifl.
\begin{proposition}[Downward closure property, \cite{va07, ho97}]\label{downward closure property}\index{downward closure property}
Let $\phi$ be a for\-mu\-la of  \df or \ifl, $\mA$  a model, and $Y\subseteq X$ teams. Then $\mA\models_X \phi$ implies $\mA\models_Y\phi$. 
\end{proposition}
As argued in \cite[Proposition~3.10]{va07}, the downward closure property suits the intuition that a true formula expressing dependence should not become false when making the team smaller, since if dependence holds in a large set then it does even more in a smaller set.

A feature of \df and \ifl that nicely accomodates the downward closure is that $\mA\models _{\emptyset}\phi$ holds for all $\mA$ and all formulas $\phi$ of \df and \ifl. This observation is important in noting that for \emph{sentences} $\phi$ and $\psi$ it holds that
\[\phi\vee \psi  \ \feqv\ \phi\nor\psi.\]

Väänänen \cite[Chapter~6.1]{va07} and Hodges \cite{ho97b} have shown that the expressive power of sentences of $\df$ and \ifl coincides with that of existential second-order sentences:
\begin{theorem}\label{d equiv if equiv eso}
$\df\leqv \ifl\leqv \eso$.
\end{theorem}

Let $X$ be a team with domain $\{x_1,\ldots,x_k\}$ and $V\subseteq \{x_1,\ldots,x_k\}$. We denote by
$X\restricted{} V$ the team $\{s\restricted{} V \mid s\in X\}$ with domain $V$ (here, $s\restricted{} V$ denotes the restriction of the function $s$ to the domain $V$).\index{$\restricted{}$}
The following proposition shows that the truth of a \df-formula only depends on the interpretations of the variables occurring free in the formula.
\begin{proposition}[\cite{va07, cadeja09}]\label{freevar} Let $\phi\in \df$ be any formula or $\phi\in \ifl$ a sentence. If 
 $V\supseteq \frr(\phi)$, then $\mA \models _X\phi$ if and only if $\mA \models _{X\upharpoonright V} \phi$.
\end{proposition}
The analogue of \cref{freevar} does not hold for open formulas of \ifl. In other words, the truth of an $\ifl$-formula may depend on the interpretations of variables that do not occur
in the formula. For example, the truth of the formula
\begin{equation*}
\phi \dfn \exists x/\{y\}(x=y)
\end{equation*}
 in a team $X$ with domain $\{x,y,z\}$ depends on the values of $z$ in $X$, although $z$ does not occur in $\phi$.

We have noted in the last chapter that for first-order formulas the usual single-assignment based semantics coincides with our team semantics, \cf \cref{fo single equals team}.
This is not always the case for \df and \ifl formulas.
\begin{definition}[Flatness]
Let $\phi$ be a \df or an \ifl formula. Then $\phi$ is said to be \cim{flat} iff for all structures $\mA$ and teams $X$ of $A$ it holds that
\[\mA \models_X \phi \quad\text{iff}\quad  \forall s\in X,\  \mA \models_{\{s\}} \phi.\]
\end{definition}
The left-to-right implication in the above definition follows from the downward closure property (\cref{downward closure property}) while the other direction does not always hold. For plain first-order formulas, however, it does.

\subsection{Two-variable dependence logic}
\begin{definition}[$\df^k$, $\ifl^k$]\label{def:dtwo}\index{dependence logic!two-variable}\index{D@$\df^k$}\index{independence-friendly logic!two-variable}\index{IF@$\iftwo$}
$\df^k$ and $\ifl^k$ are defined as restrictions of $\df$ and $\ifl$ analogous to the definition of $\fo^k$ from $\fo$ (\Cref{def:fok syntax}).
\end{definition}

\section{Modal dependence logic}
Now we will introduce the syntax and semantics of modal dependence logic and show some important properties of it.
For a more profound overview consult V\"a\"an\"anen's introduction \cite{va08}.

\begin{definition}[Syntax of \MDL]\label{def:mdl syntax}\index{dependence logic!modal}\index{MDL@$\MDL$}
The syntax of $\MDL$ extends the syntax of modal logic by the rules
\[\phi\ \ddfn\ \dep[p_1,\dots,p_{n-1},p_n] \;\mid\; \neg \dep[p_1,\dots,p_{n-1},p_n] \;\mid\; \phi \nor \phi,\]
where $n\geq 1$.
\end{definition}


The semantics of \MDL will be given by a translation to \dtwo. In the first-order side of the translation we do not use the usual dependence atoms but instead a variation of them. This is, however, not a problem since the variation turns out to be definable by the usual version.
We follow this approach because it nicely demonstrates that the embedding of modal logic into two-variable first-order logic carries over to the respective dependence logics.

\begin{definition}[Semantics of \MDL]\label{def:mdl semantics}
We extend the translation from \ML to \fotwo (\Cref{def:ml to fotwo}) by the following rules to get a translation from \MDL to \dtwo:
\[\begin{array}{rcl}
\dep[p_1,\dots,p_n]                  &\mapsto&       \dep[P_1(x),\dots,P_n(x)],\\
\neg\dep[p_1,\dots,p_n]              &\mapsto&       \false,\\
\phi \nor \psi                       &\mapsto&       \fom{\phi}(x) \nor \fom{\psi}(x).\\
\end{array}\]
\end{definition}
As stated above the resulting formula is not a strict $\df$ formula since inside the dependence atoms there are atomic formulas instead of terms. We can, however, further translate the formula by the rule
\[\dep[P_1(x),\dots,P_n(x)]  \;\mapsto\;  \exists y_1 \dots \exists y_n \big( \bigwedge_{i=1}^n (y_i=c \bimp P_i(x)) \ \wedge \  \dep[y_1,\dots,y_n]\big),\]
where $c$ is either a constant symbol from the first-order vocabulary or a variable that is existentially quantified in the beginning of the complete formula, \ie its value is the same in all subformulas.

In \cref{chap:mdl} we will investigate and classify several fragments of \MDL.

\begin{definition}[\MDLk]\label{def:MDLk}
$\MDLk$\index{MDLk@$\MDLk$|textbf}\index{dependence logic!modal!bounded arity|textbf}\index{arity!bounded|textbf} is the subset of \MDL that contains all formulas which do not contain any dependence atoms whose arity is greater than $k$.
\end{definition}

We will classify \MDL for all fragments defined by sets of operators.

\begin{definition}[$\MDL(M)$, $\PDL$]\label{def:mdl fragments}
Let $\LL$ be a modal logic and let $M$ be a set of operators and constants from the language of $\LL$. Then $\LL[M]$\index{$\LL[M]$}\index{L@$\LL[M]$}\index{MDLM@$\MDL[M]$} is the subset of $\LL$ containing all formulas built from atomic propositions using only operators and constants from $M$. We usually write $\LL[op_1,op_2,\dots]$ instead of $\LL[\{op_1,op_2,\dots\}]$.


Now, $\PDL$\index{dependence logic!propositional}\index{PDL@$\PDL$} is defined as $\MDL[{\wedge,\vee,\aneg,\dep[]}]$, \ie the restriction of \MDL to formulas without modalities and without classical disjunction $\nor$.
\end{definition}

\subsection{\texorpdfstring{Basic properties of \MDL}{Basic properties of MDL}}\label{sec:mdl properties}
Most properties of first-order dependence logic carry over to modal dependence logic, \eg neither having the law of excluded middle (\cref{no excluded middle}) nor the idempotence of disjunction or the fact that the empty team satisfies all formulas.

Furthermore, the downward closure property (\cref{downward closure property}) holds for modal dependence logic as well.
\begin{proposition}[Downward closure property, \cite{va08}]\label{downward closure property mdl}\index{downward closure property!$\MDL$}
Let $K=(S,R,\pi)$ be a Kripke structure, $T'\subseteq T\subseteq S$ and $\phi \in \MDL$. Then $W,T\models \phi$ implies $W,T'\models \phi$.
\end{proposition}

Also, the concept of flatness can be translated to modal logic in the canonical way.
\begin{definition}[Flatness]\index{flat!$\MDL$}
An $\MIDL$ formula $\phi$ is said to be \emph{flat} if for any Kripke model $K$ and any team $T$ of $K$
\[K, T \models \phi \quad\text{iff}\quad \forall s\in T,~ K, \{s\} \models \phi.\]
\end{definition}
And again, analogous to first-order logic, all plain modal logic formulas are flat.

Additionally, some more equivalences follow almost immediately from the semantics of $\MDL$.
\begin{lemma}\label{mdl equivalences}
The following holds for \MDL:
\begin{enumerate}
\item\label{true with sor} $\true\ \feqv\  p\sor\neg p$,\quad $\false\ \feqv\  p\wedge \neg p\ \feqv\ \neg \dep[p_1,\dots,p_n]$,
\item\label{dep with nor zero} $\dep[p]\ \feqv\ p\nor\neg p$\quad and
\item\label{dep with nor full} $\dep[p_1,\dots,p_n]\ \feqv\ \bigsor\limits_{i_1,\dots,i_{n-1}\in\{\true,\false\}}\, \big(p_1^{i_1}\wedge \dots \wedge p_{n-1}^{i_{n-1}} \wedge (p_n \nor \neg p_n)\big)$,\\
where $p^\true \dfn p$ and $p^\false \dfn \neg p$.
\end{enumerate}
\end{lemma}
\begin{proof}
The only difficult case is \enumref{dep with nor full}. For this note that $K,T\models \dep[p_1,\dots,p_n]$ means that for all subteams $T'\subseteq T$ where the evaluations of $p_1,\dots,p_{n-1}$ are each constant in $T'$, it holds that the value of $p_n$ is also constant in $T'$. On the other hand, the right hand side of the equivalence is true in a team $T$ if and only if $T$ can be partitioned into $2^{n-1}$ subteams $T''$ such that in each team $T''$ the evaluations of $p_1,\dots,p_{n-1}$ are as specified and the value of $p_n$ is constant. Since all possible evaluations of $p_1,\dots,p_{n-1}$ occur in the disjunction all the subteams $T'$ from above will occur as one of the teams $T''$ and vice versa.
\end{proof}
Note that the first-order equivalents of \enumref{dep with nor zero} and \enumref{dep with nor full} do not hold -- contrary to \enumref{true with sor}. Here, the crucial difference between first-order and modal logic is that in the latter case variables are always only Boolean; that especially means that the set of possible values for a variable occuring in a dependence atom is finite and fixed.


\section{Intuitionistic dependence logic}\label{sec:idl def}

Intuitionistic dependence logic was introduced in \cite{abva08}.

\begin{definition}[\IDL]\label{def:idl}\index{intuitionistic dependence logic|textbf}\index{IDL@$\IDL$|textbf}
The syntax of \IDL extends the syntax of \df by the rule
\[\phi \ \ddfn\ \phi \imp \phi.\]
Here, $\imp$ denotes \emph{intuitionistic implication}\index{implication!intuitionistic|textbf}.

For \IDL the satisfaction relation from \cref{def:ifl semantics} is extended by the rule
\[\mA \models_X \phi \imp \psi \text{ iff }\text{for all $Y\subseteq X$ with $\mA\models_Y \phi$ it holds that $\mA\models_Y\psi$},\]
\ie $\phi \imp \psi$ is satisfied in a team iff truth of $\phi$ in a subteam always implies truth of $\psi$ in that subteam.
\end{definition}

Now we introduce the modal logic version of intuitionistic dependence logic.

\begin{definition}[\MIDL]\label{def:midl}\index{intuitionistic dependence logic!modal|textbf}\index{MIDL@$\MIDL$|textbf}
The syntax of \MIDL extends the syntax of\linebreak \MDL by the rule
\[\phi \ \ddfn\ \phi \imp \phi.\]

The semantics of \MIDL is defined by the extension of the translation from \cref{def:mdl semantics} by the rule
\[\phi \imp \psi \quad \mapsto\quad \fom{\phi}(x) \imp \fom{\psi}(x).\]
\end{definition}

All the properties of \MDL from \cref{sec:mdl properties} hold for \MIDL as well.

In addition, some more equivalences hold.
\begin{lemma}\label{midl equivalences}
All equivalences from \cref{mdl equivalences} hold for \MIDL and also the following:
\begin{enumerate}
\item $\neg p\ \feqv\ p\imp\false$,
\item\label{dep-with-imp} $\dep[p_1,\dots,p_n]\ \feqv \ \big(\dep[p_1]\wedge\dots\wedge\dep[p_{n-1}]\big)\to\dep[p_n]$\, and
\item\label{imp with sor} $\phi\imp\psi\ \equiv\ \dual{\phi}\sor\psi$\quad if both $\phi$ and $\psi$ are $\ML$ formulas and $\dual{\phi}$ is the dual of $\phi$ (\cf \cref{dual modal logic formulas}), \ie $\imp$ has a conservative semantics -- in the sense that for $\ML$ formulas its semantics coincides with the semantics of the usual classical modal logic implication.
\end{enumerate}
\end{lemma}
Note that although $\neg p$ is equivalent to $p \imp \false$, of course one of $\neg$ or $\false$ is needed to achieve full \MIDL expressiveness.
The above \lcnamecref{midl equivalences} also shows that we could have equivalently defined the syntax of \MIDL without dependence atoms.
The rationale for making the constants, the $\neg$ operator and the $\dep$ operator parts of the language, although we could have left out some of them by applying \cref{midl equivalences}, is that we will investigate sublogics of \MIDL where we will still use constants and negation but not always $\wedge$, $\sor$ and $\imp$.

Finally, we define the propositional version of intuitionistic dependence logic.
\begin{definition}[$\PIDL$]\label{def:pidl}\index{intuitionistic dependence logic!propositional|textbf}\index{PIDL@$\PIDL$|textbf}
$\PIDL$ is defined as
\[\MIDL[{\wedge,\nor,\neg,\imp,\dep[]}].\]
\end{definition}

From \cref{midl equivalences} we obtain the following characterizations:
\[\begin{array}{rcl}
\PIDL  &\leqv&  \MIDL[{\wedge,\nor,\neg,\imp}]\\
       &\leqv&  \MIDL[{\wedge,\nor,\bot,\imp}] \quad \text{and}\\
\MIDL  &\leqv&  \MIDL[{\AX,\EX,\wedge,\sor,\nor,\neg,\imp}]\\
       &\leqv&  \MIDL[{\AX,\EX,\wedge,\sor,\nor,\bot,\imp}].
\end{array}\]

Note that $\PDL \nsubseteq \PIDL$ since \PDL uses dependence disjunction and \PIDL uses classical disjunction.

\Cref{fig:logics} shows the main modal logics considered in this thesis.

\begin{figure}[ht]
\begin{center}
\begin{tikzpicture}[thick]
        \node           at (0,0.5)        {$\MIDL$};
        \node           at (-0.75,-0.3)        {$\MDL$};
        \node           at (0,-1)        {$\AX \quad \EX \quad \sor \quad \wedge \quad \neg \quad \dep \quad \nor \quad \imp$};

        \node           at (1.125,-2.6)        {$\PIDL$};
        \node           at (0,-1.8)        {$\PDL$};     

    \draw (-3.0, 0.1) -- (3.0,0.1);
    \draw (-3.0,-0.1) -- (-3.0,0.1);
    \draw (3.0,-0.1) -- (3.0,0.1);
    
    \draw (-3.0, -0.6) -- (2.1,-0.6);
    \draw (-3.0,-0.6) -- (-3.0,-0.8);
    \draw (2.1,-0.6) -- (2.1,-0.8);

    \draw (-0.8,-2.3) -- (3.0,-2.3);
    \draw (-0.8,-2.3) -- (-0.8,-2.1);
    \draw (3.0,-2.3) -- (3.0,-2.1);
  
     \draw (-1.5,-1.5) -- (1.4,-1.5);
     \draw (-1.5,-1.3) -- (-1.5,-1.5);
     \draw (1.4,-1.3) -- (1.4,-1.5);

\end{tikzpicture}
\end{center}
\caption{Sublogics of \MIDL}
\label{fig:logics}
\end{figure}


\bigskip
In later chapters we will investigate the model checking and satisfiability problems of several dependence logics.
For now we note that, as pointed out in \cref{sec:expressiveness}, the above equivalences between \MIDL fragments do not necessarily imply that the computational complexity of the problems is the same for equivalent fragments, \ie it is possible to have sets $M\neq M'$ of operators such that $\MIDL[M]\leqv\MIDL[M']$ but $\MIDL[M]\hmc \not\equivpm \MIDL[M']\hmc$.
This is caused by a possible exponential blow-up in translating from one fragment to another.

\chapter{Modal Dependence Logic}\label{chap:mdl}
\section{Satisfiability}\label{sec:mdl-sat}

In this section we will classify operator fragments of \MDL according to the complexity of their satisfiability problem, \ie -- according to \cref{def:sat} -- for all $M\subseteq \{\AX,\EX,\wedge,\sor,\nor,\neg,\dep[]\}$ and all $k\geq 0$ the problem
\problemdef{
$\MDL[M]\hsat$ ($\MDLk[M]\hsat$)
}{
A $\MDL[M]$ (resp.~ $\MDLk[M]$) formula $\phi$.
}{
Is there a Kripke structure $W$ and a non-empty set $T$ of worlds in $W$ such that $W,T\models \phi$ holds?
}

Note that, because of the downward closure property (\cref{downward closure property mdl}), to check satisfiability of a formula $\phi$ it is enough to check whether there is a frame $W$ and a single world $w$ in $W$ such that $W,\{w\}\models \phi$.

Our first lemma concerns sets of operators including classical disjunction.

\begin{lemma}\label{bullet distributivity}
Let $M$ be a set of \MDL operators. Then the following hold:
\begin{enumerate}
 \item\label{bullet in front} Every $\MDL[M\cup\{\nor\}]$ ($\MDLk[M\cup\{\nor\}]$) formula $\phi$ is equivalent to a formula
 \[\bignor^{2^{|\phi|}}_{i=1}\,\psi_i\]
 with $\psi_i \in \MDL[M ]$ (resp.~$\MDLk[M ]$) for all $i\in\{1,\dots,2^{|\phi|}\}$.
 \item If $\calC$ is an arbitrary complexity class with $\P \subseteq \calC$ and $\MDL[M]\hsat\in\calC$ ($\MDLk[M]\hsat\in\calC$) then $\MDL[M\cup\{\nor\}]\hsat \in \exist\calC$ (resp.~$\MDLk[M\allowbreak\cup\{\nor\}]\hsat \in \exist\calC$).
\end{enumerate}
\end{lemma}
\begin{proof}
a) follows from the distributivity of $\nor$ with all other \MDL operators. More specifically, $\phi \star (\psi\nor\sigma)\feqv(\phi \star \psi)\nor(\phi \star\sigma)$ for $\star \in\{\wedge,\vee\}$ and $\nabla (\phi \nor \psi)\feqv \allowbreak (\nabla\phi)\nor(\nabla\phi)$ for $\nabla \in\{\Diamond,\Box\}$.\footnote{Interestingly, but not of relevance for our work, $\phi\nor(\psi\vee\sigma)\fneqv(\phi\nor\psi)\vee(\phi\nor\sigma)$.}
b) follows from a) with the observation that $\bignor^{2^{|\phi|}}_{i=1}\,\psi_i$ is satisfiable if and only if there is an $i\in\{1,\dots,2^{|\phi|}\}$ such that $\psi_i$ is satisfiable. Note that given $i\in\{1,\dots,2^{|\phi|}\}$ the formula $\psi_i$ can be computed from the original formula $\phi$ in polynomial time by choosing (for all $j\in\{1,\dots,|\phi|\}$) from the $j$th subformula of the form $\psi \nor \sigma$ the formula $\psi$ if the $j$th bit of $i$ is 0 and $\sigma$ if it is 1.
\end{proof}

We need the following simple property of monotone \MDL formulas.
\begin{lemma}\label{dep without negation}
Let $M$ be a set of \MDL operators with $\aneg\notin M$. Then an arbitrary $\MDL[M ]$ formula $\phi$ is satisfiable iff the formula generated from $\phi$ by replacing every dependence atom and every atomic proposition with the same fresh atomic proposition $t$ is satisfiable.
\end{lemma}
\begin{proof}
If a frame $W$ is a model for $\phi$, so is the frame generated from $W$ by setting all atomic propositions in all worlds to true.
\end{proof}

We are now able to classify some cases that can be easily reduced to known results.
\begin{corollary}\label{simple cases}
\hspace*{-1em}\begin{enumerate}
 \item If $\{\Box,\Diamond,\wedge,\vee,\aneg\}\subseteq M\subseteq\{\Box,\Diamond,\wedge,\vee,\aneg,\true,\false,\nor\}$ then \MDL[M]\hsat is \P\SPACE-complete.
 \item If $\{\Box,\Diamond,\wedge,\vee,\false\}\subseteq M\subseteq\{\Box,\Diamond,\wedge,\vee,\true,\false,\dep[],\nor\}$ then \MDL[M]\hsat and \MDLk[M]\hsat are \P\SPACE-complete for all $k\geq 0$.
 \item\label{simple coNP} If $\{\Box,\Diamond,\wedge,\false\}\subseteq M\subseteq\{\Box,\Diamond,\wedge,\true,\false,\dep[]\}$ then \MDL[M]\hsat and\linebreak \MDLk[M]\hsat are \co\NP-complete  for all $k\geq 0$.
 \item If $M \subseteq \{\Box,\Diamond,\wedge,\vee,\true,\dep[],\nor\}$ then every \MDL[M ] formula is satisfiable.
 \item If $M \subseteq \{\wedge,\aneg, \true,\false,\dep[]\}$ then \MDL[M]\hsat is in \P.
 \item If $M \subseteq \{\wedge,\vee, \true,\false,\dep[],\nor\}$ then \MDL[M]\hsat is in \P.
\end{enumerate}
\end{corollary}
\begin{proof}
The lower bound of a) was shown by Ladner \cite{la77}, who proves \P\SPACE-completeness for the case of full ordinary modal logic. The upper bound follows from this, \cref{bullet distributivity} and the fact that $\exist\P\SPACE=\P\SPACE$.
The lower bound for b) was shown by Hemaspaandra \cite[Theorem~6.5]{he01} and the upper bound follows from a) together with \cref{dep without negation}.

The lower bound for c) was shown by Donini et al.~\cite{dolenahonuma92} who prove \NP-hardness of the problem to decide whether an $\mathcal{ALE}$-concept is unsatisfiable. $\mathcal{ALE}$ is a description logic which essentially is nothing else than $\MDL[\Box,\Diamond,\wedge,\aneg,\true,\false]$ ($\aneg$ is not used in the hardness proof and $\true$, although it is used in the original proof, can be substituted by a fresh atomic proposition $t$). For the upper bound Ladner's \P\SPACE-algorithm \cite{la77} can be used, as in the case without disjunction it is in fact a \co\NP-algorithm, together with \cref{dep without negation}.

d) follows from \cref{dep without negation} together with the fact that every \MDL formula with $t$ as the only atomic subformula is satisfied in the transitive singleton, \ie the frame consisting of only one state which has itself as successor, in which $t$ is true.

e) follows from the polynomial time complexity of deciding satisfiability of a \cnf[1] formula.
Finally, f) reduces to Boolean formula evaluation by \cref{dep without negation}.
Note that for e) and f) dependence atoms can be replaced by \true{} because there we do not have any modality.
\end{proof}

\subsection{Poor man's dependence logic}\index{Poor Man's Logic|textbf}

We now turn to the $\SigmaP{2}$-complete cases. These include monotone poor man's logic, \ie modal logic without negation and dependence disjunction, with and without dependence atoms.
\begin{theorem}\label{poor man bullet complexity}
If $\{\Box,\Diamond,\wedge,\aneg,\nor\} \subseteq M \subseteq\{\Box,\Diamond,\wedge,\aneg,\true,\false,\allowbreak\nor\}$ or $\{\Box,\Diamond,\wedge,\false,\allowbreak \nor\} \subseteq M\subseteq\{\Box,\Diamond,\wedge,\allowbreak\true,\allowbreak\false,\allowbreak\dep[],\allowbreak\nor\}$ then $\MDL[M]\hsat$ and $\MDLk[M]\hsat$ are \SigmaP{2}-complete for all $k\geq 0$.
\end{theorem}
\begin{proof}
Proving the upper bound for the second case reduces to proving the upper bound for the first case by \cref{dep without negation}. For the first case it holds with \cref{bullet distributivity} that $\MDL[\Box,\Diamond,\wedge,\aneg,\true,\false,\nor]\hsat\in \exist\co\NP=\SigmaP{2}$ since $\MDL[\Box,\Diamond,\wedge,\aneg,\true,\false]\hsat\in \co\NP$ (\cf the proof of \cref{simple cases}\ref{simple coNP}). The latter follows directly from Ladner's \P\SPACE-algo\-rithm for modal logic satisfiability \cite{la77} which is in fact a \co\NP-algorithm in the case without disjunction.

For the lower bound we consider the quantified constraint satisfaction problem \oneInThreeQcsp shown to be \PiP{2}-complete by Bauland et al.~\cite{babocrrescvo10}. This problem can be reduced to the complement of $\MDL[\Box,\Diamond,\wedge,\aneg/\false,\nor]\hsat$ in polynomial time.

An instance of \oneInThreeQcsp consists of universally quantified Boolean variables $p_1,\dots,p_k$, existentially quantified Boolean variables $p_{k+1},\dots,p_n$ and a set of clauses each consisting of exactly three of those variables. \oneInThreeQcsp is the set of all those instances for which for every truth assignment for $p_1,\dots,p_k$ there is a truth assignment for $p_{k+1},\dots,p_n$ such that in each clause exactly one variable evaluates to true. Note that for our reduction it is necessary that in each clause the variables are pairwise different whereas in \oneInThreeQcsp this does not need to be the case. However, the \PiP{2}-hardness proof can be adapted to account for this as follows.

\begin{claim}\label{qcsp hardness}
\oneInThreeQcsp with pairwise different variables in each clause is \PiP{2}-hard.
\end{claim}
\begin{claimproof}{\cref*{qcsp hardness}}
We give a reduction from \qbfcnfneg, which is long known to be $\PiP{2}$-complete (\cf \cite{wr77}). The only differences between an instance of \oneInThreeQcsp and an instance of \qbfcnfneg are that in the latter a clause may contain negated variables, more than one literal that evaluates to true and the same variable more than once.

Our reduction works as follows. The quantification of all original variables stays unchanged. All newly introduced variables are existentially quantified. Our new set of clauses consists of the clauses
\begin{enumerate}
 \item $\{t,f,f'\},\{t,f',f''\},\{t,f'',f\}$, where $t,f,f',f''$ are new variables,
 \item for each original variable $x$:\quad$\{x,x',f\}$ and
 \item for each original clause $\{x,y,z\}$:\quad$\begin{array}[t]{@{}l@{}}\{f,\hat{x},a\},\{a',\hat{y},b\},\{b',\hat{z},f\},\\\{a,a',a''\},\{b,b',b''\},\end{array}$\\
 where $a,a',a'',b,b',b''$ are new variables (fresh for each original clause) and $\hat{u}\dfn \left\{\begin{array}{l@{\quad}l}p&\text{if $u=p$}\\p'&\text{if $u=\neg p$}\end{array}\right.$.
\end{enumerate}
a) enforces the new variable $f$ to evaluate to $\false$ and b) simulates negated variables. c) uses \oneInThree-clauses to allow for more than one of the three literals $x$, $y$ and $z$ to be true by splitting one old clause in three new clauses. But at the same time it still ensures that at least one of them is true by making the new variables interdependent in such a way that at most two of them can be true.
\end{claimproof}

For the reduction from \oneInThreeQcsp to the complement of $\MDL[\Box,\Diamond,\allowbreak\wedge,\allowbreak\aneg/\false,\allowbreak\nor]\hsat$ we extend a technique from the \co\NP-hardness proof for $\MDL[\Box,\allowbreak\Diamond,\wedge,\false]\hsat$ by Donini et al.~\cite[Theorem~3.3]{dolenahonuma92}. Let $p_1,\dots,p_k$ be the universally quantified and $p_{k+1},\dots,p_n$ the existentially quantified variables of a \oneInThreeQcsp instance and let $C_1,\dots,C_m$ be its clauses (we assume w.l.o.g.~that each variable occurs in at least one clause). Then the corresponding $\MDL[\Box,\Diamond,\wedge,\allowbreak\false,\allowbreak\nor]$ formula is
\[\vspace{-0.3ex}\begin{array}{lcclllll}
\phi\dfn &&\bigwedge\limits_{i=1}^k\;\big(&&\nabla_{i1}\dots\nabla_{im}&\nabla_{i1}\dots\nabla_{im}&\Box^{i-1}\Diamond\Box^{k-i}& p\\
&&&\nor&\Box^{m}&\Box^{m}&\Box^{i-1}\Diamond\Box^{k-i}& p\,\big)\\
&\wedge&\bigwedge\limits_{i=k+1}^n&&\nabla_{i1}\dots\nabla_{im}&\nabla_{i1}\dots\nabla_{im}&\Box^k&p\\
&\wedge&&&\Box^m&\Box^m&\Box^k&\false
\end{array}\vspace{-0.3ex}\]
where $p$ is an arbitrary atomic proposition and $\nabla_{ij} \dfn \left\{\begin{array}{l@{\quad}l}\Diamond&\text{if $p_i \in C_j$}\\\Box&\text{else}\end{array}\right.$.

For the corresponding $\MDL[\Box,\Diamond,\wedge,\aneg,\nor]$ formula replace every $\false$ with $\neg p$.

To prove the correctness of our reduction we will need two claims.

\begin{claim}\label{mdl oldclaim}
For $r,s\geq0$ a $\MDL[\Box,\Diamond, \wedge,\aneg,\true, \false]$ formula $\Diamond\phi_1\wedge\dots\wedge\Diamond\phi_r\wedge\Box\psi_1\wedge\dots\wedge\Box\psi_s$ is unsatisfiable iff there is an $i\in\{1,\dots,r\}$ such that $\phi_i\wedge\psi_1\wedge\dots\wedge\psi_s$ is unsatisfiable.
\end{claim}
\begin{claimproof}{\cref*{mdl oldclaim}}
\qte{$\Leftarrow$}: If $\phi_i\wedge\psi_1\wedge\dots\wedge\psi_s$ is unsatisfiable, so is $\Diamond\phi_i\wedge\Box\psi_1\wedge\dots\wedge\Box\psi_s$ and even more $\Diamond\phi_1\wedge\dots\wedge\Diamond\phi_r\wedge\Box\psi_1\wedge\dots\wedge\Box\psi_s$.

\qte{$\Rightarrow$}: Suppose that $\phi_i\wedge\psi_1\wedge\dots\wedge\psi_s$ is satisfiable for all $i\in\{1,\dots,r\}$. Then $\Diamond\phi_1\wedge\dots\wedge\Diamond\phi_r\wedge\Box\psi_1\wedge\dots\wedge\Box\psi_s$ is satisfiable in a frame that consists of a root state and for each $i\in\{1,\dots,r\}$ a separate branch, reachable from the root in one step, which satisfies $\phi_i\wedge\psi_1\wedge\dots\wedge\psi_s$.
\end{claimproof}

Note that $\Diamond\phi_1\wedge\dots\wedge\Diamond\phi_r\wedge\Box\psi_1\wedge\dots\wedge\Box\psi_s$ is always satisfiable if $r=0$.

\smallskip
{\bfseries Definition.}
Let $v\colon\{p_1,\dots,p_k\}\to\{0,1\}$ be a valuation of $\{p_1,\dots,p_k\}$. Then $\phi_v$ denotes the $\MDL[\Box,\Diamond,\wedge,\aneg/\false]$ formula
\[\begin{array}{cclllc}
&\bigwedge\limits_{\substack{i\in\{1,\dots,k\},\\v(p_i)=1}}&\nabla_{i1}\dots\nabla_{im}&\nabla_{i1}\dots\nabla_{im}&\Box^{i-1}\Diamond\Box^{k-i}& p\\
\wedge&\bigwedge\limits_{\substack{i\in\{1,\dots,k\},\\v(p_i)=0}}&\Box^{m}&\Box^{m}&\Box^{i-1}\Diamond\Box^{k-i}& p\\
\wedge&\bigwedge\limits_{i=k+1}^n&\nabla_{i1}\dots\nabla_{im}&\nabla_{i1}\dots\nabla_{im}&\Box^k&p\\
\wedge&&\Box^m&\Box^m&\Box^k&\neg p\,/\,\false
\end{array}\]
\smallskip

\begin{claim}\label{mdl oldclaimtwo}
Let $v\colon \{p_1,\dots,p_k\}\to\{0,1\}$ be a valuation. Then $\phi_v$ is unsatisfiable iff $v$ can be continued to a valuation $v'\colon \{p_1,\dots,p_n\}\to\{0,1\}$ such that in each of the clauses $\{C_1,\dots,C_m\}$ exactly one variable evaluates to true under $v'$.
\end{claim}
\begin{claimproof}{\cref*{mdl oldclaimtwo}}
By iterated use of \cref{mdl oldclaim}, $\phi_v$ is unsatisfiable iff there are $i_1,\dots,i_{2m}$ with
\[\begin{array}{lcl}
i_j\in&&\big\{i\in\{1,\dots,n\}\mid \nabla_{ij'}=\Diamond\big\}\setminus\big\{i\in\{1,\dots,k\}\mid v(p_i)=0\big\}\\
&=&\big\{i\in\{1,\dots,n\}\mid p_i\in C_{j'}\big\}\setminus\big\{i\in\{1,\dots,k\}\mid v(p_i)=0\big\},
\end{array}\]
where $j'\dfn
\left\{\begin{array}{l@{\;\;}l}j&\text{if $j\leq m$}\\j-m&\text{else}\end{array}\right.$,\quad
such that \emph{(i)}
\[\begin{array}{lcclc}
\phi_v(i_1,\dots,i_{2m})\dfn&&\bigwedge\limits_{\substack{i\in\{1,\dots,k\},\\i\in\{i_1,\dots,i_{2m}\},\\v(p_i)=1}}&\Box^{i-1}\Diamond\Box^{k-i}& p\\
&\wedge&\bigwedge\limits_{\substack{i\in\{1,\dots,k\},\\v(p_i)=0}}&\Box^{i-1}\Diamond\Box^{k-i}& p\\
&\wedge&\bigwedge\limits_{\substack{i\in\{k+1,\dots,n\},\\i\in\{i_1,\dots,i_{2m}\}}}&\Box^k&p\\
&\wedge&&\Box^k&\neg p\,/\,\false
\end{array}\]
is unsatisfiable and such that \emph{(ii)} there are no $a,b\in\{1,\dots,2m\}$ with $a<b$, $\nabla_{i_ba'}=\nabla_{i_bb'}=\Diamond$ (this is the case iff $p_{i_b}\in C_{a'}$ and $p_{i_b}\in C_{b'}$) and $i_a \neq i_b$.
Condition \emph{(ii)} 
simply ensures that no subformula is selected after it has already been discarded in an earlier step.
Note that $\phi_v(i_1,\dots,i_{2m})$ is unsatisfiable iff \emph{(i')} for all $i\in\{1,\dots,k\}$: $v(p_i)=1$ and $i\in\{i_1,\dots,i_{2m}\}$\quad or\quad $v(p_i)=0$ (and $i\notin\{i_1,\dots,i_{2m}\}$).

We are now able to prove the claim.

\qte{$\Leftarrow$}: For $j=1,\dots,2m$ choose $i_j\in\{1,\dots,n\}$ such that $p_{i_j}\in C_{j'}$ and $v'(p_{i_j})=1$. By assumption, all $i_j$ exist and are uniquely determined. Hence, for all $i\in\{1,\dots,k\}$ we have that $v(p_i)=0$ (and then $i\notin\{i_1,\dots,i_{2m}\}$) or $v(p_i)=1$ and there is a $j$ such that $i_j=i$ (because each variable occurs in at least one clause). Therefore condition \emph{(i')} is satisfied. Now suppose there are $a<b$ that violate condition \emph{(ii)}. By definition of $i_b$ it holds that $p_{i_b}\in C_{b'}$ and $v'(p_{i_b})=1$. Analogously, $p_{i_a}\in C_{a'}$ and $v'(p_{i_a})=1$. By the supposition $p_{i_b}\in C_{a'}$ and $p_{i_a}\neq p_{i_b}$. But since $v'(p_{i_a})=v'(p_{i_b})=1$, that is a contradiction to the fact that in clause $C_{a'}$ only one variable evaluates to true.

\qte{$\Rightarrow$}: If $\phi_v$ is unsatisfiable, there are $i_1,\dots,i_{2m}$ such that \emph{(i')} and \emph{(ii)} hold. Let the valuation $v'\colon\{p_1,\dots,p_n\}\to\{0,1\}$ be defined by
\[v'(p_i)\dfn \left\{\begin{array}{l@{\ }l}1&\text{if $i\in\{i_1,\dots,i_{2m}\}$}\\0&\text{else}\end{array}\right..\]
Note that $v'$ is a continuation of $v$ because \emph{(i')} holds.

We will now prove that in each of the clauses $C_1,\dots,C_m$ exactly one variable evaluates to true under $v'$. Therefore let $j\in\{1,\dots,m\}$ be arbitrarily chosen.

By choice of $i_j$ it holds that $p_{i_j}\in C_j$. It follows by definition of $v'$ that $v'(p_{i_j})=1$. Hence, there is at least one variable in $C_j$ that evaluates to true.

Now suppose that besides $p_{i_j}$ another variable in $C_j$ evaluates to true. Then by definition of $v'$ it follows that there is a $\ell\in\{1,\dots,2m\}$, $\ell\neq j$, such that this other variable is $p_{i_\ell}$. We now consider two cases.

\emph{Case~$j < \ell$}:
This is a contradiction to \emph{(ii)} since, by definition of $\ell$, $p_{i_\ell}$ is in $C_{j'}$ as well as, by definition of $i_\ell$, in $C_{\ell'}$ and $i_j\neq i_\ell$.

\emph{Case~$\ell< j$}:
Since $j\in\{1,\dots,m\}$ it follows that $\ell\leq m$. Since $C_{\ell'}=C_{(\ell+m)'}$ it holds that $p_{i_{\ell+m}}\in C_{\ell'}$ and $p_{i_{\ell+m}}\in C_{(\ell+m)'}$. Furthermore $\ell<\ell+m$ and thus, by condition \emph{(ii)}, it must hold that $i_\ell= i_{\ell+m}$. Therefore $p_{i_{\ell+m}}\in C_j$ and $v'(p_{i_{\ell+m}})=1$. Because $j<\ell+m$ this is a contradiction to condition \emph{(ii)} as in the first case.
\end{claimproof}

The correctness of the reduction now follows with the observation that $\phi$ is equivalent to
\[\bignor\limits_{v\colon {\{p_1,\dots,p_k\}}\to\{0,1\}} \phi_v\]
and that $\phi$ is unsatisfiable iff $\phi_v$ is unsatisfiable for all valuations $v\colon \{p_1,\dots$ \allowbreak $,p_k\} \to\{0,1\}$.

The \oneInThreeQcsp instance is true iff every valuation $v\colon\{p_1,\dots,p_k\}\to \{0,1\}$ can be continued to a valuation $v'\colon\{p_1,\dots,p_n\}\to\{0,1\}$ such that in each of the clauses $\{C_1,\dots,C_m\}$ exactly one variable evaluates to true under $v'$ iff, by \cref{mdl oldclaimtwo}, $\phi_{v}$ is unsatisfiable for all $v\colon\{p_1,\dots,p_k\}\to\{0,1\}$ iff, by the above observation, $\phi$ is unsatisfiable.
\end{proof}

Next we turn to (non-monotone) poor man's logic.

\begin{theorem}\label{poor man dep complexity}
If $\{\Box,\Diamond,\wedge,\aneg,\dep[]\}\subseteq M$ then \MDL[M]\hsat is \NEXP-complete.
\end{theorem}
\begin{proof}
Sevenster showed that the problem is in \NEXP in the case of $\nor\notin M$ \cite[Lemma~14]{se09}. Together with \cref{bullet distributivity} and the fact that $\exist\NEXP=\NEXP$ the upper bound applies.

For the lower bound we reduce \dqbfcnf, which was shown to be \NEXP-hard by Peterson et al.~\cite[Lemma~5.2.2]{pereaz01}\footnote{Peterson et al.~showed \NEXP-hardness for \dqbf without the restriction that the formulas must be in \cnf. However, the restriction does not lower the complexity since every propositional formula is satisfiability-equivalent to a formula in \cnf whose size is bounded by a polynomial in the size of the original formula.}, to our problem.

An instance of \dqbfcnf consists of universally quantified Boolean variables $p_1,\dots,p_k$, existentially quantified Boolean variables $p_{k+1},\dots,p_n$, dependence constraints $P_{k+1},\dots,P_n\subseteq\{p_1,\dots,p_k\}$ and a set of clauses each consisting of three (not necessarily distinct) literals. Here, $P_i$ intuitively states that the value of $p_i$ only depends on the values of the variables in $P_i$. Now, \dqbfcnf is the set of all those instances for which there is a collection of functions $f_{k+1},\dots,f_n$ with $f_i\colon\{0,1\}^{P_i}\to\{0,1\}$ such that for every valuation $v\colon\{p_1,\dots,p_k\}\to\{0,1\}$ there is at least one literal in each clause that evaluates to true under the valuation $v'\colon\{p_1,\dots,p_n\}\to\{0,1\}$ defined by
\[v'(p_i)\dfn\left\{\begin{array}{l@{\quad}l}
v(p_i)&\text{if $i\in\{1,\dots,k\}$}\\
f_i(v\upharpoonright P_i)&\text{if $i\in\{k+1,\dots,n\}$}
\end{array}\right.\enspace.\]

The functions $f_{k+1},\dots,f_n$ act as restricted existential quantifiers, \ie for an $i\in\{k+1,\dots,n\}$ the variable $p_i$ can be assumed to be existentially quantified dependent on all universally quantified variables in $P_i$ (and, more importantly, independent of all universally quantified variables not in $P_i$).
Dependencies are thus explicitly specified through the dependence constraints and can contain -- but are not limited to -- the traditional sequential dependencies, \eg the quantifier sequence $\forall p_1 \exists p_2 \forall p_3 \exists p_4$ can be modeled by the dependence constraints $P_2=\{p_1\}$ and $P_4=\{p_1,p_3\}$.

For the reduction from \dqbfcnf to $\MDL[{\Box,\Diamond,\allowbreak\wedge,\allowbreak\aneg,\dep[]}]\hsat$ we use an idea from Hemaspaandra \cite[Theorem~4.2]{he01}. There, \P\SPACE-hardness of $\MDL[\Box,\Diamond,\wedge,\allowbreak\aneg]\hsat$ over the class $\mathcal F_{\leq 2}$ of all Kripke structures in which every world has at most two successors is shown. The crucial point in the proof is to ensure that every Kripke structure satisfying the constructed $\MDL[\Box,\Diamond,\wedge,\aneg]$ formula adheres to the structure of a complete binary tree and does not contain anything more than this tree. In the class $\mathcal F_{\leq 2}$ this is automatically the case since in a complete binary tree all worlds already have two successors.

Although in our case there is no such an a priori restriction and therefore we cannot make sure that every satisfying structure is not more than a binary tree, we are able to use dependence atoms to ensure that everything in the structure that does not belong to the tree is essentially nothing else than a copy of a subtree. This will be enough to show the desired reducibility.

Let $p_1,\dots,p_k$ be the universally quantified and $p_{k+1},\dots,p_n$ the existentially quantified variables of a \dqbfcnf instance $\phi$ and let $P_{k+1},\dots,P_n$ be its dependence constraints and $\{l_{11},l_{12},l_{13}\},\dots,\allowbreak\{l_{m1},\allowbreak l_{m2},l_{m3}\}$ its clauses. Then the corresponding $\MDL[\Box,\Diamond,\allowbreak\wedge,\allowbreak\aneg,\dep[]]$ formula is
\[\begin{array}{lcl@{\quad}c}
g(\phi)\dfn&&\bigwedge\limits_{i=1}^n \Box^{i-1}(\Diamond \Box^{n-i}p_i \wedge \Diamond \Box^{n-i}\overline{p_i})&(i)\\
&\wedge&\bigwedge\limits_{i=1}^m \Diamond^n (\overline{l_{i1}} \wedge \overline{l_{i2}} \wedge \overline{l_{i3}} \wedge f_i)&(ii)\\
&\wedge&\bigwedge\limits_{i=1}^m \Box^n \dep[l'_{i1},l'_{i2},l'_{i3},f_i]&(iii)\\
&\wedge&\Box^k\Diamond^{n-k}\big(\overline{f_1} \wedge \dots \wedge \overline{f_m} \;\wedge\; \bigwedge_{i=k+1}^n \dep[P_i,p_i]\big)&(iv)
\end{array}\]
where $p_1,\dots,p_n,f_1,\dots,f_m$ are atomic propositions and $l'_{ij}\dfn\left\{\begin{array}{l@{\quad}l}p&\text{if $l_{ij}=p$,}\\p&\text{if $l_{ij}=\overline{p}$.}\end{array}\right.$

Now if $\phi$ is valid, consider the frame which consists of a complete binary tree with $n$ levels (not counting the root) and where each of the $2^n$ possible labelings of the atomic propositions $p_1,\dots,p_n$ occurs in exactly one leaf. Additionally, for each $i\in\{1,\dots, m\}$ $f_i$ is labeled in exactly those leaves in which $l_{i1}\vee l_{i2}\vee l_{i3}$ is false. This frame obviously satisfies $(i)$, $(ii)$ and $(iii)$. And since the modalities in $(iv)$ model the quantors of $\phi$, $\overline{f_i}$ is true exactly in the leaves in which $l_{i1}\vee l_{i2}\vee l_{i3}$ is true and the \dep[] atoms in $(iv)$ model the dependence constraints of $\phi$, $(iv)$ is also true and therefore $g(\phi)$ is satisfied in the root of the tree.

As an example see \cref{fig:generated frame} for a frame satisfying $g(\phi)$ if the first clause in $\phi$ is $\{\overline{p_1},p_n\}$.
\begin{figure}[ht]
\begin{center}
\begin{tikzpicture}[auto,every state/.style={minimum size=4mm}]
\node[state]           (z0)  at (11.5,10)                   {};
\node[state]           (z11) at (8,9)                     {};
\node                  (z12) at (15,9)                   {$\begin{array}{@{}c@{}}\phantom{\vdots}\\\vdots\end{array}$};
\node[rotate=-45]                  (zn1c) at (7,8)                  {$\vdots$};
\node[rotate=45]                  (zn2c) at (9,8)                   {$\vdots$};
\node[state]           (zn1) at (6,7)                     {};
\node[state,label=below:{$\begin{array}{c}p_1\\p_2\\\vdots\\p_n\end{array}$}]           (zn1a) at (5,6)                     {};
\node[state,label={below:{$\begin{array}{c}p_1\\p_2\\\vdots\\\overline{p_n}\\f_1\end{array}$}}] (zn1b) at (7,6)                     {};
\node                  (znn1) at (8,6)                    {$\cdots$};
\node[state]           (zn2) at (10,7)                     {};
\node[state,label=below:{$\begin{array}{c}p_1\\\overline{p_2}\\\vdots\\p_n\end{array}$}]           (zn2a) at (9,6)                     {};
\node[state,label={below:{$\begin{array}{c}p_1\\\overline{p_2}\\\vdots\\\overline{p_n}\\f_1\end{array}$}}] (zn2b) at (11,6)                     {};

\path[->]       (z0)    edge                    node {$p_1$}        (z11)
                (z0)    edge                    node {$\overline{p_1}$}        (z12)
                (zn1)   edge                    node {$p_n$}        (zn1a)
                (zn1)   edge                    node {$\overline{p_n}$}        (zn1b)
                (zn2)   edge                    node {$p_n$}        (zn2a)
                (zn2)   edge                    node {$\overline{p_n}$}        (zn2b);
\path[->]       (z11)   edge                    node[label={-180:$p_2\,$}] {}        (zn1c)
                (z11)   edge                    node[label={3:$\;\overline{p_2}$}] {}        (zn2c);
\path[->]       (zn1c)   edge                            (zn1)
                (zn2c)   edge                            (zn2);
\end{tikzpicture}
\caption{Frame satisfying $g(\phi)$}
\label{fig:generated frame}
\end{center}
\end{figure}

If, on the other hand, $g(\phi)$ is satisfiable, let $W$ be a frame and $t$ a world in $W$ such that $W,\{t\}\models g(\phi)$.
Now $(i)$ enforces $W$ to contain a complete binary tree $T$ with root $t$ such that each labeling of $p_1,\dots,p_n$ occurs in a leaf of $T$.

We can further assume w.l.o.g.~that $W$ itself is a tree since in \MDL different worlds with identical proposition labelings are indistinguishable and therefore every frame can simply be unwinded to become a tree. Since the modal depth of  $g(\phi)$ is $n$ we can assume that the depth of $W$ is at most $n$. And since $(i)$ enforces that every path in $W$ from $t$ to a leaf has a length of at least $n$, all leaves of $W$ lie at levels greater than or equal to $n$. Altogether we can assume that $W$ is a tree, that all its leaves lie at level $n$ and that it has the same root as $T$. The only difference is that the degree of $W$ may be greater than that of $T$.

But we can nonetheless assume that up to level $k$ the degree of $W$ is~2 $(*)$. This is the case because if any world up to level $k-1$ had more successors than the two lying in $T$, the additional successors could be omitted and $(i)$, $(ii)$, $(iii)$ and $(iv)$ would still be fulfilled. For $(i)$, $(ii)$ and $(iii)$ this is clear and for $(iv)$ it holds because $(iv)$ begins with $\Box^k$.

We will now show that, although $T$ may be a proper subframe of $W$, $T$ is already sufficient to fulfill $g(\phi)$. From this the validity of $\phi$ will follow immediately.

\begin{claim}\label{poor man oldclaim}
$T,\{t\}\models g(\phi)$.
\end{claim}
\begin{claimproof}{\cref*{poor man oldclaim}}
We consider sets of leaves of $W$ that satisfy $\overline{f_1} \wedge \dots \wedge \allowbreak \overline{f_m} \;\wedge\;\allowbreak \bigwedge_{i=k+1}^n \dep[P_i,p_i]$ and that can be reached from the set $\{t\}$ by the modality sequence $\Box^k\Diamond^{n-k}$. Let $S$ be such a set and let $S$ be chosen so that there is no other such set that contains less worlds outside of $T$ than $S$ does. Assume there is a $s\in S$ that does not lie in $T$.

Let $i\in\{1,\dots,m\}$ and let $s'$ be the leaf in $T$ that agrees with $s$ on the labeling of $p_1,\dots,p_n$. Then, with $W,\{s\}\models \overline{f_i}$ and $(iii)$, it follows that $W,\{s'\}\models\overline{f_i}$.

Let $S' \dfn (S\setminus\{s\})\cup\{s'\}$. Then it follows by the previous paragraph that $W,S'\models \overline{f_1} \wedge \dots \wedge \overline{f_m} $. Since $W,S\models\bigwedge_{i=k+1}^n \dep[P_i,p_i]$ and $s'$ agrees with $s$ on the propositions $p_1,\dots,p_n$ it follows that $W,S'\models\bigwedge_{i=k+1}^n \dep[P_i,p_i]$. Hence, $S'$ satisfies $\overline{f_1} \wedge \dots \wedge \overline{f_m} \;\wedge\; \bigwedge_{i=k+1}^n \dep[P_i,p_i]$ and as it only differs from $S$ by replacing $s$ with $s'$ it can be reached from $\{t\}$ by $\Box^k\Diamond^{n-k}$ because $s$ and $s'$ agree on $p_1,\dots,p_k$ and, by $(*)$, $W$ does not differ from $T$ up to level $k$. But this is a contradiction to the assumption since $S'$ contains one world less than $S$ outside of $T$. Thus, there is no $s\in S$ that does not lie in $T$ and therefore $(iv)$ is fulfilled in $T$. Since $(i)$, $(ii)$ and $(iii)$ are obviously also fulfilled in $T$, it follows that $T,\{t\}\models g(\phi)$.
\end{claimproof}

$(ii)$ ensures that for all $i\in\{1,\dots,m\}$ there is a leaf in $W$ in which $\neg(l_{i1}\vee l_{i2}\vee l_{i3})\wedge f_i$ is true. This leaf can lie outside of $T$. However, $(iii)$ ensures that all leaves that agree on the labeling of $l_{i1}$, $l_{i2}$ and $l_{i3}$ also agree on the labeling of $f_i$. And since there is a leaf where $\neg(l_{i1}\vee l_{i2}\vee l_{i3})\wedge f_i$ is true, it follows that in all leaves, in which $\neg(l_{i1}\vee l_{i2}\vee l_{i3})$ is true, $f_i$ is true. Conversely, if $\overline{f_i}$ is true in an arbitrary leaf of $W$ then so is $l_{i1}\vee l_{i2}\vee l_{i3}$ $(**)$.

The modality sequence $\Box^k\Diamond^{n-k}$ models the quantors of $\phi$ and $\bigwedge_{i=k+1}^n \dep[P_i,\allowbreak p_i]$ models its dependence constraints. And so there is a bijective correspondence between sets of worlds reachable in $T$ by $\Box^k\Diamond^{n-k}$ from $\{t\}$ and that satisfy $\bigwedge_{i=k+1}^n \dep[P_i,p_i]$ on the one hand and truth assignments to $p_1,\dots,p_n$ generated by the quantors of $\phi$ and satisfying its dependence constraints on the other hand.
Additionally, by $(**)$ follows that $\overline{f_1}\wedge\dots\wedge\overline{f_m}$ implies $\bigwedge_{i=1}^m (l_{i1}\vee l_{i2} \vee l_{i3})$ and since $T,\{t\}\models g(\phi)$, $\phi$ is valid.
\end{proof}

\subsection{Cases with only one modality}

We will now examine formulas with only one modality.

\begin{theorem}\label{only one modality}
Let $M \subseteq \{\Box, \Diamond, \wedge, \vee, \aneg, \true, \false, \nor\}$ with $\Box\notin M$ or $\Diamond \notin M$. Then the following hold:
\begin{enumerate}
 \item \MDL[{M \cup \{\dep[]\}}]\hsat $\leqpm$ \MDL[M \cup \{\true, \false\}]\hsat, \ie adding the $\dep$ operator does not increase the complexity if we only have one modality.
  \item For every \MDL[{M \cup\{\dep[]\}}] formula $\phi$ it holds that $\nor$ is equivalent to $\vee$, \ie $\phi$ is equivalent to every formula that is generated from $\phi$ by replacing some or all occurrences of $\nor$ by $\vee$ and vice versa.
\end{enumerate}
\end{theorem}
\begin{proof}
Every negation $\neg \dep$ of a dependence atom is by definition always equivalent to $\false$ and can thus be replaced by the latter. For positive $\dep$ atoms and the $\nor$ operator we consider two cases.

\emph{Case $\Diamond\notin M$.} If an arbitrary $\MDL[{\Box,\wedge, \vee, \aneg, \true, \false, \dep[], \nor}]$ formula $\phi$ is satisfiable then it is so in an intransitive singleton frame, \ie a frame that only contains one world which does not have a successor, because there every subformula that begins with a $\Box$ is automatically satisfied. In a singleton frame all $\dep$ atoms obviously hold and $\nor$ is equivalent to $\vee$. Therefore the (un-)satisfiability of $\phi$ is preserved when substituting every $\dep$ atom in $\phi$ with $\true$ and every $\nor$ with $\vee$ (or vice versa).

\emph{Case $\Box \notin M$.} If an arbitrary $\MDL[{\Diamond,\wedge, \vee, \aneg, \true, \false, \dep[], \nor}]$ formula $\phi$ is satisfiable then, by the downward closure property, there is a frame $W$ with a world $s$ such that $W,\{s\} \models \phi$. Since there is no $\Box$ in $\phi$, every subformula of $\phi$ is also evaluated in a singleton set (because a $\Diamond$ can never increase the cardinality of the evaluation set). And as in the former case we can replace every $\dep$ atom with $\true$ and every $\nor$ with $\vee$ (or vice versa).
\end{proof}

Thus we obtain the following consequences -- note that this takes care of all remaining fragments for the unbounded arity case.
\begin{corollary}\label{one modality cases}
\begin{enumerate}
 \item If $\{\wedge,\aneg\}\subseteq M\subseteq\{\Box,\Diamond,\wedge,\vee,\aneg,\true,\false,\dep[],\nor\}$, $M\cap\{\vee,\nor\}\allowbreak\neq \emptyset$ and $|M\cap\{\Box,\Diamond\}|=1$ then \MDL[M]\hsat and \MDLk[M]\hsat are \NP-complete for all $k\geq 0$.
 \item If $\{\wedge,\aneg\}\subseteq M\subseteq\{\Box,\Diamond,\wedge,\aneg,\true,\false,\dep[]\}$ and $|M\cap\{\Box,\Diamond\}|=1$ then \linebreak $\MDL[M]\hsat \in\P$.
 \item If $\{\wedge\}\subseteq M\subseteq\{\Box,\Diamond,\wedge,\vee,\true,\false,\dep[],\nor\}$ and $|M\cap\{\Box,\Diamond\}|=1$ then \linebreak $\MDL[M]\hsat \in \P$.
 \item If $\wedge \notin M$ then $\MDL[M]\hsat \in \P$.
\end{enumerate}
\end{corollary}
\begin{proof}
Part a) without the $\dep$ and $\nor$ operators is exactly \cite[Theorem~6.2(2)]{he01}. \Cref{only one modality} extends the result to the case with the new operators.
Part b) is \cite[Theorem~6.4(c,d)]{he01} together with \cref{only one modality}a for the $\dep$ operator.
Part c) is \cite[Theorem~6.4(e,f)]{he01} together with \cref{only one modality}a,b.

Part d) without $\dep$ and $\nor$ is \cite[Theorem~6.4(b)]{he01}. The proof for the case with the new operators is only slightly different: Let $\phi$ be an arbitrary \MDL[M ] formula. By the same argument as in the proof of \cref{only one modality}b we can replace all top-level (\ie not lying inside a modality) occurrences of $\nor$ in $\phi$ with $\vee$ to get the equivalent formula $\phi'$. $\phi'$ is of the form $\Box \psi_1\vee\dots\vee\Box\psi_k\vee\Diamond\sigma_1\vee\dots\vee\Diamond\sigma_m\vee a_1\vee\dots\vee a_s$ where every $\psi_i$ and $\sigma_i$ is a \MDL[M ] formula and every $a_i$ is an atomic formula. If $k>0$ or any $a_i$ is a literal, $\true$ or a dependence atom then $\phi'$ is satisfiable. Otherwise it is satisfiable iff one of the $\sigma_i$ is satisfiable and this can be checked recursively in polynomial time.
\end{proof}

\subsection{Bounded arity dependence}\label{sec:bounded sat}
Finally, we investigate the remaining fragments for the bounded arity case.

\begin{theorem}\label{bounded dep upper}
Let $k\geq 0$. Then the following hold:
\begin{enumerate}
 \item If $M \subseteq \{\Box, \Diamond, \wedge, \vee, \aneg, \true, \false, \dep[]\}$ then $\MDLk[M]\hsat \in \P\SPACE$.
 \item If $M \subseteq \{\Box, \Diamond, \wedge, \aneg, \true, \false, \dep[]\}$ then $\MDLk[M]\hsat \in \SigmaP{3}$.
\end{enumerate}
\end{theorem}
\begin{proof}
a) Let $\phi \in \MDLk[M ]$. Then by \cite[Theorem~6]{se09} there is an ordinary modal logic formula $\phi^T$ which is equivalent to $\phi$ on singleton sets of evaluation, \ie for all Kripke structures $W$ and states $w$ in $W$
\[W,\{w\}\models \phi \text{\quad{}iff\quad} W,w\models \phi^T.\]
Here $\phi^T$ is constructed from $\phi$ in the following way:
Let $\dep[p_{i_{1,1}},\dots,p_{i_{1,k_1}},\allowbreak p_{i_{1,k_1+1}}],\allowbreak\dots,\allowbreak\dep[p_{i_{n,1}},\dots,p_{i_{n,k_n}},p_{i_{n,k_n+1}}]$ (for $k_1,\dots,k_n\leq k$) be all dependence atoms occurring inside $\phi$ (in an arbitrary order and including multiple occurrences of the same atom in $\phi$ multiple times) and for all $j \geq 0$ let
\[B_j \dfn \{\alpha_f(p_1,\dots,p_j) \mid f\colon\{\true,\false\}^j \to \{\true,\false\}\text{ is a total Boolean function}\},\]
where $\alpha_f(p_1,\dots,p_j)$ is a propositional encoding of $f$, \eg
\[\alpha_f(p_1,\dots,p_j) \dfn \bigvee_{(i_1,\dots,i_j)\in f^-1(\true)}\, p_1^{i_1}\wedge\dots\wedge p_j^{i_j},\]
with $p^i\dfn\left\{\begin{array}{l@{\text{ if }}l}p&i=\true\\\neg p&i=\false\end{array}\right.$.
Note that for all $f\colon\{\true,\false\}^j \to \{\true,\false\}$ and all valuations $V\colon\{p_1,\dots,p_j\}\to \{\true,\false\}$ it holds that $V \models \alpha_f$ iff $f(V(p_1),\dots,V(p_j))=\true$.

Then $\phi^T$ is defined as
\[\bigvee_{\alpha_1\in B_{k_1}} \dots \bigvee_{\alpha_n\in B_{k_n}}\,\phi'(\alpha_1,\dots,\alpha_n),\]
where $\phi'(\alpha_1,\dots,\alpha_n)$ is generated from $\phi$ by replacing each dependence atom $\dep[p_{i_{\ell,1}},\allowbreak\dots,\allowbreak p_{i_{\ell,k_\ell}},p_{i_{\ell,k_\ell+1}}]$ with the propositional formula $\alpha_\ell(p_{i_{\ell,1}},\dots,p_{i_{\ell,k_{\ell}}}) \leftrightarrow p_{i_{\ell,k_\ell+1}}$.
Note that for all $\ell\in\{1,\dots,n\}$ we have that $|\alpha_\ell| \in \bigo{2^{k_\ell}}$ and $|B_{k_\ell}| = 2^{2^{k_\ell}}$. Therefore
\[|\phi^T| \in \prod_{1\leq \ell \leq n} 2^{2^{k_\ell}}\,\cdot\,|\phi|\cdot \bigo{2^{k_\ell}} \subseteq \bigo{\left(2^{2^k}\right)^n\cdot |\phi|}.\]
This means that $\phi^T$ is an exponentially (in the size of $\phi$) large disjunction of terms of linear size (remember that $k$ is fixed). $\phi^T$ is satisfiable if and only if at least one of its terms is satisfiable. Hence we can nondeterministically guess in polynomial time which one of the exponentially many terms should be satisfied and then check in deterministic polynomial space whether this one is satisfiable. The latter is possible because $\phi'(\alpha_1,\dots,\alpha_n)$ is an ordinary modal logic formula and the satisfiability problem for this logic is in \P\SPACE \cite{la77}. Altogether this leads to $\MDLk[M]\hsat \in \exist \P\SPACE = \P\SPACE$.

\smallskip
b) In this case we cannot use the same argument as before without modifications since that would only lead to a \P\SPACE upper bound again. The problem is that in the contruction of $\phi^T$ we introduce the subformulas $\alpha_\ell$ and these may contain the $\vee$ operator. We can, however, salvage the construction by looking inside Ladner's \P\SPACE algorithm \cite[Theorem~5.1]{la77}. For convenience we restate (a slightly modified form of) it in \cref{algo:sat}. It holds for all ordinary modal logic formulas $\phi$ that $\phi$ is satisfiable (by a Kripke structure from the class K) if and only if \lstinline!satisfiable($\{\phi\}$, $\emptyset$, $\emptyset$)!$=$\lstinline!true!.

\Needspace{12\baselineskip}
\begin{lstlisting}[caption=\lstinline!satisfiable($T$\, $A$\, $E$)!,label=algo:sat]
if $T\nsubseteq \atom$ then  $\quad$//$\atom$ denotes the set of atomic propositions, their negations
                    $\quad$//and the constants $\true$ and $\false$
  choose $\psi\in T\setminus \atom\quad$   //deterministically (but arbitrarily)
  set $T' := T\setminus \{\psi\}$
  if $\psi = \psi_1 \wedge \psi_2$ then
    return satisfiable($T' \cup \{\psi_1, \psi_2\}$, $A$, $E$)
  elseif $\psi = \psi_1 \vee \psi_2$ then
    nondeterministically existentially guess $i\in\{1,2\}$
    return satisfiable($T' \cup \{\psi_i\}$, $A$, $E$)
  elseif $\psi = \Box\psi_1$ then
    return satisfiable($T'$, $A\cup\{\psi_1\}$, $E$)
  elseif $\psi = \Diamond\psi_1$ then
    return satisfiable($T'$, $A$, $E\cup\{\psi_1\}$)
  end
else
  if $T$ is consistent then
    if $E \neq \emptyset$
      nondeterministically universally guess $\psi \in E$
      return satisfiable($A \cup \{\psi\}$, $\emptyset$, $\emptyset$)
    else
      return true
    end
  else
    return false
  end
end
\end{lstlisting}

The algorithm works in a top-down manner and runs in alternating polynomial time. It universally guesses when encountering a $\Box$ operator and existentially guesses when encountering a $\vee$ operator -- in all other cases it is deterministic.
Now, to check whether $\phi^T$ is satisfiable we first existentially guess which of the exponentially many terms should be satisfied and then check whether this term $\phi'(\alpha_1,\dots,\alpha_n)$ is satisfiable by invoking \lstinline!satisfiable($\{\phi'(\alpha_1,\dots,\alpha_n)\}$, $\emptyset$, $\emptyset$)!.

To see that this in fact gives us a $\SigmaP{3}$-algorithm note that $\phi$ does not contain any disjunctions. Hence also $\phi'(\alpha_1,\dots,\alpha_n)$ contains no disjunctions apart from the ones that occur inside one of the subformulas $\alpha_1,\dots,\alpha_n$. Therefore \cref{algo:sat} does not do any nondeterministic existential branching apart from when processing an $\alpha_i$. But in the latter case it is impossible to later nondeterministically universally branch because univeral guessing only occurs when processing a $\Box$ operator and these cannot occur inside an $\alpha_i$, since these are purely propositional formulas. Therefore the algorithm, if run on a formula $\phi'(\alpha_1,\dots,\alpha_n)$ as input, is essentially a $\PiP{2}$ algorithm. Together with the existential guessing of the term in the beginning we get that $\MDLk[M]\hsat \in \exist \PiP{2} = \SigmaP{3}$.
\end{proof}

\begin{theorem}\label{poor man bounded}
If $\{\Box, \Diamond, \wedge, \aneg, \dep[]\} \subseteq M$ then $\MDLpara[M]{3}\hsat$ is \SigmaP{3}-hard.
\end{theorem}
\begin{proof}
We use the same construction as in the hardness proof for \cref{poor man dep complexity} to reduce the problem \qbfcnf, which was shown to be \SigmaP{3}-complete by\linebreak{} Wrathall~\cite[Corollary~6]{wr77}, to our problem. \qbfcnf is the set of all propositional sentences of the form
\[\exists p_1\dots\exists p_k \forall p_{k+1}\dots \forall p_\ell \exists p_{\ell+1}\dots \exists p_n \bigwedge_{i=1}^m (l_{i1}\vee l_{i2} \vee l_{i3}),\]
where the $l_{ij}$ are literals over $p_1,\dots,p_n$, which are valid.

Now let $\phi$ be a \qbfcnf instance, let $p_1,\dots,p_n$ be its variables and let $k$, $\ell$, $m$, $(l_{ij})_{\substack{i=1,\dots,m\\j=1,2,3}}$ be as above. Then the corresponding $\MDLpara[{\Box, \Diamond, \wedge, \aneg, \dep[]}]{3}$ formula is
\[\begin{array}{lcl@{\hspace*{2.9em}}c}
g(\phi)\dfn&&\bigwedge\limits_{i=1}^n \Box^{i-1}(\Diamond \Box^{n-i}p_i \wedge \Diamond \Box^{n-i}\overline{p_i})&(i)\\
&\wedge&\bigwedge\limits_{i=1}^m \Diamond^n (\overline{l_{i1}} \wedge \overline{l_{i2}} \wedge \overline{l_{i3}} \wedge f_i)&(ii)\\
&\wedge&\bigwedge\limits_{i=1}^m \Box^n \dep[l'_{i1},l'_{i2},l'_{i3},f_i]&(iii)\\
&\wedge&\Diamond^k\Box^{\ell-k}\Diamond^{n-\ell}(\dep[p_1]\wedge \dots \wedge \dep[p_k]\ \wedge\ \overline{f_1} \wedge \dots \wedge \overline{f_m})&(iv)
\end{array}\]
where $p_1,\dots,p_n,f_1,\dots,f_m$ are atomic propositions and $l'_{ij}\dfn\left\{\begin{array}{l@{\quad}l}p&\text{if $l_{ij}=p$,}\\p&\text{if $l_{ij}=\overline{p}$.}\end{array}\right.$

The proof that $g$ is a correct reduction is essentially the same as for \cref{poor man dep complexity} (also see \cref{fig:generated frame} for an example of a typical Kripke structure satisfying $g(\phi)$). The only difference is that there we had arbitrary dependence atoms in part $(iv)$ of $g(\phi)$ whereas here we only have 0-ary dependence atoms. This difference is due to the fact that there we had to be able to express arbitrary dependencies because we were reducing from \dqbfcnf whereas here we only have two kinds of dependencies for the existentially quantified variables: either complete constancy (for the variables that get quantified before any universal variable does) or complete freedom (for the variables that get quantified after all universal variables are already quantified). The former can be expressed by 0-ary dependence atoms and for the latter we simply omit any dependence atoms.

Note that it might seem as if with the same construction even $\Sigma^p_k$-hardness for arbitrary $k$ could be proved by having more alternations between the two modalities in part $(iv)$ of $g(\phi)$. The reason that this does not work is that we do not really ensure that a structure fulfilling $g(\phi)$ is not more than a binary tree, \eg it can happen that the root node of the tree has three successors: one in whose subtree all leaves on level $n$ are labeled with $p_1$, one in whose subtree no leaves are labeled with $p_1$ and one in whose subtree only some leaves are labeled with $p_1$. Now, the first diamond modality can branch into this third subtree and then the value of $p_1$ is not yet determined. Hence the modalities alone are not enough to express alternating dependencies and hence we need the $\dep[p_i]$ atoms in part $(iv)$ to ensure constancy.
\end{proof}

\begin{corollary}\label{bounded dep concrete}
\begin{enumerate}
 \item Let $k\geq 0$ and $\{\Box, \Diamond, \wedge, \vee, \aneg\} \subseteq M$. Then $\MDLk[M]\hsat$ is \P\SPACE-complete.
 \item Let $k\geq 3$ and $\{\Box, \Diamond, \wedge, \aneg, \dep[]\} \subseteq M \subseteq \{\Box, \Diamond, \wedge, \aneg, \true, \false, \dep[], \nor\}$. Then $\MDLk[M]\hsat$ is \SigmaP{3}-complete.
\end{enumerate}
\end{corollary}
\begin{proof}
The lower bound for a) is due to the \P\SPACE-completeness of ordinary modal logic satisfiability which was shown in \cite{la77}. The upper bound follows from \cref{bounded dep upper}a, \cref{bullet distributivity}b and the fact that $\exist \P\SPACE = \P\SPACE$.

The lower bound for b) is \cref{poor man bounded}. The upper bound follows from \cref{bounded dep upper}b, \cref{bullet distributivity} and $\exist \SigmaP{3} = \SigmaP{3}$.
\end{proof}

Note that in \cref{bounded dep concrete}b we require that $k\geq 3$. This is due to the fact that the reduction in our proof of \cref{poor man bounded} needs ternary dependence atoms.
With $k\leq 2$ this reduction does not work and thus the lower bound remains open in that case.

\section{Model checking}\label{sec:mdl-mc}
Now that we have classified the operator fragments of \MDL with respect to the complexity of satisfiability we turn over to model checking and do the same. We will investigate -- according to \cref{def:mc} -- for all $M\subseteq \{\AX,\EX,\wedge,\sor,\nor,\neg,\dep[]\}$ and all $k\geq 0$ the problem
\problemdef{
$\MDL[M]\hmc$ ($\MDLk[M]\hmc$)
}{
An $\MDL[M]$ (resp.~$\MDLk[M]$) formula $\phi$, a Kripke structure $K=(S,R,\pi)$ and a team $T\subseteq S$.
}{
$K,T\models \phi$?
}

The first lemma shows that whether to include $\top$, $\bot$ or $\neg$ in a sublogic \MDL[M] of \MDL does not affect the complexity of \MDL[M]\hmc.
\begin{lemma}\label{neg-dont-matter}
Let $M$ be an arbitrary set of \MDL operators, \ie $M\subseteq\{\allMDL,\allowbreak\bot,\top\}$. Then it holds that 
\[\MDL[M]\hmc\ \equivpm\ \MDL[M\setminus \{\true,\false,\neg\}]\hmc.\]
\end{lemma}
\begin{proof}
It suffices to show $\leqpm$.
So let $K=(S,R,\pi)$ be a Kripke structure, $T\subseteq S$, $\phi \in \MDL[M]$ and $\var{\phi}=\{p_1,\dots,p_n\}$.
Let $p_1',\dots,p_n',t,f$ be fresh propositional variables.
Then $K,T \models \phi$ iff $K',T\models \phi'$, where
$K' \dfn (S,R,\pi')$ 
with $\pi'$ defined by
\[\begin{array}{rcl}
\pi'(s)\cap \{t,f\}            &\dfn& \{t\},\\
\pi'(s)\cap \{p_i,p_i'\}      &\dfn& \left\{
                         \begin{array}{ll}
                           \{p_i\}          & \text{if }p_i\in\pi(s),\\
                           \{p_i'\}       & \text{if }p_i\notin\pi(s),
                         \end{array}
                       \right.
\end{array}\]
for all $i\in\{1,\dots,n\}$ and $s\in S$, and $\phi \in \MDL[M\setminus \{\true,\false,\neg\}]$ defined by
\[\phi' \ \dfn\  \phi(p_1'/\neg p_1)(p_2'/\neg p_2)\dots(p_n'/\neg p_n)(t / \true)(f / \false).\]
\end{proof}

Now we will show that the most general of our problems is in \NP and therefore all model checking problems investigated later are as well.

\begin{proposition}\label{mdlmc-in-np}
Let $M$ be an arbitrary set of \MDL operators. Then \MDLMC[M] is in \NP. And hence also $\MDLMCk[M]$ is in  $\NP$ for every $k\geq0$.
\end{proposition}
\begin{proof}
The following non-deterministic top-down algorithm 
clearly checks the truth of the formula $\phi$ on the Kripke structure $W$ in the evaluation set $T$ in polynomial time.

\Needspace{9\baselineskip}
\begin{lstlisting}[caption={\lstinline!check($W=(S,R,\pi)$, $\phi$, $T$)!}, label={algo:mdl check}]
case $\phi$
§\Needspace{2\baselineskip}§
when $\phi=\true$
  return true
§\Needspace{2\baselineskip}§
when $\phi=\false$
  return false
§\Needspace{5\baselineskip}§
when $\phi=p$
  foreach $s \in T$
    if not $p \in \pi(s)$ then
      return false
  return true
§\Needspace{5\baselineskip}§
when $\phi=\neg p$
  foreach $s \in T$
    if $p \in \pi(s)$ then
      return false
  return true
§\Needspace{7\baselineskip}§
when $\phi = \,\depp{p_1,\ldots,p_n}$
  foreach $(s,s')\in T\times T$
    if $\pi(s)\cap\{p_1,\dots,p_{n-1}\}\,=\,\pi(s')\cap\{p_1,\dots,p_{n-1}\}$ then
          // \ie $s$ and $s'$ agree on the valuations of the propositions $p_1,\dots, p_{n-1}$
      if ($q\in\pi(s)$ and not $q\in \pi(s')$) or (not $q\in\pi(s)$ and $q\in \pi(s')$) then
        return false
  return true
§\Needspace{6\baselineskip}§
when $\phi = \psi\vee \theta$
  guess two sets of states $A$, $B$ with $A\cup B = T$
  return (check($W,A,\psi$) and check($W,B,\theta$))
§\Needspace{2\baselineskip}§
when $\phi = \psi\nor \theta$
  return (check($W,T,\psi$) or check($W,T,\theta$))
§\Needspace{2\baselineskip}§
when $\phi= \psi\wedge \theta$
  return (check($W,T,\psi$) and check($W,T,\theta$))
§\Needspace{8\baselineskip}§
when $\phi= \AX \psi$
  $T':=\emptyset$
  foreach $s' \in S$
    foreach $s \in T$
      if $(s,s') \in R$ then
        $T':= T' \cup \{s'\}$
            // $T'$ is the set of all successors of all states in $T$, i.e. $T'=R(T)$
  return check($W,T',\psi$)
§\Needspace{7\baselineskip}§
when $\phi= \EX \psi$
  guess set of states $T'\subseteq S$
    foreach $s\in T$
      if there is no $s'\in T'$ with $(s,s')\in R$ then
        return false
            // $T'$ contains at least one successor of every state in T, i.e. $T'\in\sucteams{T}$
  return check($W,T',\psi$)
\end{lstlisting}

The algorithm runs in polynomial time since it processes each subformula of $\phi$ in polynomial time and exactly once.
\end{proof}

\subsection{Unbounded arity fragments}\label{sec:unbounded-fragments}

We now focus on the cases with unbounded arity dependence atoms -- though sometimes our results directly carry over to the bounded cases.

We first show that the model checking problem is \NP-hard in general and that this still holds without modalities.

\begin{theorem}\label{wedge-vee-np-complete}
Let $\{\wedge, \vee, \dep[]\}\subseteq M$. 
Then \MDLMC[M] is \NP-complete. Furthermore, \MDLMCk[M] is \NP-complete for every $k\geq 0$.
\end{theorem}
\begin{proof}
Membership in \NP follows from \cref{mdlmc-in-np}. For the hardness proof we reduce from \ThreeSAT to $\MDLpara[{\wedge, \vee, \dep[]}]{0}\hmc$.

For this purpose let $\phi = C_{1}\wedge\ldots\wedge C_{m}$
be an arbitrary $\cnf[3]$ formula
with variables $x_{1},\ldots,x_{n}$. Let $W$ be the Kripke structure $\left(S,R,\pi\right)$ over the atomic propositions $r_1,\dots,r_n,p_1,\dots,p_n$ defined by
\[\begin{array}{rcl}
S       &       \dfn & \{ s_{1}, \ldots, s_{m} \},\\
R & \dfn & \emptyset,\\
\pi(s_{i}) \cap \{r_j,p_j\} & \dfn & \begin{cases}
\{r_j,p_j\} &\mbox{ iff $x_j$ occurs in $C_{i}$ positively,}\\
\{r_j\} &\mbox{ iff $x_j$ occurs in $C_{i}$ negatively,}\\
\emptyset &\mbox{ iff $x_j$ does not occur in $C_{i}$.}\\
\end{cases}
\end{array}\]

Let $\psi$ be the $\MDLpara[{\wedge,\vee,\dep[]}]{0}$ formula
\[\bigvee_{j=1}^{n}\ r_j\wedge \dep[p_j]\]
and let $T\dfn\left\{s_{1},\dots,s_{m}\right\}$ be the evaluation set.

We will show that $\phi \in \ThreeSAT$ iff $W, T \models \psi$. Then it follows that $\ThreeSAT\leqpm \MDLMCpara[M]{0}$ and therefore $\MDLMCpara[M]{0}$ is \NP-hard.

Now assume that $\phi \in \ThreeSAT$ and that $\theta$ is a satisfying valuation for $\phi$. 
From the valuations $\theta(x_j)$ of all $x_j$ we construct subteams $T_1,\ldots, T_n$ such that for all $j \in \{1,\ldots, n\}$ it holds that $W, T_j \models r_j \wedge \depp{p_j}$. The $T_j$ are constructed as follows:
\begin{align*}
&T_j \dfn \begin{cases}
\big\{s_i \in S \mid \pi(s_i)\cap \{r_j, p_j\} = \{r_j, p_j\}\big\} &\text{ iff }\ \theta(x_j) = 1,\\
\big\{s_i \in S \mid \pi(s_i)\cap \{r_j, p_j\} = \{r_j\}\big\} &\text{ iff }\ \theta(x_j) = 0,
\end{cases}
\end{align*}
\ie $T_j$ is the team consisting of exactly the states corresponding to clauses satisfied by $\theta(x_j)$.

Since every clause in $\phi$ is satisfied by some valuation $\theta(x_j) = 1$ or $\theta(x_j) = 0$ we have that $T_1 \cup \ldots \cup T_n = T$ such that $W, T \models \phi$.

On the other hand, assume that $W, T \models \psi$, therefore we have $T = T_1 \cup T_2 \cup \ldots \cup T_n$ such that for all $j \in \{1,\ldots,n\}$ it holds that $T_j \models r_j\wedge \dep[p_j]$. Therefore $\pi(s_i)\cap\{p_j\}$ is constant for all elements $s_i \in T_j$. From this we can construct a satisfying valuation $\theta$ for $\phi$. 

For all $j$ let $I_j \dfn \{i \mid s_i \in T_j\}$. For every $j\in \{1,\ldots,n\}$ we consider $T_j$. If for every element $s_i \in T_j$ it holds that $\pi(s_i)\cap \{p_j\} = \{p_j\}$ then we have for all $i \in I_j$ that $x_j$ is a literal in $C_i$. In order to satisfy those $C_i$ we set $\theta(x_j)=1$. If for every element $s_i \in T_j$ it holds that $\pi(s_i) \cap \{p_j\} = \emptyset$ then we have for every $i \in I_j$ that $\neg {x_j}$ is a literal in $C_i$. In order to satisfy those $C_i$ we set $\theta(x_j) = 0$.\\
Since for every $s_i \in T$ there is a $j$ with $s_i \in T_j$ we have an evaluation $\theta$ that satisfies every clause in $\phi$. Therefore we have $\theta \models \phi$.
\end{proof}

Instead of not having modalities at all, we can also allow nothing but the $\EX$ modality, \ie we disallow propositional connectives and the $\AX$ modality, and model checking is \NP-complete as well.

\begin{theorem}
\label{diamond-np-complete}
Let $\{\EX,\depp{}\}\subseteq M$. Then \MDLMC[M] is \NP-complete.
\end{theorem}
\begin{proof}
Membership in \NP follows from \cref{mdlmc-in-np} again.

For hardness we again reduce from $\ThreeSAT$.
Let $\phi = \bigwedge_{i=1}^{m}C_i$ be an arbitrary \cnf[3] formula built from the variables $x_1,\ldots,x_n$. Let $W$ be the Kripke structure $(S,R,\pi)$, over the atomic propositions $p_1,\dots,p_n,q$, shown in \cref{figure:diamond-dep1} 
and formally defined by
\[\begin{array}{rcl}
S       &       \dfn & \{c_1,\ldots,c_m,s_1^1,\ldots,s_n^1, s_1^0 ,\ldots, s_n^0  \},\\
R \cap\{(c_i,s_j^1),(c_i,s_j^0)\} & \dfn & \begin{cases}
\{(c_i, s_j^1)\}  & \text{ iff $x_j$ occurs in $C_{i}$ positively,}\\
\{(c_i, s_j^0 )\} & \text{ iff $x_j$ occurs in $C_{i}$ negatively,}\\
\emptyset         & \text{ iff $x_j$ does not occur in $C_{i}$,}\\
\end{cases}\\
\pi(c_i) & \dfn & \emptyset,\\
\pi(s_j^1)  &\dfn & \{p_j,q\},\\
\pi(s_j^0 ) &\dfn & \{p_j\}.
\end{array}\]

\begin{figure}[ht]
\begin{center}
\begin{tikzpicture}[->,>=stealth',shorten >=1pt,auto,node distance=1.8cm,every state/.style={minimum size=7mm,inner sep=0pt},thick]
                    
        \node[state]                             (C1)                      {$c_i$};
  \node[state]                                               (x1_) [below left of=C1, label=left:{$p_j$}]  {$s_j^0$};
  \node[state]                                           (x1)  [below right of=C1, label=right:{$p_j,q$}] {$s_j^1$};

  \path (C1) edge     node {}(x1);
\end{tikzpicture}
\caption{Kripke structure part corresponding to the \cnf[3] clause $C_i = x_j$.}
\label{figure:diamond-dep1}
\end{center}
\end{figure}
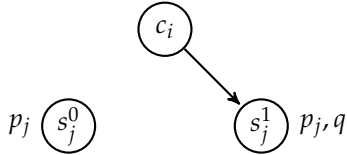

%
%
%

Let $\psi$ be the $\MDL[\EX,\depp{}]$ formula $\EX  \depp[p_1,\ldots, p_n]{q}$ and let $T\dfn\{c_1,\ldots,c_m\}$.
Then we will show that $\phi \in \ThreeSAT$ iff $W, T \models \psi$. Hence, $\ThreeSAT\leqpm \MDLMC[M]$ and $\MDLMC[M]$ is \NP-hard.

First suppose we have a satisfying valuation $\theta$ for $\phi$. From $\theta$ we will construct a successor team $T'$ of $T$, \ie for all $s\in T$ there is an $s'\in T'$ s.t. $(s,s')\in R$ with $W, T' \models \depp{p_1,\ldots, p_n,q}$. $T'$ is defined by:
\[
T' \dfn \{s_j^z \mid \theta(x_j) = z, j \in \{1,\ldots,n\}\}
\]
Since $\theta$ satisfies every clause $C_i$ of $\phi$ we have that for every $C_i$ there is an $x_j$ with 
\[
\theta(x_j) = \begin{cases}
1, &\text{iff } x_j \in C_i\\
0, &\text{iff } \neg x_j \in C_i.
\end{cases}
\]
It follows that for every $s \in T$ there is an $s' \in T'$ such that $(s,s')\in R$.

By construction of $T'$ it is not possible to have both $s_j^0$ and $s_j^1$ in $T'$. Hence for all elements $s_j^0, s_{j'}^{1} \in T'$ it follows that $j\neq j'$ and therefore $\pi(s_j^0)\cap \{p_1,\ldots,p_n\} \neq \pi(s_{j'}^{1})\cap\{p_1,\ldots,p_n\}$. Thus $W, T' \models \depp{p_1,\ldots, p_n,q}$.


On the other hand assume $W,T\models \psi$. Then there is a successor set $T'$ of $T$ s.t.~for every $s \in T$ there is an $s' \in T'$ with $(s,s')\in R$ and $T'\models \depp{p_1,\ldots, p_n,q}$. We construct $\theta$ as follows:
\[
\theta(x_j) \dfn \begin{cases}
1, &\text{iff } s_j^1\in T'\\
0, &\text{iff } s_j^0 \in T'\\
0, &\text{iff } s_j^0,s_j^1 \notin T'.
\end{cases}
\]
Note that in the latter case it does not matter if 0 or 1 is chosen.

Since $W,T' \models \depp{p_1,\ldots,\allowbreak p_n,\allowbreak q}$ and for every $j$ it holds that $W,\{s_j^0,s_j^1\}\not\models \depp{p_1,\ldots,\allowbreak p_n,\allowbreak q}$ we have that for every $j$ at most one of $s_j^0$ or $s_j^1$ is in $T'$. It follows that $\theta$ is well-defined.

Since for every $c_i\in T$ there is an $s_j^z \in T'$ s.t. $(c_i,s_j^z) \in R$ with $\theta(x_j) = z$, we have by contruction of $W$ that $\theta$ satisfies every clause $C_i$ of $\phi$. From this follows $\phi \in \ThreeSAT$.
\end{proof}

If we disallow $\EX$ but allow $\AX$ instead we have to also allow $\vee$ to get \NP-hardness.
\begin{theorem}
\label{box-vee-np-complete}
Let $\{\AX,\vee,\dep[]\}\subseteq M$. 
Then \MDLMC[M] is \NP-complete. Also, \MDLMCk[M] is \NP-complete for every $k\geq 0$.
\end{theorem}
\begin{proof}
Membership in \NP follows from \cref{mdlmc-in-np} again.
To prove hardness, we reduce \ThreeSAT to $\MDLpara[{\AX,\sor,\dep[]}]{0}\hmc$.
Before we formally define the reduction and prove its correctness let us demonstrate its general idea with a concrete example.

\begin{example}\label{box-vee-np-complete-example}
Let $\phi$ be the \cnf[3] formula
\[\underbrace{(\neg x_1 \vee x_2 \vee x_3)}_{C_1}\,\wedge\,\underbrace{(x_2\vee \neg x_3 \vee x_4)}_{C_2}\,\wedge\,\underbrace{(x_1 \vee \neg x_2)}_{C_3}.\]
Now we have to construct a corresponding Kripke structure $W=(S,R,\pi)$, a team $T\subseteq S$ and a $\MDL[{\AX,\sor,\dep[]}]$ formula $\psi$ such that $\phi \in \ThreeSAT$ iff $W,T \models \psi$.

The structure $W$ is shown in \cref{fig:box-vee-np-complete-example}. It has \emph{levels} 0 to 4 where the $j$th level (for  $j\geq 1$ corresponding to the variable $x_j$ in the formula $\phi$) is the set of nodes reachable via exactly $j$ transitions from the set of nodes $s_1$, $s_2$ and $s_3$ (corresponding to the clauses of $\phi$).

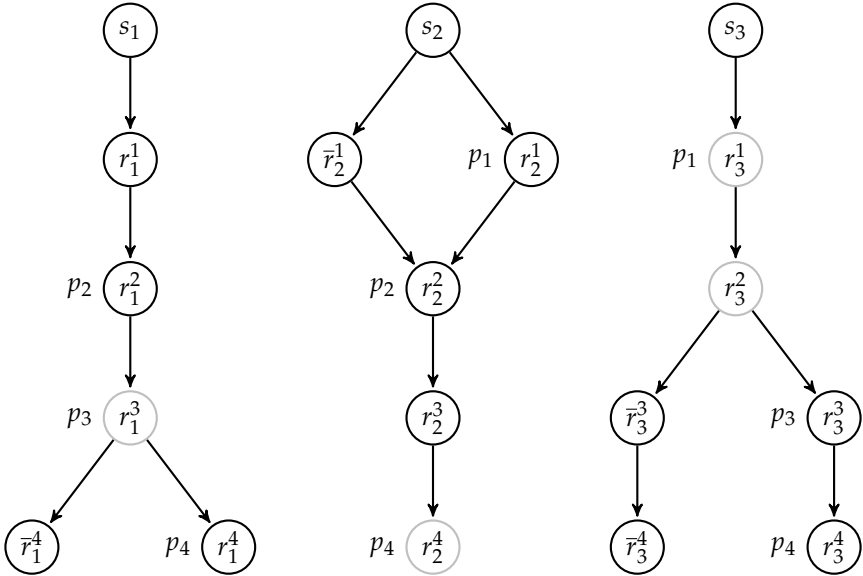
\begin{figure}[ht]
\begin{center}
\begin{tikzpicture}[->,>=stealth',shorten >=1pt,auto,node distance=1.7cm and 1.3cm,thick,on grid=true,every state/.style={minimum size=7mm,inner sep=1pt}]
        \node[state]                                         (C1)                                                                                                                           {$s_1$};
  \node[state]               (C2)    [right =4cm of C1]                                                                                                                 {$s_2$};
  \node[state]                 (C3)    [right =4cm of C2]                                                                                                               {$s_3$};
  \node[state]                                                                                   (s1x1)  [below =of C1, label=left:{}]              {$r_1^1$};
  \node[state]                                                           (s2x1_) [below left= of C2]                                                                                             {$\overline r_2^1 $};
  \node[state]                                                                                   (s2x1)  [below right= of C2, label=left:$p_1$]                  {$r_2^1$};
  \node[color=lightgray, state]                                                                      (s3x1)  [below of=C3, label=left:$p_1$]        {\textcolor{black}{$r_3^1$}};
  
  \node[state]                                                           (s1x2)  [below of=s1x1, label=left:$p_2$]                                      {$r_1^2$};
  \node[state]                                                           (s2x2)  [below left= of s2x1, label=left:$p_2$]                                         {$r_2^2$};
  \node[color=lightgray, state]                                                              (s3x2)  [below of=s3x1]                                                                                                                {\textcolor{black}{$r_3^2$}};
  
  \node[color=lightgray, state]                                                              (s1x3)  [below of=s1x2, label=left:$p_3$]                                      {\textcolor{black}{$r_1^3$}};
  \node[state]                                                           (s2x3)  [below of=s2x2]                                                                                                                {$r_2^3$};
  \node[state]                                                                                   (s3x3_) [below left= of s3x2]                                                                                                           {$\overline r_3^3 $};
  \node[state]                                                           (s3x3)  [below right= of s3x2, label=left:$p_3$]                                        {$r_3^3$};  
  \node[state]                                                                   (s1x4)  [below right= of s1x3, label=left:$p_4$]        {$r_1^4$};
  \node[state]                                                                   (s1x4_)  [below left= of s1x3]           {$\overline r_1^4 $};
  \node[color=lightgray, state]              (s2x4)  [below of=s2x3, label=left:$p_4$]      {\textcolor{black}{$r_2^4$}};
  \node[state]           (s3x4)  [below of=s3x3, label=left:$p_4$]      {$r_3^4$}       ;
  \node[state]           (s3x4_)  [below of=s3x3_]       {$\overline r_3^4 $};

  \path (C1)                    edge            node {}(s1x1)
        (s1x1)                  edge            node {}(s1x2)
        (s1x2)                  edge            node {}(s1x3)
        (s1x3)                  edge            node {}(s1x4_)
                                                edge            node {}(s1x4)
        (C2)                            edge            node {}(s2x1)
                                                edge            node     {}(s2x1_)
        (s2x1)                  edge            node {}(s2x2)
        (s2x1_)                 edge            node {}(s2x2)
        (s2x2)                  edge            node {}(s2x3)
        (s2x3)                  edge            node {}(s2x4)
        (C3)                            edge            node {}(s3x1)
        (s3x1)                  edge            node {}(s3x2)
        (s3x2)                  edge            node    {}(s3x3)
        (s3x2)                  edge            node    {}(s3x3_)
        (s3x3_)                 edge            node {}(s3x4_)
        (s3x3)                  edge            node {}(s3x4);
\end{tikzpicture}
\end{center}
\caption{Kripke structure corresponding to $\phi\,=\,(\neg x_1 \vee x_2 \vee x_3)\wedge(x_2\vee \neg x_3 \vee x_4)\wedge(x_1 \vee \neg x_2)$}
\label{fig:box-vee-np-complete-example}
\end{figure}

The formula $\psi$ is
\[\underbrace{\AX \depp{p_1}}_{\gamma_1} \vee \underbrace{\AX\AX \depp{p_2}}_{\gamma_2} \vee \underbrace{\AX\AX\AX \depp{p_3}}_{\gamma_3} \vee \underbrace{\AX\AX\AX\AX \depp{p_4}}_{\gamma_4}.\]

And the team $T$ is $\{s_1,s_2,s_3\}$.

Now suppose that $W,T \models \psi$. Then, for all $j \in \{1,\ldots,4\}$ let $T_j \subseteq T$ with $T_j \models \gamma_j$ and $T_1 \cup \ldots \cup T_4 = T$.
By comparing the formulas $\gamma_j$ with the chains in the Kripke structure one can easily verify that $T_1\subsetneqq \{s_1,s_3\}$, \ie there can at most be one of $s_1$ and $s_3$ in $T_1$ since $\pi(r_1^1)\cap{p_1} \neq \pi(r_3^1)\cap\{p_1\}$ and $s_2$ cannot be in $T_1$ since its direct successors $\overline r_{2}^1, r_2^1$ do not agree on $p_1$. In this case $T_1 = \{s_1\}$ means that $C_1$ is satisfied by setting $\theta(x_1) = 0$ and the fact that $\{s_2\} \not\models \gamma_1$ corresponds to the fact that there is no way to satisfy $C_2$ via $x_1$, because $x_1$ does not occur in $C_2$.
Analogously, $T_2\subseteq\{s_1,s_2\}$ or $T_2\subseteq\{s_3\}$, and $T_3 \subsetneqq \{s_1,s_2\}$ and $T_4 \subseteq \{s_2\}$.

Now, for example, the valuation $\theta$ where $x_1$, $x_3$ and $x_4$ evaluate to true and $x_2$ to false satisfies $\phi$. From this valuation one can construct sets $T_1, \dots, T_4$ with $T_1\cup\dots\cup T_4=\{s_1,s_2,s_3\}$ such that $T_j \models \gamma_j$ for all $j=1,\dots,4$ by defining $T_j\dfn\{s_i \mid \text{$x_j$ satisfies clause $C_i$ under }\allowbreak\theta\allowbreak\}$ for all $j$. This leads to $T_1=T_2=\{s_3\}$, $T_3=\{s_1\}$ and $T_4=\{s_2\}$.

The gray colourings indicate which chains (resp.~clauses) are satisfied on which levels (resp.~by which variables).
$\psi$ (resp.~$\phi$) is satisfied because there is a gray coloured state in each chain.
\end{example}

Now, in general, let $\phi = \bigwedge_{i=1}^{m}C_i$ be an arbitrary $\cnf[3]$ formula over the variables $x_1,\ldots,x_n$.
Let $W$ be the Kripke structure $(S,R,\pi)$ over the atomic propositions $p_1,\dots,p_n$ shown in \cref{chain1,chain2,chain3,chain4,chain5,chain6} and formally defined as follows: 
\[\begin{array}{lcl}
S & \dfn & \big\{s_i | i \in \{1,\ldots,m\}\big\}\\
  &      & \cup \ \big\{r_k^j | k\in \{1,\ldots,m\}, j \in \{1,\ldots,n\}\big\} \\
  &      & \cup \ \big\{\overline r_k^j | k\in \{1,\ldots,m\}, j \in \{1,\ldots,n\}\big\}\\[2ex]

\multicolumn{3}{l}{R\, \cap\bigcup\limits_{j\in\{1,\dots,n\}}\,\{(s_i,r_i^j),(s_i,\overline r_i^j)\}\ \dfn}\\[3ex]
\multicolumn{3}{l}{\;\;\;\left\{\!\!\begin{array}{l@{\;}l@{\;\;}l}
\{(s_i,r_i^1)\}                                         & \text{iff $x_1$ occurs in $C_i$ (positively or negatively)} & \text{\emph{(Fig.~\ref{chain1})}}\\
\{(s_i,r_i^1),(s_i, \overline r_i^1 )\}                 & \text{iff $x_1$ does not occur in $C_i$} & \text{\emph{(Fig.~\ref{chain2})}}\\
\end{array}\right.}\\[4ex]

\multicolumn{3}{l}{R \,\cap \bigcup\limits_{k\in\{1,\dots,n\}}\,\{(r_i^j,r_i^{k}),(r_i^j,\overline r_i^{k}),(\overline r_i^j,r_i^{k}),(\overline r_i^j,\overline r_i^{k})\}\ \dfn}\\[3ex]
\multicolumn{3}{l}{\quad\left\{\begin{array}{lll}
\{(r_i^j,r_i^{j+1})\}                                   & \parbox[t]{15em}{iff $x_{j}$ and $x_{j+1}$ both occur in $C_i$} & \text{\emph{(Fig.~\ref{chain3})}}\\
\{(r_i^j,r_i^{j+1}),(r_i^j, \overline r_i^{j+1} )\}     & \parbox[t]{15em}{iff $x_{j}$ occurs in $C_i$ but $x_{j+1}$ does not occur in $C_i$} & \text{\emph{(Fig.~\ref{chain4})}}\\[3ex]
\{(r_i^j,r_i^{j+1}),(\overline r_i^j ,r_i^{j+1})\}      & \parbox[t]{15em}{iff $x_j$ does not occur in $C_i$ but $x_{j+1}$ does occur in $C_i$} & \text{\emph{(Fig.~\ref{chain5})}}\\[3ex]
\{(r_i^j,r_i^{j+1}),(\overline r_i^j , \overline r_i^{j+1} )\} & \parbox[t]{15em}{iff neither $x_j$ nor $x_{j+1}$ occur in $C_i$} & \text{\emph{(Fig.~\ref{chain6})}}
\end{array}\right.}
\end{array}\]

\[\begin{array}{lcl}
\pi(s_i) & \dfn & \emptyset\\
\pi(r_i^j) & \dfn & \begin{cases}\{p_j\}   &\text{iff $x_j$ occurs in $C_i$ positively or not at all}\\
                                 \emptyset &\text{iff $x_j$ occurs in $C_i$ negatively}\end{cases}\\
\pi(\overline r_i^j ) & \dfn & \emptyset
\end{array}\]

\begin{figure}
\begin{minipage}[b]{0.48\textwidth}
\centering
\begin{tikzpicture}[->,>=stealth',shorten >=1pt,auto,node distance=2.8cm,every state/.style={minimum size=9mm,inner sep=1pt},
                    thick]
        \node[state]                                                                                     (C1)                                   {$s_i$};
        \node[state]                                                                                     (s1x1)  [below of=C1]                  {$r_i^1$};
        
        \path (C1)                              edge       node {}(s1x1);
        
\end{tikzpicture}
\caption{$x_1$ occurs in $C_i$.}
\label{chain1}
\end{minipage}
\hspace{0.02\textwidth}
\begin{minipage}[b]{0.48\textwidth}
\centering
\begin{tikzpicture}[->,>=stealth',shorten >=1pt,auto,node distance=2.8cm,every state/.style={minimum size=9mm,inner sep=1pt},
                    thick]
        \node[state]                                                                                     (C1)                                                   {$s_i$};
        \node[state]                                                                                     (s1x1)  [below right of=C1]            {$r_i^1$};
        \node[state]                                                                                     (s1x1_) [below left of=C1]             {$\overline r_i^1 $};
        
        \path (C1)                              edge       node {}(s1x1_)
                                                                                edge       node {}(s1x1);
\end{tikzpicture}
\caption{$x_1$ does not occur in $C_i$.}
\label{chain2}
\end{minipage}
\end{figure}

\begin{figure}
\begin{minipage}[b]{0.48\textwidth}
\begin{center}
\begin{tikzpicture}[->,>=stealth',shorten >=1pt,auto,node distance=2.8cm,every state/.style={minimum size=9mm,inner sep=1pt},
                    thick]
        \node[state]                                                                                     (s1x2)                                                                                                         {$r_i^{j+1}$};
        \node[state]                                                                                     (s1x1)  [above of=s1x2]                        {$r_i^j$};
        
        \path (s1x1)                            edge       node {}(s1x2);
\end{tikzpicture}
\caption{$x_j$ and $x_{j+1}$ occur in $C_i$.}
\label{chain3}
\end{center}
\end{minipage}
\hspace{0.02\textwidth}
\begin{minipage}[b]{0.48\textwidth}
\begin{center}
\begin{tikzpicture}[->,>=stealth',shorten >=1pt,auto,node distance=2.8cm,every state/.style={minimum size=9mm,inner sep=1pt},
                    thick]
        \node[state]                                                                                     (s1x1)                                                                                         {$r_i^j$};
        \node[state]                                                                                     (s1x2_) [below left of=s1x1]           {$\overline r_i^{j+1 }$};
        \node[state]                                                                                     (s1x2)  [below right of=s1x1]          {$r_i^{j+1}$};
        
        \path (s1x1)                            edge       node {}(s1x2);
  \path (s1x1)                          edge       node {}(s1x2_);
\end{tikzpicture}
\caption{$x_j$ occurs in $C_i$ but $x_{j+1}$ does not occur in $C_i$.}
\label{chain4}
\end{center}
\end{minipage}
\end{figure}

\begin{figure}
\begin{minipage}[b]{0.48\textwidth}
\begin{center}
\begin{tikzpicture}[->,>=stealth',shorten >=1pt,auto,node distance=2.8cm,every state/.style={minimum size=9mm,inner sep=1pt},
                    thick]
        \node[state]                                                                                     (s1x2)                                                                                                         {$r_i^{j+1}$};
        \node[state]                                                                                     (s1x1)  [above right of=s1x2]    {$r_i^j$};
        \node[state]                                                                                     (s1x1_) [above left of=s1x2]           {$\overline r_i^j $};

        \path (s1x1)                            edge       node {}(s1x2);
  \path (s1x1_)                         edge       node {}(s1x2);
\end{tikzpicture}
\caption{$x_j$ does not occur in $C_i$ but $x_{j+1}$ does occur in $C_i$.}
\label{chain5}
\end{center}
\end{minipage}
\hspace{0.02\textwidth}
\begin{minipage}[b]{0.48\textwidth}
\begin{center}
\begin{tikzpicture}[->,>=stealth',shorten >=1pt,auto,node distance=2.8cm,every state/.style={minimum size=9mm,inner sep=1pt},
                    thick]
        \node[state]                                                                                     (s1x1)                                                                                             {$r_i^j$};
        \node[state]                                                                                     (s1x1_) [left of=s1x1]                                         {$\overline r_i^j $};
        \node[state]                                                                                     (s1x2)  [below of=s1x1]                                        {$r_i^{j+1}$};  
        \node[state]                                                                                     (s1x2_) [below of=s1x1_]                                       {$\overline r_i^{j+1} $};       
        
        \path (s1x1)                            edge       node {}(s1x2);
  \path (s1x1_)                         edge       node {}(s1x2_);
\end{tikzpicture}
\caption{$x_j$ and $x_{j+1}$ do not occur in $C_i$.}
\label{chain6}
\end{center}
\end{minipage}
\end{figure}

Let $\psi$ be the $\MDL[\AX,\vee,\depp{}]$ formula
\[\bigvee_{j=1}^n\ \AX^{j} \depp{p_j}\]
and let $T \dfn \{s_1,\ldots,s_m\}$.

Then, as we will show, $\phi \in \ThreeSAT$ iff $W,T \models \psi$ and therefore $\MDLMCpara[\AX,\allowbreak\vee,\allowbreak\depp{}]{0}$ is \NP-complete.
Intuitively, the direction from left to right holds because the disjunction splits the team $\{s_1,\dots,s_m\}$ of all starting points of chains of length $n$ into $n$ subsets (one for each variable) in the following way: $s_i$ is in the subset that belongs to $x_j$ iff $x_j$ satisfies the clause $C_i$ under the variable valuation that satisfies $\phi$. Then the team that belongs to $x_j$ collectively satisfies the disjunct $\AX^j \depp{p_j}$ of $\psi$. For the reverse direction the $\overline r_i^j$ states are needed to ensure that a state $s_i$ can only satisfy a disjunct $\AX^j \depp{p_j}$ if there is a variable $x_j$ that occurs in clause $C_i$ (positively or negatively) and satisfies $C_i$.

More precisely, assume that $\theta$ is a satisfying valuation for $\phi$. From $\theta$ we construct subteams $T_1,\ldots, T_n$ with $T_1 \cup \ldots \cup T_n = T$ s.t.~for all $j$ it holds that $T_j \models \AX^j \depp{p_j}$. $T_j$ is defined by
\[
T_j \dfn \begin{cases}
\{s_i  \mid \{s_i\} \models \AX^j p_j\} &\text{ iff } \theta(x_j) = 1\\
\{s_i  \mid \{s_i\} \models \AX^j \neg p_j\} &\text{ iff } \theta(x_j) = 0
\end{cases}
\] for all $j\in\{1,\dots,n\}$.
Obviously, for all $j$ it holds that $T_j \models \AX^j \depp{p_j}$.
Now we will show that for all $i\in\{1,\dots,m\}$ there is a $j\in\{1,\dots,n\}$ such that $s_i \in T_j$.
For this purpose let $i\in\{1,\dots,m\}$ and suppose $C_i$ is satisfied by $\theta(x_j)=1$ for a $j\in\{1,\dots,n\}$. Then, by definition of $W$, $\pi(r_i^j)=p_j$, hence $\{s_i\}\models \AX^j p_j$ and therefore $s_i \in T_j$. If, on the other hand, $C_i$ is satisfied by $\theta(x_j)=0$ then we have that $\pi(r_i^j) = \emptyset$, hence $\{s_i\}\models \AX^j \neg p_j$ and again it follows that $s_i \in T_j$.
Altogether we have that for all $i$ there is a $j$ such that $s_i \in T_j$. It follows that $T_1 \cup \ldots \cup T_n = T$ and therefore $W,T \models \psi$.

On the other hand assume $W,T \models \psi$. Therefore we have $T = T_1 \cup \ldots \cup T_n$ with $T_j \models \AX^j \depp{p_j}$ for all $j\in\{1,\dots,n\}$. We define a valuation $\theta$ by
\[
\theta(x_j) \dfn \begin{cases}
1 &\text{iff } T_j \models \AX^j p_j\\
0 &\text{iff } T_j \models \AX^j \neg p_j.
\end{cases}
\]
Since every $s_i$ is contained in a $T_j$ we know that for all $i\in\{1,\dots,m\}$ there is a $j\in\{1,\dots,n\}$ with $\{s_i\} \models \AX^j \depp{p_j}$. From this it follows that $x_j$ occurs in $C_i$ (positively or negatively) since otherwise, by definition of $W$, both $r_i^j$ and $\overline r_i^j$ would be reachable from $s_i$.

It also holds that $\{s_i\} \models \AX^j p_j$ or $\{s_i\} \models \AX^j \neg p_j$. In the former case we have that $\pi(r_i^j) = p_j$, hence, by definition of $W$, $x_j$ is a literal in $C_i$. By construction of $\theta$ it follows that $C_i$ is satisfied. In the latter case it holds that $\overline x_j$ is a literal in $C_i$. Again, by construction of $\theta$ it follows that $C_i$ is satisfied. Hence, $\phi \in \ThreeSAT$.
\end{proof}

If we disallow both $\EX$ and $\vee$ the problem becomes tractable since the non-deterministic steps in the model checking algorithm are no longer needed.

\begin{theorem}
\label{box-wedge-in-p}
Let $M \subseteq \{\AX,\wedge, \nor, \aneg, \dep[]\}$. Then \MDLMC[M] is in \P.
\end{theorem}
\begin{proof}
\Cref{algo:mdl check} is a non-deterministic algorithm that checks the truth of an arbitrary \MDL formula in a given structure in polynomial time. Since $M\subseteq \{\AX,\wedge, \nor, \aneg, \depp{}\}$ it holds that $\EX,\,\vee\notin M$. Therefore the non-deterministic steps are never used and the algorithm is in fact deterministic in this case.
\end{proof}

Note that this deterministic polynomial time algorithm is a top-down algorithm and therefore works in a fundamentally different way than the usual deterministic polynomial time bottom-up model checking algorithm for plain modal logic.

Now we have seen that $\MDLMC[M]$ is tractable if $\vee\notin M$ and $\EX \notin M$ since these two operators are the only source of non-determinism.
On the other hand, $\MDLMC[M]$ is \NP-complete if $\dep \in M$ and either $\EX\in M$ (\Cref{diamond-np-complete}) or $\vee,\,\AX\in M$ (\Cref{box-vee-np-complete}). The remaining question is what happens if only $\vee$ (but not $\AX$) is allowed. Unfortunately this case has to remain open for now.

\subsection{Bounded arity fragments}\label{sec:bounded-fragments}

We will now show that $\MDLk[{\vee, \aneg, \deppp}]\hmc$ is in $\P$ for all $k\geq 0$. To prove this statement we will decompose it into two smaller propositions.

First we show that even the whole $\MDL[{\vee, \aneg, \dep[]}]$ fragment with unrestricted $\dep$ atoms is in $\P$ as long as it is guaranteed that in every input formula at least a specific number of dependence atoms -- depending on the size of the Kripke structure -- occur.

We will need the following lemma stating that a dependence atom is always satisfied by a team containing at least half of all the worlds.
\begin{lemma}\label{split-in-two}
Let $W=(S,R,\pi)$ be a Kripke structure, $\phi \dfn\dep[p_1,\dots,p_n,q]$ ($n\geq 0$) an atomic formula and $T\subseteq S$ an arbitrary team. Then there is a set $T'\subseteq T$ such that $|T'| \geq \frac{|T|}{2}$ and $T'\models \phi$.
\end{lemma}
\begin{proof}
Let $T_0\dfn\{s\in T\mid q\notin\pi(s)\}$ and $T_1\dfn\{s\in T\mid q\in\pi(s)\}$. Then $T_0\cup T_1=T$ and $T_0\cap T_1=\emptyset$. Therefore there is an $i\in\{1,2\}$ such that $|T_i| \geq \frac{|T|}{2}$. Let $T'\dfn T_i$. Since $q$ is either labeled in every state of $T'$ or in no one, it holds that $W,T'\models \phi$.
\end{proof}


We will now formalize a notion of \qte{many dependence atoms in a formula}.

\begin{definition}\label{defMDLMCl}
For $\phi\in\MDL$ let $\sigma(\phi)$ be the number of positive dependence atoms in $\phi$. Let $\ell\colon \mathbb{N} \to \mathbb{R}$ be an arbitrary function and $\star\in\{<,\leq,>,\geq,=\}$.
Then $\MDLMCparas[M]{}{\star\,\ell(n)}$ (resp.~$\MDLMCparas[M]{k}{\star \ell(n)}$) is the problem $\MDLMC[M]$ (resp.~$\MDLMCk[M]$) restricted to inputs $(W=(S,R,\pi),T,\phi)$ that satisfy the condition $\sigma(\phi)\, \star\, \ell(|S|)$.
\end{definition}

If we only allow $\vee$ and we are guaranteed that there are many dependence atoms in each input formula then model checking becomes trivial -- even for the case of unbounded dependence atoms.

\begin{proposition}\label{many-dep-atoms-trivial}
Let $M\subseteq\{\vee,\aneg,\depp{}\}$. Then $\MDLMCparas[M]{}{> \log_2(n)}$ is trivial, \ie for all Kripke structures $W=(S,R,\pi)$ and all $\phi \in \MDL[M]$ such that the number of positive dependence atoms in $\phi$ is greater than $\log_2(|S|)$ it holds for all $T\subseteq S$ that $W, T \models \phi$.
\end{proposition}
\begin{proof}
Let $W=(S,R,\pi)$, $\phi \in \MDL[M]$, $T\subseteq S$ be an arbitrary instance with $\ell > \log_2(|S|)$ positive dependence atoms in $\phi$. Then it follows that
\[\phi\ =\ \bigvee_{i=1}^\ell \underbrace{\depp{p_{j_{i,1}},\dots,p_{j_{i,k_i}}}}_{\psi_i}\,\vee\,\bigvee_i l_i,\]
where each $l_i$ is either a (possibly negated) atomic proposition or a negated dependence atom.

\begin{claim}\label{mdl many atoms claim}
For all $k\in\{0,\dots, \ell\}$ there is a set $T_k\subseteq T$ such that $W,T_k\models \bigvee_{i=1}^k \psi_i$ and $|T\setminus T_k| < 2^{\ell-k}$.
\end{claim}

The main proposition follows immediately from case $k=\ell$ of this claim: From $|T\setminus T_\ell| < 2^{\ell-\ell} = 1$ follows that $T=T_\ell$ and from $W,T_\ell \models \bigvee_{i=1}^\ell \psi_i$ follows that $W,T \models \phi$.

\begin{claimproof}{\cref*{mdl many atoms claim} (inductive)}
For $k=0$ we can choose $T_k \dfn \emptyset$. For the inductive step let the claim be true for all $k'<k$. By \cref{split-in-two} there is a set $T'_k\,\subseteq\,T\setminus T_{k-1}$ such that $W,T'_k\models \psi_k$ and $|T'_K| \geq \frac{|T\setminus T_{k-1}|}{2}$. Let $T_k\dfn T_{k-1}\cup T'_k$. Since $W,T_{k-1}\models \bigvee_{i=1}^{k-1} \psi_i$ it follows by definition of the semantics of $\vee$ that $W,T_k\models \bigvee_{i=1}^{k} \psi_i$. Furthermore,
\[\begin{array}{rcccl}
|T\setminus T_k|&=&|(T\setminus T_{k-1}) \setminus T'_k| & = & |T\setminus T_{k-1}| - |T'_k|\\
&\leq& |T\setminus T_{k-1}| - \frac{|T\setminus T_{k-1}|}{2} &=& \frac{|T\setminus T_{k-1}|}{2}\\
&<& \frac{2^{\ell-(k-1)}}{2} &=& 2^{\ell-k}.
\end{array}\]
\end{claimproof}
~
\end{proof}
Note that $\MDLMCparas[M]{}{> \log_2(n)}$ is only trivial, \ie all instance structures satisfy all instance formulas, if we assume that only valid instances, \ie where the number of dependence atoms is guaranteed to be large enough, are given as input. However, if we have to verify this number the problem clearly remains in \P.

Now we consider the case in which we have very few dependence atoms (which have bounded arity) in each formula. We use the fact that there are only a few dependence atoms by searching through all possible determining functions for the dependence atoms. Note that in this case we do not need to restrict the set of allowed \MDL operators as we have done above.

\begin{proposition}
\label{few-dep-atoms-in-p}
Let $k\geq 0$. Then $\MDLMCparas{k}{\leq \log_2(n)}$ is in \P.
\end{proposition}
\begin{proof}
From the semantics of $\dep$ it follows that $\depp[p_1,\dots,p_k]{p}$ is equivalent to
\begin{equation}\label{eq:dep-as-function}
\exists f\,f(p_1,\dots,p_{k})\leftrightarrow p\quad \dfn\quad \exists f\,\big( (\neg f(p_1,\dots,p_k) \vee  p) \wedge (f(p_1,\dots,p_k) \vee \neg p) \big)
\end{equation}
where $f(p_1,\dots,p_k)$ and $\exists f \phi$ -- both introduced by Sevenster \cite[Section~4.2]{se09} -- have the following semantics:
\[\begin{array}{l@{\;\;\text{iff}\;\;}p{6.3cm}}
W,T\models \exists f \phi            & there is a Boolean function $f^W$ such that $(W,f^W),\allowbreak T\models \phi$\\
(W,f^W),T\models f(p_1,\dots,p_k)       & for all $s\in T$ and for all $x_1,\dots,x_k\in\{0,1\}$ with $x_i=1$ iff $p_i\in\pi(s)$ $(i=1,\dots,k)$: $f^W(x_1,\dots,x_k)=1$\\
(W,f^W),T\models \neg f(p_1,\dots,p_k)  & for all $s\in T$ and for all $x_1,\dots,x_k\in\{0,1\}$ with $x_i=1$ iff $p_i\in\pi(s)$ $(i=1,\dots,k)$: $f^W(x_1,\dots,x_k)=0$\\
\end{array}\]

Now let $W=(S,R,\pi)$, $T\subseteq S$ and $\phi\in \MDL_k$ be a problem instance. First, we count the number $\ell$ of dependence atoms in $\phi$. If $\ell > \log_2(|S|)$ we reject the input instance.
Otherwise we replace every dependence atom by its translation according to \cref{eq:dep-as-function} (each time using a new function symbol).
Since the dependence atoms in $\phi$ are at most $k$-ary we have from the transformation \cref{eq:dep-as-function} that the introduced function variables $f_1,\dots,f_\ell$ are also at most $k$-ary. From this it follows that the upper bound for the number of interpretations of each of them is $2^{2^k}$.
For each possible tuple of interpretations $f_1^W,\dots,f_\ell^W$ for the function variables we obtain an \ML formula $\phi^*$ by replacing each existential quantifier $\exists f_i$ by a Boolean formula encoding of the interpretation $f_i^W$ (for example by encoding the truth table of $f_i$ with a formula in disjunctive normal form).
For each such tuple we model check $\phi^*$. That is possible in polynomial time in $|S| + |\phi^*|$ as shown by Clarke et al.~\cite{clemsi86}. Since the encoding of an arbitrary $k$-ary Boolean function has length at most $2^{k}$ and $k$ is constant this is a polynomial in $|S| + |\phi|$.

Furthermore, the number of tuples over which we have to iterate is bounded by
\[\begin{array}{rclcl}
\left(2^{2^k}\right)^{\log_2(|S|) } & = & 2^{2^k\cdot \log_2(|S|)}\\
 & = & \left(2^{\log_2(|S|)}\right)^{2^k}\\
 & = & |S|^{2^k} &\in & |S|^{\mathCommandFont{O}(1)}.
\end{array}\]
\end{proof}

With \cref{many-dep-atoms-trivial,few-dep-atoms-in-p} we have shown the following theorem.
\begin{theorem}
\label{vee-bounded-in-p}
Let $M\subseteq \{\vee,\aneg,\depp{}\}$, $k\geq 0$. Then \MDLMCk[M] is in \P.
\end{theorem}
\begin{proof}
Given a Kripke structure $W=(S,R,\pi)$ and a $\MDL_k(\vee, \aneg,\depp{})$ formula $\phi$ the algorithm counts the number $m$ of positive dependence atoms in $\phi$. If $m> \log_2(|S|)$ the input is accepted (because by \cref{many-dep-atoms-trivial} the formula is always fulfilled in this case). Otherwise the algorithm from the proof of \cref{few-dep-atoms-in-p} is used.
\end{proof}

And there is another case where we can use the exhaustive determining function search.
\begin{theorem}
\label{no-wedge-no-vee}
Let $M\subseteq\{\AX, \EX, \aneg, \depp{}\}$. Then $\MDLMCk[M]$ is in $\P$ for every $k\geq 0$.
\end{theorem}
\begin{proof}
Let $\phi\in \MDL_k(M)$. Then there can be at most one dependence atom in $\phi$ because $M$ only contains unary operators. Therefore we can once again use the algorithm from the proof of \cref{few-dep-atoms-in-p}.
\end{proof}

In \cref{diamond-np-complete} we saw that $\MDLMC[\EX, \depp{}]$ is \NP-complete.
The previous theorem includes $\MDLMCk[\EX,\depp{}] \in \P$ as a special case.
Hence, the question remains which are the minimal supersets $M$ of $\{\EX,\depp{}\}$ such that $\MDLMCk[M]$ is \NP-complete.

We will now see that adding either $\wedge$ (\Cref{diamond-wedge-bounded}) or $\vee$ (\Cref{diamond-vee-bounded}) is already enough to get \NP-completeness again.
But note that in the case of $\vee$ we need $k\geq 1$ while for $k=0$ the question remains open.
\begin{theorem}\label{diamond-wedge-bounded}
Let $\{\EX, \wedge, \dep[]\}\subseteq M$. Then $\MDLMCk[M]$ is \NP-complete for every $k\geq 0$.
\end{theorem}
\begin{proof}
Membership in \NP follows from \cref{mdlmc-in-np}. For hardness we once again reduce \ThreeSAT to our problem.

For this purpose let $\phi \dfn \bigwedge_{i=1}^m C_i$ be an arbitrary \cnf[3] formula built from the variables $x_1, \dots, x_n$. Let $W$ be the Kripke structure $(S, R, \pi)$ shown in \cref{figure:diamond-wedge-bounded} and formally defined by
\[\begin{array}{lcl}
S & \dfn & \{c_i \mid i\in\{1,\dots, m\}\} \cup \{s_{j,j'}, \overline{s}_{j,j'} \mid j,j'\in\{1,\dots,n \}\}\\
  &    & \cup\ \{t_j,\overline{t}_j\mid j\in\{1,\dots,n\}\}\\
R & \dfn & \{(c_i, s_{1,j}) \mid x_j \in C_i\} \cup \{(c_i, \overline{s}_{1,j}) \mid \overline x_j  \in C_i\}\\
  &    & \cup \ \{(s_{k,j}, s_{k+1,j}) \mid j\in \{1,\dots,n\}, k \in \{1,\dots, n-1\}\}\\
  &    & \cup \ \{(\overline{s}_{k,j}, \overline{s}_{k+1,j}) \mid j\in \{1,\dots,n\}, k \in \{1,\dots, n-1\}\}\\
  &    & \cup \ \{(s_{k,j}, t_j), (\overline{s}_{k,j},\overline{t}_j) \mid j\in\{1,\dots,n\}, k\in \{1,\dots,n\}\}\\
  &    & \cup \ \{(s_{k,j}, \overline{t}_j), (\overline{s}_{k,j}, t_j) \mid j\in\{1,\dots,n\}, k\in \{1,\dots,n\}, j\neq k\}\\
\pi(c_i) & \dfn & \emptyset\\
\pi(s_{j,j'}) & \dfn & \emptyset
\\
\pi(\overline{s}_{j,j'}) & \dfn & \emptyset
\\
\pi(t_j) & \dfn & \{r_j, p_j\}\\
\pi(\overline{t}_j) & \dfn & \{r_j\}.\\
\end{array}\]

\begin{figure}[ht]
\begin{center}
\begin{tikzpicture}[->,>=stealth',shorten >=1pt,auto,node distance=1.5cm,on grid,thick,every state/.style={minimum size=7mm,inner sep=1pt}]
\node[state]                             (C1)                      {$c_1$};
\node[state]                             (C2)       [right = 3cm of C1]               {$c_2$};
\node[state]                             (C3)       [right = 3cm of C2]               {$c_3$};
\node                            (Cx)       [right = 1 of C3]               {$\dots$};

\node[state]                             (x1)       [below = 2.2 of C1]               {$s_{1,1}$};
\node[state]                             (tx1)       [right of=x1]               {$\overline{s}_{1,1}$};
\node[state]                             (x2)       [right of=tx1]               {$s_{1,2}$};
\node[state]                             (tx2)       [right of=x2]               {$\overline{s}_{1,2}$};
\node[state]                             (x3)       [right of=tx2]               {$s_{1,3}$};
\node                            (xx)       [right = 1 of x3]               {$\dots$};
\node[state]                             (tt1)       at  (-1.5,-2.8)  [label=270:{$r_1$}]               {$\overline{t}_{1}$};
\node[state]                             (t1)        at  (-3,-1.5)    [label=270:{$r_1,p_1$}]           {$t_{1}$};

\node[state]                             (xx1)       [below of=x1]               {$s_{2,1}$};
\node                            (xx)       at (-2.5,-3.5)               {$\vdots$};
\node[state]                             (txx1)       [below of=tx1]               {$\overline{s}_{2,1}$};
\node[state]                             (xx2)       [below of=x2]               {$s_{2,2}$};
\node[state]                             (txx2)       [below of=tx2]              {$\overline{s}_{2,2}$};
\node[state]                             (xx3)       [below of=x3]               {$s_{2,3}$};
\node                            (xx)       [right = 1 of xx3]               {$\dots$};

\node[state]                             (xxx1)       [below of=xx1]               {$s_{3,1}$};
\node[state]                             (txxx1)       [below of=txx1]               {$\overline{s}_{3,1}$};
\node[state]                             (xxx2)       [below of=xx2,label=below:{$\vdots$}]               {$s_{3,2}$};
\node[state]                             (txxx2)       [below of=txx2]              {$\overline{s}_{3,2}$};
\node[state]                             (xxx3)       [below of=xx3]               {$s_{3,3}$};
\node                            (xxx)       [right = 1 of xxx3]               {$\dots$};


\path (C1) edge  node {}(x1);
\path (C1) edge  node {}(tx2);
\path (C2) edge  node {}(x1);
\path (C2) edge  node {}(x2);
\path (C2) edge  node {}(x3);
\path (C3) edge  node {}(tx1);
\path (C3) edge  node {}(x3);

\path (x1) edge  node {}(xx1);
\path (tx1) edge  node {}(txx1);
\path (x2) edge  node {}(xx2);
\path (tx2) edge  node {}(txx2);
\path (x3) edge  node {}(xx3);
\path (x1) edge              node {}(t1);
\path (x2) edge [bend right=14,out=-15]  node {}(t1);
\path (tx2) edge [bend right=14,out=-15]  node {}(t1);
\path (x3) edge [bend right=14,out=-15]  node {}(t1);
\path (tx1) edge [bend left=10,out=20]  node {}(tt1);
\path (x2) edge [bend left=10,out=20]  node {}(tt1);
\path (tx2) edge [bend left=10,out=20]  node {}(tt1);
\path (x3) edge [bend left=10,out=20]  node {}(tt1);

\path (xx1) edge  node {}(xxx1);
\path (txx1) edge  node {}(txxx1);
\path (xx2) edge  node {}(xxx2);
\path (txx2) edge  node {}(txxx2);
\path (xx3) edge  node {}(xxx3);
\end{tikzpicture}
\caption{Kripke structure corresponding to a \cnf[3] formula containing the clauses $C_1=x_1\vee\neg x_2$, $C_2=x_1 \vee x_2\vee x_3$ and $C_3=\neg x_1\vee x_3$}
\label{figure:diamond-wedge-bounded}
\end{center}
\end{figure}

And let $\psi$ be the \MDL[\EX,\wedge,\depp{}] formula
\[\begin{array}{cl}
& \EX\left(\bigwedge\limits_{j=1}^n\, \EX^{j} (r_j \wedge \dep[p_j]) \right)\\
= & \EX\big(\EX(r_1\wedge \dep[p_1])\,\wedge\, \EX\EX (r_2\wedge\dep[p_2]) \,\wedge\, \dots\, \wedge\, \EX^{n}(r_n\wedge \dep[p_n])\big).
\end{array}\]

We again show that $\phi \in \ThreeSAT$ iff $W,\{c_1,\dots, c_m\} \models \psi$.
First assume that $\phi \in \ThreeSAT$ and that $\theta$ is a satisfying valuation for the variables in $\phi$.
Now let
\[s_j\dfn\begin{cases}
               s_{1,j} & \text{if $x_j$ evaluates to true under $\theta$}\\
               \overline s_{1,j} & \text{if $x_j$ evaluates to false under $\theta$}
              \end{cases}\]
for all $j=1,\dots,n$.
Then it holds that $W, \{s_1,\dots,s_n\} \models \bigwedge\limits_{j=1}^n\, \EX^{j} (r_j \wedge \dep[p_j])$.

Furthermore, since $\theta$ satisfies $\phi$ it holds for all $i=1,\dots,m$ that there is a $j_i\in\{1,\dots,n\}$ such that $(c_i,s_{j_i})\in R$.
Hence,
\[W, \{c_1,\dots,c_m\} \models \EX\left(\bigwedge\limits_{j=1}^n\, \EX^{j} (r_j \wedge \dep[p_j])\right).\]

%

For the reverse direction assume that $W, \{c_1,\dots,c_m\}  \models \psi$.
Now let $T\subseteq \{s_{1,1},\allowbreak\overline s_{1,1},\allowbreak s_{1,2},\allowbreak\dots,\overline s_{1,n}\}$ such that $T\models \bigwedge\limits_{j=1}^n\, \EX^{j} (r_j \wedge \dep[p_j])$ and for all $i=1,\dots,m$ there is a $s\in T$ with $(c_i,s)\in R$.

Since $T\models \EX^{j} (r_j \wedge \dep[p_j])$ there is no $j\in\{1,\dots,n\}$ with $s_{1,j}\in T$ and also $\overline s_{1,j}\in T$.
Now let $\theta$ be the valuation of $x_1,\dots,x_n$ defined by
\[\theta(x_j)\dfn\begin{cases}
  1 & \text{if $s_{1,j}\in T$}\\
  0 & \text{else}.
\end{cases}\]

Since for each $i=1,\dots,m$ there is a $j\in\{1,\dots,n\}$ such that either $(c_{i},s_{1,j})\in R$ and $s_{1,j}\in T$ or $(c_{i},\overline s_{1,j})\in R$ and $\overline s_{1,j}\in T$ it follows that for each clause $C_i$ of $\phi$ there is a $j\in\{1,\dots,n\}$ such that $x_j$ satisfies $C_i$ under $\theta$.
\end{proof}

\begin{theorem}
\label{diamond-vee-bounded}
Let $\{\EX, \sor, \dep[]\}\subseteq M$. Then $\MDLMCk[M]$ is \NP-complete for every $k\geq 1$.
\end{theorem}
\begin{proof}
As above membership in \NP follows from \cref{mdlmc-in-np} and for hardness we reduce \ThreeSAT to our problem.

For this purpose let $\phi \dfn \bigwedge_{i=1}^m C_i$ be an arbitrary \cnf[3] formula built from the variables $p_1, \dots, p_n$. Let $W$ be the Kripke structure $(S, R, \pi)$ shown in \cref{figure:diamond-vee-bounded} and formally defined by
\[\begin{array}{lcl}
S & \dfn & \{c_{i,j} \mid i\in\{1,\dots, m\}, j\in\{1,\dots,n\}\}\\
  &    &  \cup\  \{x_{j,j'} \mid j,j'\in\{1,\dots,n \}, j'\leq j\}\\
R & \dfn & \{(c_{i,j}, c_{i,j+1})\mid i\in\{1,\dots, m\}, j\in\{1,\dots, n-1\}\}\\
  &    &  \cup\  \{(x_{j,j'},x_{j,j'+1}) \mid j\in\{1,\dots,n\}, j'\in\{1,\dots, j-1\}\}\\
\pi(x_{j,j'}) & \dfn & \left \{ \begin{array}{ll}
  \{q, p_j\} & \text{iff } j'=j\\
  \{q\} & \text{iff } j'<j
\end{array}\right.\\
\pi(c_{i,j}) & \dfn & \left \{ \begin{array}{lll}
  \{q\} & \text{iff } p_j, \neg p_j \notin C_i\\[0.5ex]
  \{p_j\} & \text{iff } p_j\in C_i\\
  \emptyset & \text{iff } \neg p_j \in C_i
\end{array}\right.
\end{array}\]

\begin{figure}[ht]
\begin{center}
\begin{tikzpicture}[->,>=stealth',shorten >=1pt,auto,node distance=1.5cm,on grid,every state/.style={minimum size=7mm,inner sep=1pt},thick]
\node[state]                             (c1)       [label=left:{$q$}]               {$c_{1,1}$};
\node[state]                             (c2)       [right of=c1,label=left:{$q$}]               {$c_{2,1}$};
\node[state]                             (c3)       [right of=c2]               {$c_{3,1}$};
\node[state]                             (c12)      [below of=c1]                {$c_{1,2}$};
\node[state]                             (c22)      [below of=c2,label=left:{$p_2$}]                {$c_{2,2}$};
\node[state]                             (c32)      [below of=c3,label=left:{$q$}]                {$c_{3,2}$};
\node[state]                             (c13)      [below of=c12,label={[label distance=-2mm]below:{$\vdots$}},label=left:{$q$}]                {$c_{1,3}$};
\node[state]                             (c23)      [below of=c22,label={[label distance=-2mm]below:{$\vdots$}}]                {$c_{2,3}$};
\node[state]                             (c33)      [below of=c32,label={[label distance=-2mm]below:{$\vdots$}},label=left:{$q$}]                {$c_{3,3}$};
\node                            (cx)       [right = 8mm of c3]               {$\dots$};
\node                            (cx2)       [below = of cx]               {$\dots$};
\node                            (cx3)       [below = of cx2]               {$\dots$};

\node[state]                             (x2)       [right = 4.7cm of c3,label=left:{$q$}]               {$x_{2,1}$};
\node[state]                             (x1)       [left of=x2,label=left:{$q,p_1$}]               {$x_{1,1}$};
\node[state]                             (x3)       [right of=x2,label=left:{$q$}]               {$x_{3,1}$};

\node[state]                             (xx2)       [below of=x2,label=left:{$q,p_2$}]               {$x_{2,2}$};
\node[state]                             (xx3)       [below of=x3,label=left:{$q$}]               {$x_{3,2}$};
\node                            (cxx1)       [right = 8mm of x3]               {$\dots$};
\node                            (cxx2)       [below = of cxx1]               {$\dots$};
\node                            (cxx3)       [below = of cxx2]               {$\dots$};

\node[state]                             (xxx3)       [below of=xx3,label=left:{$q,p_3$}]               {$x_{3,3}$};

\path (c1) edge  node {}(c12);
\path (c2) edge  node {}(c22);
\path (c3) edge  node {}(c32);
\path (c12) edge  node {}(c13);
\path (c22) edge  node {}(c23);
\path (c32) edge  node {}(c33);

\path (x2) edge  node {}(xx2);
\path (x3) edge  node {}(xx3);

\path (xx3) edge  node {}(xxx3);
\end{tikzpicture}
\caption{Kripke structure corresponding to a \cnf[3] formula containing the clauses $C_1=\neg p_2$, $C_2=p_2\vee \neg p_3$ and $C_3=\neg p_1$}
\label{figure:diamond-vee-bounded}
\end{center}
\end{figure}
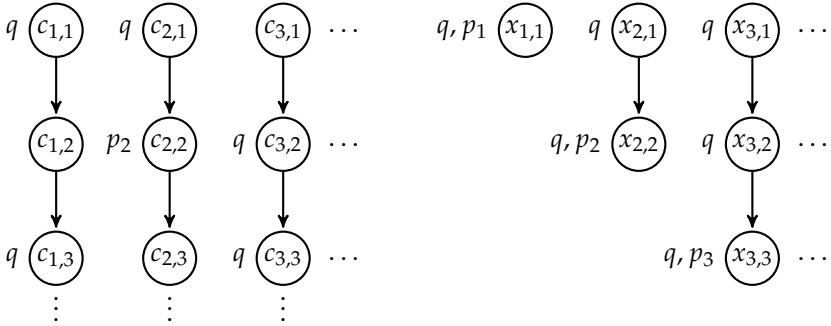

Let $\psi$ be the \MDL formula
\[\begin{array}{cl}
  & \bigvee\limits_{j=1}^n \EX^{j-1} \depp[q]{p_j}\\
= & \depp[q]{p_1} \vee \EX \depp[q]{p_2} \vee \EX \EX \depp[q]{p_3} \vee \dots \vee \EX^{n-1}\depp[q]{p_n}.
\end{array}\]

Once again we show that $\phi \in \ThreeSAT$ iff $W, \{c_{1,1},\dots, c_{m,1},\, x_{1,1}, x_{2,1},\dots, x_{n,1}\}\models \psi$.
Intuitively, the part of the structure comprised of the states $x_{j,j'}$ (pictured in the right part of \cref{figure:diamond-vee-bounded}) together with the subformulas $\EX^{j-1} \dep[q,p_j]$ for $j=1,\dots,n$ ensures that (for the subteam corresponding to the $j$th disjunct of $\psi$) at the $j$th level (where the levels are counted from $1$) $p_j$ is labeled wherever $q$ is labeled. But since in the chains in the $c_{i,j}$ part of the structure it never happens that $p_j$ is labeled if $q$ is labeled it follows that the $i$th chain cannot belong to the subteam corresponding to the $j$th disjunct of $\psi$ if $q$ is labeled at its $j$th level, and, by definition of $W$, the latter is the case iff neither $p_j$ nor $\neg p_j$ occurs in $C_i$.

More precisely, first assume that $\phi \in \ThreeSAT$ and that $\theta$ is a satisfying valuation for the variables in $\phi$.
Now let $P_j\dfn\{c_{i,1}\mid \text{$C_i$ is satisfied by $p_j$ under $\theta$}\}$ for all $j=1,\dots,n$. Then it follows that $\bigcup\limits_{j=1}^n P_j = \{c_{1,1},\dots,c_{m,1}\}$ and that
\[W,P_j \models \EX^{j-1} (\neg q \wedge \depp{p_j})\]
for all $j=1,\dots,n$.
Additionally, it holds that $W,\{x_{j,1}\}\models \EX^{j-1} (q \wedge \depp{p_j})$ for $j=1,\dots,n$.

Together it follows that $W,P_j\cup\{x_{j,1}\}\models \EX^{j-1} \depp[q]{p_j}$ for all $j=1,\dots,n$. This implies
\[W,\bigcup_{j=1}^n (P_j\cup \{x_{j,1}\}) \models \bigvee_{j=1}^n \EX^{j-1} \depp[q]{p_j}\]
which is equivalent to
\[W,\{c_{1,1},\dots, c_{m,1},\, x_{1,1}, x_{2,1},\dots, x_{n,1}\} \models \psi.\]

For the reverse direction assume that $W,T \models \psi$ with $T\dfn\{c_{1,1},\dots, c_{m,1},\allowbreak\, x_{1,1},\allowbreak x_{2,1},\allowbreak\dots, x_{n,1}\}$.
Let $T_1,\ldots,T_n$ be subsets of $T$ with $T_1 \cup \dots \cup T_n = T$ such that for all $j\in\{1,\dots,n\}$ it holds that $T_j\models \EX^{j-1} \depp[q]{p_j}$.
Then it follows that $x_{1,1}\in T_1$ since the chain starting in $x_{1,1}$ consists of only one state. From $\pi(x_{1,1})=\{q,p_1\}$ and $\pi(x_{2,1})=\{q\}$ it follows that $x_{2,1}\notin T_1$ and hence (again because of the length of the chain) $x_{2,1}\in T_2$. Inductively, it follows that $x_{j,1}\in T_j$ for all $j=1,\dots,n$.

Now, it follows from $x_{j,1}\in T_j$ that for all $i\in\{1,\dots,m\}$ with $c_{i,1}\in T_j$: $q\notin \pi(c_{i,j})$ (because $q,p_j\in\pi(x_{j,j}$, $p_j\notin \pi(x_{i,j})$). Since $T_j\models \EX^{j-1} \depp[q]{p_j}$, it then holds that $T_j\setminus\{x_{j,1}\}\models \EX^{j-1} (\neg q \wedge \depp{p_j})$.

Now let $\theta$ be the valuation of $p_1,\dots,p_n$ defined by
\[\theta(p_j)\dfn\begin{cases}
  1 & \text{if $T_j\setminus\{x_{j,1}\}\models \EX^{j-1} (\neg q \wedge p_j)$}\\
  0 & \text{if $T_j\setminus\{x_{j,1}\}\models \EX^{j-1} (\neg q \wedge \neg p_j)$}
\end{cases}.\]

Since for each $i=1,\dots,m$ there is a $j\in\{1,\dots,n\}$ such that $c_{i,1}\in T_j$ it follows that for each clause $C_i$ of $\phi$ there is a $j\in\{1,\dots,n\}$ such that $p_j$ satisfies $C_i$ under $\theta$.
\end{proof}

\subsection{Classical disjunction}\label{sec:mdl-mc-nor}
Finally, we investigate some remaining fragments which allow the classical disjunction operator.
Therefore we first show that classical disjunction can substitute zero-ary dependence atoms.
\begin{lemma}\label{nor replaces dep}
Let $\dep[],\nor \notin M$. Then
\[\MDLpara[{M\cup\{\dep[]\}}]{0}\hmc \leqpm \MDL[{M \cup \{\nor\}}]\hmc.\]
\end{lemma}
\begin{proof}
Follows immediately from \cref{mdl equivalences}\ref{dep with nor zero} and \cref{neg-dont-matter}.
\end{proof}

The following surprising result shows that both kinds of disjunctions together are already enough to get \NP-completeness.
\begin{theorem}\label{nor vee}
Let $\{\sor, \nor\}\subseteq M$. Then $\MDLMCk[M]$ is \NP-complete for every $k\geq 0$.
\end{theorem}
\begin{proof}
As above membership in \NP follows from \cref{mdlmc-in-np} and for hardness we reduce \ThreeSAT to $\MDLk[\sor,\nor]\hmc$ -- using a construction that bears some similarities with the one used in the proof of \cref{diamond-vee-bounded}.

For this purpose let $\phi \dfn \bigwedge_{i=1}^m C_i$ be an arbitrary \cnf[3] formula built from the variables $p_1, \dots, p_n$. Let $W$ be the Kripke structure $(S, R, \pi)$ shown in \cref{figure:nor vee} and formally defined by
\[\begin{array}{lcl}
S & \dfn & \{c_{i} \mid i\in\{1,\dots, m\}\}\\
R & \dfn & \emptyset\\
\pi(c_{i}) & \dfn & \{p_j \mid p_j \in C_i\} \cup \{q_j \mid \neg p_j \in C_i\}.
\end{array}\]

\begin{figure}[ht]
\begin{center}
\begin{tikzpicture}[->,>=stealth',shorten >=1pt,auto,node distance=2cm,every state/.style={minimum size=7mm,inner sep=1pt},thick]
\node[state]                             (c1)       [label=below:{$p_1$, $q_2$}]               {$c_{1}$};
\node[state]                             (c2)       [right of=c1,label=below:{$p_2$, $q_3$}]               {$c_{2}$};
\node[state]                             (c3)       [right of=c2,label=below:{$q_1$, $p_4$}]               {$c_{3}$};
\node                            (cx)       [right = 1cm of c3]               {$\dots$};
\end{tikzpicture}
\caption{Kripke structure corresponding to a \cnf[3] formula containing the clauses $C_1=p_1\vee \neg p_2$, $C_2=p_2\vee \neg p_3$ and $C_3=\neg p_1\vee p_4$}
\label{figure:nor vee}
\end{center}
\end{figure}

Let $\psi$ be the \MDL formula
\[\bigsor\limits_{j=1}^n (p_j \nor q_j).\]

Once again we show that $\phi \in \ThreeSAT$ iff $W, \{c_{1},\dots, c_{m}\}\models \psi$.
First assume that $\phi \in \ThreeSAT$ and that $\theta$ is a satisfying valuation for $\phi$.
Now let
\[P_j\dfn\{c_{i}\mid \text{$C_i$ is satisfied by $p_j$ under $\theta$}\}\]
for all $j=1,\dots,n$. Then it follows that $\bigcup_{j=1}^n P_j = \{c_{1},\dots,c_{m}\}$ and that
\[W,P_j \models p_j \nor q_j\]
for all $j=1,\dots,n$.
Together it follows that
\[W,\{c_{1},\dots, c_{m}\} \models \bigsor\limits_{j=1}^n (p_j \nor q_j).\]

For the reverse direction assume that $W,T \models \psi$ with $T\dfn\{c_{1},\dots, c_{m}\}$.
Let $T_1,\dots,T_n$ be subsets of $T$ with $T_1 \cup \dots \cup T_n = T$ such that for all $j\in\{1,\dots,n\}$ it holds that $T_j\models p_j\nor q_j$.
Now let $\theta$ be the valuation of $p_1,\dots,p_n$ defined by
\[\theta(p_j)\dfn\begin{cases}
  1 & \text{if }T_j\models p_j,\\
  0 & \text{if }T_j\models q_j.\\
\end{cases}\]

Since for each $i=1,\dots,m$ there is a $j\in\{1,\dots,n\}$ such that $c_{i}\in T_j$ it follows that for each clause $C_i$ of $\phi$ there is a $j\in\{1,\dots,n\}$ such that $p_j$ satisfies $C_i$ under $\theta$.
\end{proof}

Now we show that \cref{no-wedge-no-vee} still holds if we additionally allow classical disjunction.
\begin{theorem}\label{nor and unary}
Let $M\subseteq \{\AX,\EX, \nor, \neg, \dep[]\}$. Then $\MDLk[M]\hmc$ is in \P for every $k\geq 0$.
\end{theorem}
\begin{proof}
Let $\phi \in \MDL[M]$.
Because of the distributivity of $\nor$ with all other \MDL operators (\cf \cref{bullet distributivity}\ref{bullet in front}) there is a formula $\psi$ equivalent to $\phi$ which is of the form
\[\bignor_{i=1}^{|\phi|} \psi_i\]
with $\psi_i \in \MDL[{M\setminus \{\nor\}}]$ for all $i\in\{1,\dots,|\phi|\}$.
Note that there are only linearly many formulas $\psi_i$ (in contrast to exponentially many in \cref{bullet distributivity}\ref{bullet in front}) because $\phi$ does not contain any binary operators aside from $\nor$.
Further note that $\psi$ can be easily computed from $\phi$ in polynomial time.

Now it is easy to check for a given structure $W$ and team $T$ whether $W,T\models \psi$ by simply checking whether $W,T\models \psi_i$ (which can be done in polynomial time by \cref{no-wedge-no-vee}) consecutively for all $i\in\{1,\dots,|\phi|\}$.
\end{proof}

\section{Conclusion}

\Cref{mdl sat results,mdl sat results bounded,mdl mc results,mdl mc results bounded} give a complete overview of our complexity results. Note that all possible combinations of operators are included in each table.

\fragmenttableMDLsat{
+&+ & +&*&+&*&* & +&* & \NEXP&		\Cref{poor man dep complexity}\\
+&+ & +&+&+&*&* & -&* & \P\SPACE&	\Cref{simple cases}a\\
+&+ & +&+&-&*&+ & *&* & \P\SPACE&	\Cref{simple cases}b\\
+&+ & +&-&+&*&* & -&+ & \SigmaP{2}&	\Cref{poor man bullet complexity}\\
+&+ & +&-&-&*&+ & *&+ & \SigmaP{2}&	\Cref{poor man bullet complexity}\\
+&+ & +&-&+&*&* & -&- & \co\NP&		\cite{la77}, \cite{dolenahonuma92}\\
+&+ & +&-&-&*&+ & *&- & \co\NP&		\Cref{simple cases}c\\
\hline
+&- & +&+&+&*&* & *&* & \NP&		\Cref{one modality cases}a\\
-&+ & +&+&+&*&* & *&* & \NP&		\Cref{one modality cases}a\\
+&- & +&-&+&*&* & *&+ & \NP&		\Cref{one modality cases}a\\
-&+ & +&-&+&*&* & *&+ & \NP&		\Cref{one modality cases}a\\
+&- & +&-&+&*&* & *&- & \P&		\Cref{one modality cases}b\\
-&+ & +&-&+&*&* & *&- & \P&		\Cref{one modality cases}b\\
+&- & +&*&-&*&* & *&* & \P&		\Cref{one modality cases}c\\
-&+ & +&*&-&*&* & *&* & \P&		\Cref{one modality cases}c\\
*&* & -&*&*&*&* & *&* & \P&		\Cref{one modality cases}d\\
*&* & *&*&-&*&- & *&* & \Trivial&	\Cref{simple cases}d\\
\hline
-&- & +&+&+&*&* & *&* & \NP&		\cite{co71}\\
-&- & +&*&+&*&* & *&+ & \NP&		\cite{co71}, $\nor \equiv \vee$\\
-&- & *&-&*&*&* & *&- & \P&		\Cref{simple cases}e\\
-&- & *&*&-&*&* & *&* & \P&		\Cref{simple cases}f
}{Classification of complexity for fragments of $\MDL\hsat$}{mdl sat results}

\fragmenttableMDLsat{
+&+ & +&+&+&*&* & *&* & \P\SPACE&	\Cref{bounded dep concrete}a\\
+&+ & +&+&-&*&+ & *&* & \P\SPACE&	\Cref{simple cases}b\\
+&+ & +&-&+&*&* & +&* & \SigmaP{3}&	\Cref{bounded dep concrete}b\\
+&+ & +&-&+&*&* & -&+ & \SigmaP{2}&	\Cref{poor man bullet complexity}\\
+&+ & +&-&-&*&+ & *&+ & \SigmaP{2}&	\Cref{poor man bullet complexity}\\
+&+ & +&-&+&*&* & -&- & \co\NP&		\cite{la77}, \cite{dolenahonuma92}\\
+&+ & +&-&-&*&+ & *&- & \co\NP&		\Cref{simple cases}c\\
\hline
+&- & +&+&+&*&* & *&* & \NP&		\Cref{one modality cases}a\\
-&+ & +&+&+&*&* & *&* & \NP&		\Cref{one modality cases}a\\
+&- & +&-&+&*&* & *&+ & \NP&		\Cref{one modality cases}a\\
-&+ & +&-&+&*&* & *&+ & \NP&		\Cref{one modality cases}a\\
+&- & +&-&+&*&* & *&- & \P&		\Cref{one modality cases}b\\
-&+ & +&-&+&*&* & *&- & \P&		\Cref{one modality cases}b\\
+&- & +&*&-&*&* & *&* & \P&		\Cref{one modality cases}c\\
-&+ & +&*&-&*&* & *&* & \P&		\Cref{one modality cases}c\\
*&* & -&*&*&*&* & *&* & \P&		\Cref{one modality cases}d\\
*&* & *&*&-&*&- & *&* & \Trivial&	\Cref{simple cases}d\\
\hline
-&- & +&+&+&*&* & *&* & \NP&		\cite{co71}\\
-&- & +&*&+&*&* & *&+ & \NP&		\cite{co71}, $\nor \equiv \vee$\\
-&- & *&-&*&*&* & *&- & \P&		\Cref{simple cases}e\\
-&- & *&*&-&*&* & *&* & \P&		\Cref{simple cases}f
}{Classification of complexity for fragments of $\MDLk\hsat$ for $k\geq 3$}{mdl sat results bounded}

\fragmenttableMDLmc{
  *&* & +&+&* & +&* & \NP\text{-complete} & \Cref{wedge-vee-np-complete}\\\hline
  +&* & *&+&* & +&* & \NP\text{-complete} & \Cref{box-vee-np-complete}\\\hline 
  *&* & *&+&* & *&+ & \NP\text{-complete} & \Cref{nor vee}\\\hline
  *&+ & *&*&* & +&* & \NP\text{-complete} & \Cref{diamond-np-complete}\\\hline 
  *&+ & +&*&* & *&+ & \NP\text{-complete} & \hbox{\Cref{diamond-wedge-bounded},} \Cref{nor replaces dep}\\\hline
  -&- & -&+&* & +&- & \text{in }\NP & \Cref{mdlmc-in-np}\\\hline
  *&* & -&-&* & -&* & \text{in }\P & \Cref{nor and unary}\\\hline
  *&- & *&-&* & *&* & \text{in }\P & \Cref{box-wedge-in-p}\\\hline 
  *&* & *&*&* & -&- & \text{in }\P & \cite{clemsi86}\\\hline
}{Classification of complexity for fragments of $\MDL\hmc$}{mdl mc results}

\fragmenttableMDLmc{
  *&* & +&+&* & +&* & \NP\text{-complete} & \Cref{wedge-vee-np-complete}\\\hline
  +&* & *&+&* & +&* & \NP\text{-complete} & \Cref{box-vee-np-complete}\\\hline 
  *&* & *&+&* & *&+ & \NP\text{-complete} & \Cref{nor vee}\\\hline
  *&+ & +&*&* & +&* & \NP\text{-complete} & \Cref{diamond-wedge-bounded}\\\hline 
  *&+ & +&*&* & *&+ & \NP\text{-complete} & \hbox{\Cref{diamond-wedge-bounded},} \Cref{nor replaces dep}\\\hline
  *&+ & *&+&* & +&* & \NP\text{-complete} & \Cref{diamond-vee-bounded}\\\hline 
  *&* & -&-&* & *&* & \text{in }\P & 
  \Cref{nor and unary} \\\hline 
  *&- & *&-&* & *&* & \text{in }\P & \Cref{box-wedge-in-p}\\\hline 
  -&- & -&*&* & *&- & \text{in }\P & \Cref{vee-bounded-in-p} \\\hline
  *&* & *&*&* & -&- & \text{in }\P & \cite{clemsi86}\\\hline
}{Classification of complexity for fragments of $\MDLk\hmc$ for $k\geq 1$}{mdl mc results bounded}

\subsection{Satisfiability}
In this thesis we completely classified the complexity of the
satisfiability problem of modal dependence logic for all fragments of the
language defined by restricting the modal and propositional operators to a
subset of those considered by V\"a\"an\"anen and Sevenster.
Our results show a dichotomy for the unbounded arity $\dep$ operator; either the complexity jumps to \NEXP-completeness when introducing $\dep$ or it does not increase at all -- and in the latter case the $\dep$ operator does not increase the expressiveness of the logic.
Intuitively, the \NEXP-completeness can be understood as the complexity of guessing Boolean functions of unbounded arity.

An interesting question is whether there are natural fragments of modal dependence logic where adding the dependence operator does not let the complexity of satisfiability testing jump up to \NEXP but still increases the expressiveness of the logic.
This question can be answered by restricting the arity of the \dep operator. In this case dependence becomes too weak to increase the complexity beyond \P\SPACE. However, in the case of poor man's logic, \ie only disjunctions are fobidden, the complexity increases to $\SigmaP{3}$ when introducing dependence but still it is not as worse as in the case of full modal logic.
Intuitively,  the complexity drops below \NEXP because the Boolean functions which have to be guessed are now of a bounded arity.

\subsection{Model checking}
In this thesis we showed that $\MDLMC$ is \NP-complete (\Cref{wedge-vee-np-complete}). Furthermore we have systematically analyzed the complexity of model checking for fragments of \MDL defined by restricting the set of modal and propositional operators. It turned out that there are several fragments which stay \NP-complete, \eg the fragment obtained by restricting the set of operators to only $\AX, \vee$ and $\depp{}$ (\Cref{box-vee-np-complete}) or only $\EX$ and $\depp{}$ (\Cref{diamond-np-complete}). Intuitively, in the former case the \NP-hardness arises from existentially guessing partitions of teams while evaluating disjunctions and in the latter from existentially guessing successor teams while evaluating $\EX$ operators.
Consequently, if we allow all operators except $\EX$ and $\vee$ the complexity drops to \P (\Cref{box-wedge-in-p}).


For the fragment only containing $\vee$ and $\dep$ on the other hand we were not able to determine whether its model checking problem is tractable.
Our inability to prove either \NP-hardness or containment in \P led us to restrict the arity of the dependence atoms. For the aforementioned fragment the complexity drops to $\P$ in the case of bounded arity (\Cref{vee-bounded-in-p}). Furthermore, some of the cases which are known to be \NP-complete for the unbounded case drop to \P in the bounded arity case as well (\Cref{nor and unary}) while others remain \NP-complete but require a new proof technique (\Cref{diamond-wedge-bounded,diamond-vee-bounded}).
Most noteworthy in this context are probably the results concerning the $\EX$ operator. With unbounded dependence atoms this operator alone suffices to get \NP-completeness whereas with bounded dependence atoms it needs the additional expressiveness of either $\wedge$ or $\vee$ to reach \NP-hardness.

Considering the classical disjunction operator $\nor$, we showed that the complexity of $\MDLpara[{M\cup \{\dep[]\}}]{k}\hmc$ is never higher than the complexity of\linebreak $\MDLk[{M\cup \{\nor\}}]\hmc$, \ie $\nor$ is at least as bad as $\dep$ with respect to the complexity of model-checking (in contrast to the complexity of satisfiability; \cf \cref{mdl sat results bounded}). And in the case where only $\sor$ is allowed we even have a higher complexity with $\nor$ (\Cref{nor vee}) than with $\dep[]$ (\Cref{vee-bounded-in-p}). The case of $\MDL[\sor,\nor]\hmc$ is also our probably most surprising result since the non-determinism of the $\sor$ operator turned out to be powerful enough to lead to \NP-completeness although neither conjunction nor dependence atoms (which also, in a sense, contain a kind of \qte{semi-atomic conjunction}) are allowed.

Interestingly, in none of our reductions to show \NP-hardness the \MDL formula depends on anything else but the number of propositional variables of the input \cnf[3] formula. The structure of the input formula is always encoded by the Kripke structure alone.
So it might seem that even for a fully fixed formula the model checking problem could still be hard.
This, however, cannot be the case since, by \cref{few-dep-atoms-in-p}, model checking for a fixed formula is always in \P. 

Another open question, apart from the unclassified unbounded arity case, is related to a case with bounded arity dependence atoms. In \cref{diamond-vee-bounded} we were only able to prove \NP-hardness for arity at least one and it is not known what happens in the case where the arity is zero.
Additionally, it might be interesting to determine the exact complexity for the cases which are in \P since we have not shown any lower bounds in these cases so far.


\subsection{Open problems}
In a number of precursor papers, \eg \cite{le79} on propositional
logic or \cite{hescsc10} on modal logic, not only subsets of the classical
operators $\{\Box,\Diamond,\allowbreak \wedge,\allowbreak \vee,\aneg\}$ were considered but also
propositional connectives given by arbitrary Boolean functions.
%
Contrary to classical propositional or modal logic, however, the
semantics of such generalized formulas in dependence logic is not clear a priori -- for
instance, how should exclusive-or be defined in dependence logic? Even for
simple implication there seem to be several reasonable definitions,
\cf \cref{sec:idl intro} and \cite{abva08}.

A further possibly interesting restriction of dependence logic might be to
restrict the type of functional dependence beyond simply restricting the arity. Right now, dependence just
means that there is some function whatsoever that determines the value of
a variable from the given values of certain other variables. Also here it
might be interesting to restrict the function to be taken from a fixed
class in Post's lattice, \eg to be monotone or self-dual.

Finally, it seems natural to investigate the possibility of enriching classical temporal logics such as \LTL, \CTL or \CTLs with dependence as some of them are extensions of classical modal logic. The questions here are of the same kind as for \MDL: expressivity, complexity, fragments (\cf \cite{me11} for a systematic study of several classical temporal logics with respect to the complexity of the satisfiability and the model checking problem for fragments).

\chapter{Modal Intuitionistic Dependence Logic}\label{chap:midl}
\section{Model checking}\label{sec:midl mc}
In this section we study the complexity of model checking for fragments of \MIDL, \ie we extend the results from \cref{sec:mdl-mc}.

For this purpose first note that Boolean constants as well as negation do not influence the complexity of the model checking problem for \MIDL, \ie \cref{neg-dont-matter} holds for \MIDL as well as for \MDL.

We begin by stating a $\P\SPACE$ algorithm for $\MIDL\hmc$.
\begin{theorem}\label{MIDL-MC-inPSPACE}
$\MIDL\hmc$ is in $\P\SPACE$.
\end{theorem}
\begin{proof}
We state an $\AP$ algorithm to prove the theorem.

\Needspace{12\baselineskip}
\begin{lstlisting}[caption={\lstinline!check($K=(S,R,\pi)$, $\phi$, $T$)!}, label={algo:midl check}, escapechar={§}]
case $\phi$
when $\phi=\true,\ \false,\ p,\ \neg p,\ \dep[p_1,\dots,p_n],\ \psi \sor \chi,\  \psi \nor \chi,\  \psi \wedge \chi,\ \AX \psi,\ \EX \psi$
  proceed according to §\Cref{algo:mdl check}§ §(guessing steps are \emph{existential} guessing steps)§
when $\phi = \psi \imp \chi$
  universally guess a set of states $T' \subseteq T$
  if not check($K$, $\psi$, $T'$) or check($K$, $\chi$, $T'$)
    return true
  return false
\end{lstlisting}
Most of the cases in the algorithm are deterministic. Only the cases $\phi = \psi \sor \chi$ and $\phi = \Diamond \psi$ include non-deterministic existential  branching and the case $\phi = \psi \imp \chi$ includes non-deterministic universal branching. Altogether, the algorithm can be implemented on an alternating Turing machine running in polynomial time or -- equivalently (\cf \cref{def:ph} and \cite{chkost81}) -- on a deterministic machine using polynomial space.
\end{proof}

If we forbid $\sor$ and $\EX$ the complexity of the above algorithm drops to $\co\NP$.
\begin{corollary}\label{PIDL-MC-incoNP}
$\MIDL[{\AX,\wedge,\nor,\neg,\imp,\dep[]}]\hmc$ is in \co\NP. In particular, $\PIDL\hmc$ is in $\co\NP$.
\end{corollary}
\begin{proof}
In \cref{algo:midl check}, existential guessing only applies to the cases $\phi = \psi \sor \chi$ and $\phi = \EX \psi$.
\end{proof}

If neither $\dep$ nor $\nor$ are allowed in the logic the model checking problem can even be decided in deterministic polynomial time.
\begin{theorem}\label{ML-imp-inP}
$\MIDL[{\AX, \EX, \wedge,\sor,\neg, \imp}]\hmc$ is in $\P$.
\end{theorem}
\begin{proof}
Let $\phi$ be an arbitrary $\MIDL[{\AX, \EX, \wedge,\sor,\neg, \imp}]$ formula. Starting from the innermost intuitionistic implication $\imp$, by applying \cref{midl equivalences}\ref{imp with sor}, we may eliminate all the connectives $\imp$ in $\phi$ and obtain an equivalent $\ML$ formula $\phi^*$.

The translation from $\phi$ to $\phi^*$ can clearly be done in polynomial time and, by \cite{clemsi86}, $\ML\hmc$ is in $\P$. Hence, $\MIDL[{\AX, \EX, \wedge,\allowbreak\sor,\allowbreak\neg, \imp}]\hmc$ is in $\P$ as well.
\end{proof}

In the remaining part of this section we provide hardness proofs for model checking problems for various sublogics of \MIDL. We first consider the sublogic without $\EX$ and $\sor$.
\begin{theorem}\label{PIDL-MC-hard}
$\PIDL[\wedge,\nor,\imp]\hmc$ is \co\NP-hard.
\end{theorem}
\begin{proof}
We give a polynomial-time reduction from
\[\taut \dfn \{\phi\in \CL \mid \phi \text{ is a tautology}\}\]
to $\PIDL\hmc = \PIDL[{\wedge,\nor,\neg,\imp,\dep[]}]\hmc$, which implies the desired result since 
\[\begin{array}{rcl}
\PIDL[{\wedge,\nor,\neg,\imp,\dep[]}]\hmc &\equivpm& \PIDL[{\wedge,\nor,\neg,\imp}]\hmc\\
&\equivpm& \PIDL[\wedge,\nor,\imp]\hmc,
\end{array}\]
by \cref{midl equivalences} and \cref{neg-dont-matter}.


For this purpose let $\phi$ be an arbitrary propositional formula -- \wlg in negation normal form -- and let $\var{\phi}=\{p_1,\dots,p_n\}$.
Let $K=(S,R,\pi)$ be the Kripke structure shown in \cref{fig:pidl hardness structure} and formally defined by
\[\begin{array}{lcl}
S   & \dfn &\{s_1,\dots,s_n, \overline{s_1}, \dots, \overline{s_n}\}, \\
R   &\dfn & \emptyset ,\\
\pi(s_i) &\dfn & \{r_i, p_i\}, \\
\pi(\overline{s_i}) &\dfn & \{r_i\}. 
\end{array}\]

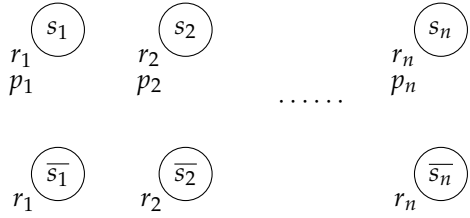
\begin{figure}[ht]
\begin{center}
\newcommand{\mynode}[5]{\node[state,label={[label distance=-10]-135:{$\begin{array}{c}#4\\[-0.5ex]#5\end{array}$}}]    (#1)  at (#2)               {$#3$};}
\begin{tikzpicture}[-stealth', auto, x=2.4em, y=2.7em,every state/.style={minimum size=7mm,inner sep=1pt}]

\mynode{s1}{0,2}{s_1}{r_1}{p_1}

\mynode{s1b}{0,0}{\overline{s_1}}{r_1}{}

\mynode{s2}{2,2}{s_2}{r_2}{p_2}

\mynode{s2b}{2,0}{\overline{s_2}}{r_2}{}

\node           (sd)  at (4,1)                   {$\cdots\cdots$};

\mynode{sn}{6,2}{s_{n}}{r_{n}}{p_{n}}
\mynode{snb}{6,0}{\overline{s_{n}}}{r_{n}}{}

\end{tikzpicture}
\end{center}
\caption{Kripke structure in the proof of \cref{PIDL-MC-hard}}
\label{fig:pidl hardness structure}
\end{figure}

Finally, let $\psi\in\PIDL$ be defined by
\[\psi\ \dfn\ \alpha_n \imp \phi^{\imp},\]
with
\[\alpha_n \dfn \bigwedge_{i=1}^{n}(r_i \imp \dep[p_i])\]
and $\phi^{\imp}$ defined by the following inductive translation:
\begin{align*}
p_i^{\imp} &\dfn r_i\imp p_i,\\
(\neg p_i)^{\imp}  &\dfn r_i\imp \neg p_i,\\
(\phi \wedge \psi)^{\imp} &\dfn \phi^{\imp} \wedge \psi^{\imp},\\
(\phi \vee \psi)^{\imp} &\dfn \phi^{\imp} \nor \psi^{\imp}.
\end{align*}

Now we will show that $\phi\in \taut$ iff $K,S \models \psi$.

Suppose $\phi\in \taut$. Let $T\subseteq S$ be an arbitrary team such that $K,T \models \alpha_n$. Then there is a $T'\supseteq T$ such that for all $1\leq i\leq n$ exactly one of the states $s_i$ and $\overline{s_i}$ is in $T'$.
Now let $\sigma$ be the truth assignment defined by
\[\sigma(p_i)\,\dfn\, \left\{
                         \begin{array}{ll}
                           \true & \mbox{if }s_i\in T', \\
                           \false & \mbox{if }\overline{s_i}\in T'.
                         \end{array}
                       \right.\]

\begin{claim}\label{sigma midl equivalence}
For all subformulas $\chi$ of $\phi$ it holds that $\sigma$ satisfies $\chi$ iff $K,T'\models \chi^{\imp}$.
\end{claim}
\begin{claimproof}{\cref*{sigma midl equivalence} (inductive)}

Case $\chi=p_i$: First suppose $\sigma(p_i)=\true$. Then for all teams $T''\subseteq T'$ with $K,T''\models r_i$ it holds that $T''\subseteq \{s_i,\overline{s_i}\}$. Furthermore, by definition of $T'$, $T'\cap\{s_i,\overline{s_i}\}=\{s_i\}$. Hence, $T''\subseteq\{s_i\}$ and $K,T''\models p_i$. Since $T''$ was chosen arbitrarily this implies $K,T'\models r_i\imp p_i$. Conversely suppose that $\sigma(p_i)=\false$. Then $T'\cap\{s_i,\overline{s_i}\}=\{\overline{s_i}\}$ and therefore $T'\not\models r_i\imp p_i$.

Case $\chi=\neg p_i$: Analogous to $\chi=p_i$.

Case $\chi=\tau\vee\theta$:
\begin{tabular}[t]{@{\,}l@{\;\;}l}
&$\sigma$ satisfies $\chi$\\
iff& $\sigma$ satisfies $\tau$ or $\sigma$ satisfies $\theta$\\
iff, by induction hypothesis,& $K,T'\models \tau^{\imp}$ or $K,T'\models \theta^{\imp}$\\
iff, by the semantics of $\nor$,& $K,T' \models \tau^{\imp} \nor \theta^{\imp}$.
\end{tabular}

Case $\chi=\tau\wedge\theta$: Analogous to $\chi=\tau\vee\theta$.
\end{claimproof}

Since $\phi\in\taut$, it holds that $\sigma$ satisfies $\phi$. By \cref{sigma midl equivalence} it follows that $K,T' \models \phi^{\imp}$ and, by the downward closure property (\Cref{downward closure property mdl}), $K,T\models \phi^{\imp}$. Since $T$ was chosen arbitrarily with $K,T\models \alpha_n$, this implies $K,S\models \psi$.

Conversely suppose that $K,S\models \psi$ and let $\sigma$ be an arbitrary truth assignment for $p_1,\dots,p_n$. Let $T\dfn \set{s_i}{$\sigma(p_i)=\true$} \cup \set{\overline{s_i}}{$\sigma(p_i)=\false$}$. Then, obviously, $K,T \models \alpha_n$, thus $K,T \models \phi^{\imp}$ and, by \cref{sigma midl equivalence}, $\sigma$ satisfies $\phi$. Since $\sigma$ was chosen arbitrarily, $\phi\in\taut$.
\end{proof}

\begin{theorem}\label{PIDL-MC}
$\PIDL\hmc$ is \co\NP-complete. Furthermore, $\MIDL[M]\hmc$ is \co\NP-complete for all $\{\wedge,\nor,\imp\}\subseteq M \subseteq \{\AX,\wedge,\nor,\neg,\imp,\dep[]\}$.
\end{theorem}
\begin{proof}
Follows directly from \cref{PIDL-MC-incoNP} and \cref{PIDL-MC-hard}.
\end{proof}

Next, we analyze the complexity of model checking for fragments of \MIDL containing $\sor$ and $\imp$.
\begin{theorem}\label{pidls-mc-lower}
Let $\{\wedge, \sor, \imp, \dep[]\} \subseteq M$. Then $\MIDL[M]\hmc$
is \P\SPACE-com\-plete.\footnote{Note that $\MIDL[{\wedge, \sor,\neg, \imp, \dep[]}]$ is a variation of \PIDL where $\nor$ has been replaced by $\sor$.}
\end{theorem}
\begin{proof}
The upper bound follows from \cref{MIDL-MC-inPSPACE}. For the lower bound we give a reduction from $\qbfcnfgeneral$, which is well-known to be $\P\SPACE$-complete. Let $\psi=\forall x_1\exists x_2 \dots \forall x_{n-1} \exists x_n\, \phi$ be a $\qbfcnfgeneral$ instance with $\phi=C_1\wedge\cdots\wedge C_m$ and
\[C_j=\alpha_{j0}\vee\alpha_{j1}\vee\alpha_{j2} \quad (1\leq j\leq m)\]
for, \wlg, distinct $\alpha_{j0},\alpha_{j1},\alpha_{j2}$.  We further assume \wlg that $n$ is even.
Then the correspoding $\MIDL[{\wedge, \sor,  \imp, \dep[]}]\hmc$ instance is defined as $(K=(S,R,\pi),S,\theta)$, where 
\begin{itemize}
\item $S\dfn \{s_1,\dots,s_n,\overline{s_1},\dots,\overline{s_n}\}$,
\item $R\dfn\emptyset$,
\item $\ap\dfn \{r_1,\dots,r_n,p_1,\dots,p_n,c_1,\dots,c_m,c_{10},\dots,c_{m0},c_{11},\dots,c_{m1}, c_{12},\linebreak \dots,c_{m2}\}$
\item $\pi(s_i)=\begin{array}[t]{rl}
\{r_i,p_i\}\ \cup&\{c_j,c_{j0} \mid \alpha_{j0}=x_i,~1\leq j\leq m\}\\
\cup&\{c_j,c_{j1} \mid \alpha_{j1}=x_i,~1\leq j\leq m\}\\
\cup&\{c_j,c_{j2} \mid \alpha_{j2}=x_i,~1\leq j\leq m\},
\end{array}$
\item $\pi(\overline{s_i})=\begin{array}[t]{rl}
\{r_i\}\ \cup&\{c_j,c_{j0} \mid \alpha_{j0}=\neg x_i,~1\leq j\leq m\}\\
\cup&\{c_j,c_{j1} \mid \alpha_{j1}=\neg x_i,~1\leq j\leq m\}\\
\cup&\{c_j,c_{j2} \mid \alpha_{j2}=\neg x_i,~1\leq j\leq m\},
\end{array}$\\
see \cref{example_K'} for an example of the construction of $K$,

\item $\theta\dfn \delta_1$, where
\begin{align*}
\delta_{2k-1}\dfn\ & (r_{2k-1}\imp \dep[p_{2k-1}])\imp \delta_{2k} \quad (1\leq k\leq n/2),\\
\delta_{2k}\dfn\  & (r_{2k}\wedge\dep[p_{2k}])\sor\delta_{2k+1} \quad (1\leq k< n/2),\\
\delta_n\dfn\ &(r_n\wedge\dep[p_n])\sor\phi',
\end{align*}
\begin{align*}
\text{\ie}\ \theta=\ \big(&r_1\to \dep[p_1]\big) \to \Big( \big(r_2\wedge \dep[p_2]\big) \sor \\
&\quad\quad\quad\quad\quad\quad\quad\cdots\quad\cdots\quad \cdots~\sor \\
&\quad\Big(\big(r_{n-1}\to \dep[p_{n-1}]\big) \to\Big(\big(r_n\wedge\dep[p_n]\big)\sor\phi'\Big)\Big)\cdots  \Big)\Big)\Big),
\end{align*}
and
\[\begin{array}{l@{}l@{}l@{}l}
\phi' \dfn \bigwedge\limits_{j=1}^m \bigg(c_j\imp \Big(&& \big(\dep[c_{j0}]\wedge \dep[c_{j1}] \wedge \dep[c_{j2}] \big)\\
&\sor &\big(\dep[c_{j0}]\wedge \dep[c_{j1}] \wedge \dep[c_{j2}]\big)&\Big)\bigg).
\end{array}\]
\end{itemize}
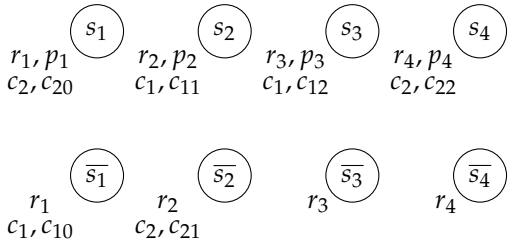
\begin{figure}[ht]
\begin{center}
\newcommand{\mynode}[5]{\node[state,label={[label distance=-10]-135:{$\begin{array}{c}#4\\[-0.5ex]#5\end{array}$}}]    (#1)  at (#2)               {$#3$};}
\begin{tikzpicture}[-stealth', auto, x=2.4em, y=2.7em, every state/.style={minimum size=7mm, inner sep=1pt}]
\mynode{s1}{0,2}{s_1}{r_1,p_1}{c_2,c_{20}}

\mynode{s1b}{0,0}{\overline{s_1}}{r_1}{c_1,c_{10}}

\mynode{s2}{2,2}{s_2}{r_2,p_2}{c_1,c_{11}}

\mynode{s2b}{2,0}{\overline{s_2}}{r_2}{c_2,c_{21}}

\mynode{sn}{4,2}{s_{3}}{r_{3},p_{3}}{c_1, c_{12}}
\mynode{snb}{4,0}{\overline{s_{3}}}{r_{3}}{}

\mynode{sn}{6,2}{s_{4}}{r_{4},p_{4}}{c_2, c_{22}}
\mynode{snb}{6,0}{\overline{s_{4}}}{r_{4}}{}
\end{tikzpicture}

\caption{Kripke structure corresponding to $\psi = \forall x_1 \exists x_2 \forall x_3 \exists x_4 \big((\neg x_1\vee x_2\vee x_3)\wedge (x_1\vee \neg x_2\vee x_4)\big)$}
\label{example_K'}
\end{center}
\end{figure}

We will now show that $\psi\in \qbfcnfgeneral$ iff $K,S \models \theta$.

The general idea is that the alternating $\imp$ and $\sor$ operators simulate the \qbfcnfgeneral formula's $\forall$ and $\exists$ quantifiers, respectively. Therefore we start with the team of all states and then for every nesting level $i$ of $\imp$ or $\sor$ we drop one of the states $s_i$ and $\overline{s_i}$. We do this until we arrive at a team that contains exactly one of the states $s_i$ and $\overline{s_i}$ for each $i\in\{1,\dots,n\}$. This team corresponds to a truth assignment $\sigma$ for $x_1,\dots,x_n$ in a natural way and it satisfies $\phi'$ iff $\sigma$ satisfies $\phi$.

For the universal quantifiers we have that $\delta_{2k-1}$ is satisfied in a team $T$ iff $\delta_{2k}$ is satisfied in all maximal subteams $T_{2k-1}\subseteq T$ which satisfy $r_{2k-1}\imp \dep[p_{2k-1}]$, \ie all maximal subteams containing only one of the states $s_{2k-1}$ and $\overline{s_{2k-1}}$.
For the existential quantifiers $\delta_{2k}$ is satisfied in $T$ iff $T$ can be split into $T_{2k}$ and $T'_{2k}$ such that $K,T_{2k}\models \delta_{2k+1}$ and $K,T'_{2k}\models r_{2k}\wedge \dep[p_{2k}]$, \ie $\delta_{2k+1}$ has to be satisfied in a team with only one of the states $s_{2k}$ and $\overline{s_{2k}}$ since $T'_{2k} \subseteq \{s_{2k},\overline{s_{2k}}\}$ and $|T'_{2k}| \leq 1$.

\smallskip
More precisely, first suppose that $\psi\in \qbfcnfgeneral$.
Then for every partial truth assignment $\sigma\restricted{x_1,x_3,\dots,x_{n-1}} \colon \{x_1,x_3,\dots,x_{n-1}\}\to \{\true,\false\}$ there is a complete truth assignment $\sigma \colon \{x_1,x_2,\dots,x_{n}\}\to \{\true,\false\}$ such that $\sigma$ satisfies $\phi$ and $\sigma(x_{2k})$ only depends on $\sigma(x_1),\sigma(x_3),\dots,\sigma(x_{2k-1})$ (for all $k\in\{1,\dots,n/2\}$).

We now have to show that
\[K,S \models (r_1\imp\dep[p_1])\imp \delta_2.\]
By the downward closure property (\Cref{downward closure property mdl}), it suffices to show that for the maximal teams $T_1\subseteq S$ such that $K,T_1\models r_1\imp \dep[p_1]$, namely the teams $T_1=S \setminus\{s_1\}$ and $T_1=S \setminus \{\overline{s_1}\}$, it holds that
\[K,T_1\models\delta_2 \text{, \ie }K,T_1\models (r_2\wedge \dep[p_2]) \sor\delta_3.\]

Choose the value of $\sigma(x_1)$ according to $T_1$ by letting
\[\sigma(x_1) \dfn \left\{
                         \begin{array}{ll}
                           \true  & \mbox{if }T_1=S \setminus\{s_1\}, \\
                           \false & \mbox{if }T_1=S \setminus\{\overline{s_1}\}.
                         \end{array}
                       \right.\]
Note that $\sigma(x_1)$ is defined as the \emph{complement} of the canonical truth assignment corresponding to $T_1$ -- which was used in the proof of \cref{PIDL-MC-hard}. We are going to continue to define the truth assignment as the complement of the canonical one. The reason for this will become clear when we show the connection between $\phi$ and $\phi'$ in the end.

Now we split the team $T_1$ into $T_2$ and $T_2'$ according to the value of $\sigma(x_2)$ (which is defined by the above assumption following from $\psi\in\qbfcnfgeneral$) as follows:
\[T_2' \dfn \left\{
                         \begin{array}{ll}
                           \{s_2\} & \mbox{if }\sigma(x_2)=\true, \\
                           \{\overline{s_2}\} & \mbox{if }\sigma(x_2)=\false,
                         \end{array}
                       \right.\]
and $T_2=T_1\setminus T_2'$. Clearly, $K,T_2'\models r_2\wedge \dep[p_2]$ and it suffices to check that $K,T_2\models\delta_3$.

We know that to show $K,S\models \delta_1$ it is enough to show that for every $T_1$ (as constructed above) there is a $T_2$ (as constructed above) such that $K,T_2\models \delta_3$.
By repeating the same arguments and constructions $n/2$ times, it remains to show that $K,T_n\models \phi'$.
Now, $T_n$ and $\sigma$ are interdependently defined by
\begin{equation}\label{Ti-xi}
\begin{array}{ll}
  \sigma(x_i)\dfn \left\{\begin{array}{rcl}
      \true  &\text{if}& T_n\cap\{s_i,\overline{s_i}\}=\{\overline{s_i}\}\\
      \false &\text{if}& T_n\cap\{s_i,\overline{s_i}\}=\{{s_i}\}
    \end{array}\right. &
  \text{for }i\in\{1,3,\dots,n-1\},\\
  T_n\cap \{s_i,\overline{s_i}\} \dfn \left\{\begin{array}{rcl}
      \{s_i \}           &\text{if}& \sigma(x_i) = \false\\
      \{\overline{s_i}\} &\text{if}& \sigma(x_i) = \true
    \end{array}\right. &
  \text{for }i\in\{2,4,\dots,n\},
\end{array}
\end{equation}
and $\sigma$ satisfies $\phi$, \ie for all $j\in\{1,\dots,m\}$ it holds that $\sigma(\alpha_{j0})=\true$, $\sigma(\alpha_{j1})=\true$ or $\sigma(\alpha_{j2})=\true$.

Now let $j\in\{1,\dots,m\}$ be arbitrarily chosen, let, for example, $\alpha_{j0}=\neg x_{i_0}$, $\alpha_{j1}=\neg x_{i_1}$, $\alpha_{j2}=x_{i_2}$ (for some $i_0,i_1,i_2\in\{1,\dots,n\}$) and let \wlg $\sigma(\alpha_{j1})=\true$, \ie $\sigma(x_{i_1})=\false$.
Let $T'\subseteq T_n$ be arbitrarily chosen such that $K,T'\models c_j$. Then, by construction of $K$, it holds that $T'\subseteq \{\overline{s_{i_0}},\overline{s_{i_1}},s_{i_2}\}$ and, by \cref{Ti-xi} and $T'\subseteq T_n$, we further obtain that $\overline{s_{i_1}}\notin T'$. Hence, $T'\subseteq \{\overline{s_{i_0}},s_{i_2}\}$, \ie $|T'| \leq 2$, and therefore \[K,T'\models \big(\dep[c_{j0}]\wedge \dep[c_{j1}] \wedge \dep[c_{j2}] \big) \sor \big(\dep[c_{j0}]\wedge \dep[c_{j1}] \wedge \dep[c_{j2}]\big).\]
Since $j$ was chosen arbitrarily it follows that $K,T_n\models \phi'$.

Conversely, suppose $K,S \models \theta$.
By reversing the arguments and constructions from above and again repeating them $n/2$ times, we arrive at the interdependence from \cref{Ti-xi} again. The crucial observation is that when evaluating $(r_{2k-1}\imp \dep[p_{2k-1}])\imp \delta_{2k}$ we only need to consider the maximal teams satisfying $r_{2k-1}\imp \dep[p_{2k-1}]$ and there are exactly two of those, one without $s_{2k-1}$ and the other one without $\overline{s_{2k-1}}$. And when evaluating $(r_{2k}\wedge\dep[p_{2k}])\sor\delta_{2k+1}$ we have to consider only the complements of the maximal teams satisfying $r_{2k}\wedge\dep[p_{2k}]$ and again there are exactly two, one without $s_{2k}$ and the other one without $\overline{s_{2k}}$.

It remains to show that $\sigma$ satisfies $\phi$. That is to show that $\sigma$ satisfies $\alpha_{j_0}\vee\alpha_{j_1}\vee\alpha_{j_2}$ for an arbitrarily chosen $j\in\{1,\dots,m\}$. Suppose, for example,
\[\alpha_{j_0}=x_{i_0},\ \alpha_{j_1}=x_{i_1}\text{ and }\alpha_{j_2}=\neg x_{i_2}.\]
Now let $T'\subseteq T_n$ be the maximal team such that $K,T' \models c_j$. Then, by construction of $K$, we have that $T'\subseteq \{s_{i_0}, s_{i_1},\overline{s_{i_2}}\}$. Since $K,T_n\models\phi'$ it holds that
\[K,T'\models \big(\dep[c_{j0}]\wedge \dep[c_{j1}] \wedge \dep[c_{j2}] \big) \sor \big(\dep[c_{j0}]\wedge \dep[c_{j1}] \wedge \dep[c_{j2}]\big)\]
and thus, by construction of $K$, it follows that $|T'|\leq 2$. Say $s_{i_1}\notin T'$, then, by maximality of $T'$, we obtain that $s_{i_1}\notin T_n$ which, by \cref{Ti-xi}, means that $\sigma(x_{i_1})=\true$, \ie $\sigma(\alpha_{j_1})=\true$. Hence, as $j$ was chosen arbitrarily, it follows that $\sigma$ satisfies $\phi$.
\end{proof}

Finally, we study the model checking problem for sublogics of \MIDL including $\EX$ and $\nor$.
\begin{theorem}\label{midl-wos-mc-lower}
Let $\{\EX, \wedge,\allowbreak  \nor,\allowbreak  \imp\} \subseteq M$. Then $\MIDL[M]\hmc$ is \P\SPACE-com\-plete.
\end{theorem}
\begin{proof}
The upper bound again follows from \cref{MIDL-MC-inPSPACE}. For the lower bound we give a polynomial-time reduction from \qbf to $\MIDL[{\EX, \wedge, \nor,\allowbreak \neg,\allowbreak\imp,\allowbreak\dep[]}]\hmc$, which implies the desired result since
\[\begin{array}{rcl}
\MIDL[{\EX, \wedge, \nor, \neg,\imp,\dep[]}]\hmc &\equivpm& \MIDL[{\EX, \wedge, \nor, \neg,\imp}]\hmc\\
&\equivpm& \MIDL[{\EX, \wedge, \nor, \imp}]\hmc,
\end{array}\]
by \cref{midl equivalences} and \cref{neg-dont-matter}.
 
Let $\psi=\forall x_1\exists x_2 \dots \forall x_{n-1} \exists x_n\, \phi$ with $n$ even and $\phi$ quantifier-free. The corresponding $\MIDL[{\EX, \wedge, \nor, \neg,\imp,\dep[]}]\hmc$ instance is defined as $(K=(S,R,\pi),T,\theta)$ where

\begin{itemize}
\item $S\dfn\bigcup\limits_{1\leq i\leq n}S_i$,\quad $R\dfn\bigcup\limits_{1\leq i\leq n}R_i$ and for $1\leq i\leq n/2$
\[
\begin{array}{rcl}
S_{2i-1}&\dfn&\{s_{2i-1},\overline{s_{2i-1}}\} \\
S_{2i}&\dfn&\{s_{2i},\overline{s_{2i}}\}\cup\{t_{i}\}\cup\{t_{i1},\cdots,t_{i(i-1)}\} \\
R_{2i-1}&\dfn&\{(s_{2i-1},s_{2i-1}),(\overline{s_{2i-1}},\overline{s_{2i-1}})\}\\
R_{2i}&\dfn&\{(t_{i},t_{i1}),(t_{i1},t_{i2}),\cdots,(t_{i(i-2)},t_{i(i-1)})\}\\
&&\cup\{(t_{i(i-1)},s_{2i}),(t_{i(i-1)},\overline{s_{2i}})\}\\
&&\cup\{(s_{2i},s_{2i}),(\overline{s_{2i}},\overline{s_{2i}})\}\\
\end{array}
\]
\item $\pi(s_j) \dfn \{r_{j},p_{j}\}$,
\item $\pi(\overline{s_j}) \dfn \{r_{j}\}$,
\item $\pi(t)  \dfn \emptyset \text{, for }t\notin \{s_j,\overline{s_j}\mid 1\leq j\leq n\}$,\\[1ex]
see \cref{fig:midl-wos} for the construction of $K$,

\item $T \dfn \{s_i,\overline{s_i}\mid 1\leq i\leq n,~i\text{ odd}\}\cup\{t_i\mid 1\leq i\leq n/2\}$;
\item $\theta=\delta_1$, where
\begin{align*}
\delta_{2k-1} \dfn & (r_{2k-1}\imp\dep[p_{2k-1}])\imp \delta_{2k} \quad (1\leq k\leq n/2),\\
\delta_{2k} \dfn & \EX\delta_{2k+1} \quad (1\leq k< n/2),\\
\delta_n      \dfn & \EX\phi^{\imp},
\end{align*}
\begin{align*}
\text{\ie}\ \theta=\ \Big( &\big(r_1\imp\dep[p_1]\big) \imp\EX\\
&\quad\quad\quad\cdots \quad \quad\cdots\imp\EX \\
&\quad\Big(\big(r_{n-1}\imp\dep[p_{n-1}]\big)\imp \EX\phi^{\to} \Big)\Big)\dots\Big)\Big),
\end{align*}
and $\phi^{\imp}$ is generated from $\phi$ by the same inductive translation as in the proof of \cref{PIDL-MC-hard}.
\end{itemize}
\begin{figure}[ht]
\begin{center}
\newcommand{\mynode}[5]{\node[state,label={[label distance=-4]45:{$#4\ifthenelse{\equal{#5}{}}{}{,#5}$}}]    (#1)  at (#2)               {$#3$};}
\begin{tikzpicture}[-stealth', auto, x=2.4em, y=2.7em, every state/.style={minimum size=4mm,inner sep=2pt}]
\mynode{s1}{0,0}{s_1}{r_1}{p_1}
\mynode{s1a}{0,-1}{\overline{s_1}}{r_1}{}
\mynode{t1}{4,-0.5}{t_1}{}{}
\mynode{s3}{0,-2.5}{s_3}{r_3}{p_3}
\mynode{s3a}{0,-3.5}{\overline{s_3}}{r_3}{}
\mynode{t2}{4,-3}{t_2}{}{}
\mynode{t21}{6,-3}{}{}{}

\node (initialdots) at (0,-4.5) {$\vdots$};
\mynode{sm}{0,-6}{s_{n-1}}{r_{n-1}}{p_{n-1}}
\mynode{sma}{0,-7.5}{\overline{s_{n-1}}}{r_{n-1}}{}
\mynode{tn}{4,-6.75}{t_{\frac{n}{2}}}{}{}

\mynode{s2}{6,0}{s_2}{r_2}{p_2}
\mynode{s2a}{6,-1}{\overline{s_2}}{r_2}{}
\mynode{s4}{8,-2.5}{s_4}{r_4}{p_4}
\mynode{s4a}{8,-3.5}{\overline{s_4}}{r_4}{}
\node (seconddots) at (4,-4.5) {$\vdots$};
\mynode{tn1}{6,-6.75}{}{}{}
\node (ndots) at (7,-6.75) {$\cdots$};
\mynode{tnf}{8,-6.75}{}{}{}
\mynode{tnff}{10,-6.75}{}{}{}
\mynode{snp}{12,-6.25}{s_n}{r_n}{p_{n}}
\mynode{snn}{12,-7.25}{\overline{s_n}}{r_n}{}

\path (s1)      edge    [loop right]            (s1);
\path (s1a)      edge    [loop right]            (s1a);
\path (t1)      edge                            (s2);
\path (t1)      edge                            (s2a);
\path (t2)      edge                            (t21);
\path (t21)      edge                            (s4);
\path (t21)      edge                            (s4a);
\path (s3)      edge         [loop right]                    (s3);
\path (s3a)      edge      [loop right]                       (s3a);
\path (s4)      edge         [loop right]                    (s4);
\path (s4a)      edge      [loop right]                       (s4a);
\path (sm)      edge    [loop right]            (sm);
\path (sma)    edge    [loop right]            (sma);
\path (tn)      edge                            (tn1);
\path (tn1)     edge                            (ndots);
\path (ndots)   edge                            (tnf);
\path (tnf)     edge                            (tnff);
\path (tnff)     edge                            (snp);
\path (tnff)    edge                            (snn);
\path (s2)     edge    [loop right]            (s2);
\path (s2a)     edge    [loop right]            (s2a);
\path (snp)     edge    [loop right]            (snp);
\path (snn)     edge    [loop right]            (snn);

\end{tikzpicture}
\caption{Kripke structure in the proof of \cref{midl-wos-mc-lower}}
\label{fig:midl-wos}
\end{center}
\end{figure}
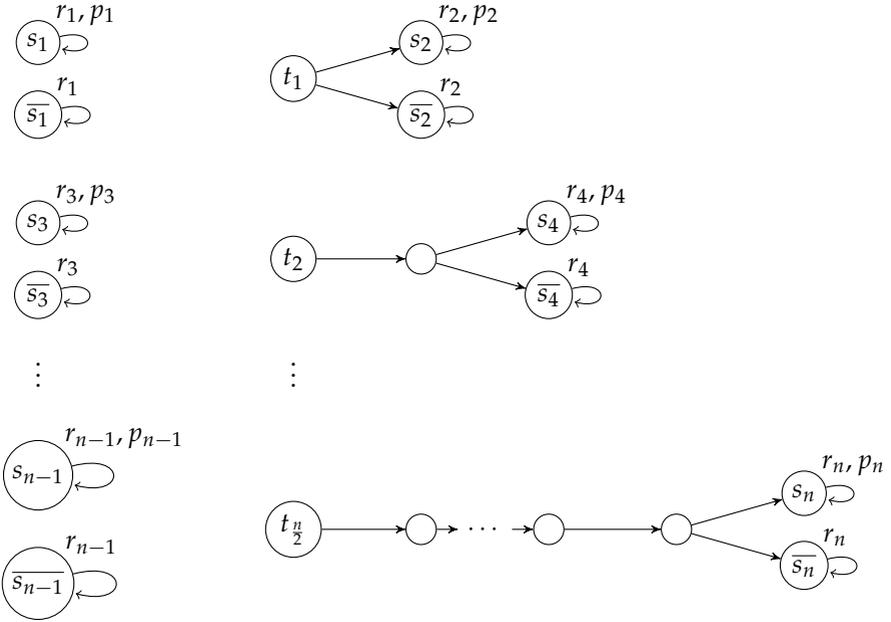

Analogously to the proofs of \cref{pidls-mc-lower} and \cref{PIDL-MC-hard} we will show that $\psi\in \qbf$ iff $K,T \models \theta$. The idea, analogous to the proof of \cref{pidls-mc-lower}, is that the alternating $\imp$ and $\EX$ operators simulate the \qbf formula's $\forall$ and $\exists$ quantifiers, respectively. Therefore we start with the team of all states in the beginning of the chains and then for every nesting level $i$ of $\imp$ we drop one of the states $s_{2i+1}$ and $\overline{s_{2i+1}}$ while for every nesting level $i$ of $\EX$ we simultaneously move forward in the chains and thereby choose one of the states $s_{2i}$ and $\overline{s_{2i}}$. We do this until we arrive at a team that contains exactly one of the states $s_i$ and $\overline{s_i}$ for each $i\in\{1,\dots,n\}$. This team corresponds to a truth assignment $\sigma$ for $x_1,\dots,x_n$ in the canonical way (as in the proof of \cref{PIDL-MC-hard}) and it satisfies $\phi^{\imp}$ iff $\sigma$ satisfies $\phi$.

\smallskip
Now suppose that $\psi\in \qbf$. Then by applying the same arguments as in the proof of \cref{pidls-mc-lower} (but with the canonical instead of the complementary mapping between truth assignments and teams) we get a team $T_1$ and values for $\sigma(x_1)$ and $\sigma(x_2)$. Depending on this we choose $T_2\in \sucteams{T_1}$ as follows:
\[T_2 \dfn \left\{
                         \begin{array}{ll}
                           R(T_1)\setminus\{\overline{s_2}\}            & \mbox{if }\sigma(x_2)=\true, \\
                           R(T_1)\setminus\{s_2\} & \mbox{if }\sigma(x_2)=\false,
                         \end{array}
                       \right.\]

Now it suffices to check that $K,T_2\models \delta_3$. And, again analogous to the proof of \cref{pidls-mc-lower}, by repeating the universal and the existential arguments $n/2$ times, it remains to show that $K,T_n\models \phi^{\imp}$.
And, analogous to \cref{Ti-xi} (except for not taking the complement), $T_n$ and $\sigma$ are interdependently defined by
\begin{equation}\label{Ti-xi-two}
\begin{array}{ll}
  \sigma(x_i)\dfn \left\{\begin{array}{rcl}
      \true  &\text{if}& T_n\cap\{s_i,\overline{s_i}\}=\{{s_i}\}\\
      \false &\text{if}& T_n\cap\{s_i,\overline{s_i}\}=\{\overline{s_i}\}
    \end{array}\right. &
  \text{for }i\in\{1,3,\dots,n-1\},\\
  T_n\cap \{s_i,\overline{s_i}\} \dfn \left\{\begin{array}{rcl}
      \{s_i \}           &\text{if}& \sigma(x_i) = \true\\
      \{\overline{s_i}\} &\text{if}& \sigma(x_i) = \false
    \end{array}\right. &
  \text{for }i\in\{2,4,\dots,n\}.
\end{array}
\end{equation}
Now, by \cref{sigma midl equivalence} in the proof of \cref{PIDL-MC-hard} ($T'$ there is of the same structure as $T_n$ here and $\phi^{\imp}$ does not contain any modalities), it holds that $K,T_n\models \phi^{\imp}$ since, by assumption, $\sigma$ satisfies $\phi$.

Conversely, suppose that $K,T\models \theta$. As in the proof of \cref{pidls-mc-lower} we can reverse the above constructions and arrive at the interdependence from \cref{Ti-xi-two} again. The crucial point is that when evaluating $\EX \delta_{2k+1}$ we only have to consider minimal successor teams.

Now, as above, by \cref*{sigma midl equivalence} in the proof of \cref{PIDL-MC-hard}, it holds that $\sigma$ satisfies $\phi$ since, by assumption and by the construction of $T_n$, $K,T_n\models \phi^{\imp}$.
\end{proof}

\section{Conclusion}

We have shown that model checking for \MIDL in general is \P\SPACE-com\-plete and that this still holds if we forbid $\AX$ and $\dep$ together with either $\EX$ or $\sor$. For \PIDL (where neither $\AX$, $\EX$ nor $\sor$ are allowed) the complexity drops to \co\NP.
All our results are listed in \cref{midl mc results}.

\fragmenttableMIDL{
  *&* & +&+&*&* & + & + & \P\SPACE & \Cref{pidls-mc-lower} \\\hline
  *&* & +&+&+&* & + & * & \P\SPACE & \mbox{\Cref{pidls-mc-lower}}, \mbox{\Cref{midl equivalences}\ref{dep-with-imp}}, \mbox{\Cref{mdl equivalences}\ref{dep with nor zero}} \\\hline
  *&+ & +&*&+&* & + & * & \P\SPACE & \Cref{midl-wos-mc-lower} \\\hline
  *&- & +&-&+&* & + & * & \co\NP   & \Cref{PIDL-MC} \\\hline
  *&* & *&*&-&* & * & - & \P       & \Cref{ML-imp-inP} \\\hline
  *&* & *&*&*&* & - & * & \P\,/\ \NP  & see \Cref{mdl mc results} \\\hline
}{Classification of complexity for fragments of \MIDL\hmc}{midl mc results}


Note that some cases are missing in \cref{midl mc results}, \eg the one where only conjunction is forbidden, the one where only both disjunctions are forbidden and the one from \cref{midl-wos-mc-lower} but with classical disjunction forbidden instead of dependence atoms.
Also, the expressiveness and the satisfiability problem of \MIDL need to be studied.


\chapter{Two-variable Dependence Logic}\label{chap:dtwo}
In this chapter we will assume that all first-order vocabularies are purely relational, \ie they do not contain function or constant symbols. And \wlg we assume that all relation symbols are at most binary.

\section{\texorpdfstring{Comparison of \iftwo and \dtwo}{Comparison of IF\textasciicircum2 and D\textasciicircum2}}\label{comparison}

In this section we show that
\[\dtwo \llneq \iftwo \lleq \df^3,\]
using results from \cref{ifsatsection,sec:satd2} in advance.
We also further discuss the expressive powers and other logical properties of $\dtwo$ and $\iftwo$.

\begin{lemma}\label{dtoif} For any formula $\phi\in \dtwo$
 there is a formula $\phi^*\in \iftwo$ such that for all structures
  $\mA$ and  teams $X$, where $\dom{X}=\{x,y\}$, it holds that  
\[\mA\models _X\phi \quad\text{iff}\quad \mA\models _X\phi^*.\]
\end{lemma}
\vspace*{-1ex}
\begin{proof}
The translation $\phi\mapsto \phi^*$ is defined as follows. For  first-order literals the translation is the identity, and negations of dependence atoms are translated by $\neg x=x$. The remaining cases are defined as follows:
\begin{eqnarray*}
\deppp(x) &\mapsto& \exists y/\{x,y\}(x=y)\\
\deppp(x,y) &\mapsto& \exists x/\{y\}(x=y)\\
\phi \wedge \psi   &\mapsto& \phi^*\! \wedge \psi^*\\
\phi \vee \psi   &\mapsto& \phi^*\! \vee \psi^*\\
\exists x \phi   &\mapsto& \exists x  \phi^* \\
\forall  x \phi   &\mapsto& \forall x  \phi^*
\end{eqnarray*}
 The claim of the lemma can now be proved using induction on $\phi$.
The only non-trivial cases are the dependence atoms. We consider the case
where $\phi$ is of the form $\deppp(x,y)$. 

Let us assume that $\mA\models_X \phi$. Then there is a function $F\colon A\to A$ such that
\begin{equation}\label{dtoif6}
\text{for all }s\in X:\ s(y)=F(s(x)).
\end{equation}
Define now  $F'\colon X \to A$ as follows: 
\begin{equation}\label{dtoif3}
F'(s)\dfn F(s(x)).
\end{equation} 
$F'$ is  $\{y\}$-independent since, if $s(x)=s'(x)$, then
\[F'(s)= F(s(x))
{=}F(s'(x))=F'(s').\]
It remains to show that 
\begin{equation}\label{dtoif2}
 \mA \models_{X(F'/x)}(x=y).
\end{equation}
Let $s\in X(F'/x)$. Then
\begin{equation}\label{dtoif5}
s= s'(F'(s')/x)\text{ for some }s'\in X.
\end{equation}
Now
\[s(x)\stackrel{\eqref{dtoif5}}{=}F'(s')\stackrel{\eqref{dtoif3}}{=}F(s'(x))\stackrel{\eqref{dtoif6}}{=}s'(y)\stackrel{\eqref{dtoif5}}{=}s(y).\]
Therefore, \cref{dtoif2} holds, and hence also
\[ \mA \models_X \exists x/\{y\}(x=y).\]

Suppose then that $\mA\not \models_X \phi$. Then there must be $s,s'\in X$ such that $s(x)=s'(x)$ and $s(y)\neq s'(y)$.
We claim now that 
\begin{equation}\label{dtoif1} 
\mA\not \models_X \exists x/\{y\}(x=y).
\end{equation}
Let $F\colon X\rightarrow A$ be an arbitrary $\{y\}$-independent function. Then, by 
$\{y\}$-independence, $F(s)=F(s')$ and since additionally $s(y)\neq s'(y)$, we have
\[s(F(s)/x)(x)=F(s)\neq s(y)=s(F(s)/x)(y)\]
or
\[s'(F(s')/x)(x)=F(s')\neq s'(y)=s'(F(s')/x)(y).\]
This implies that
\[\mA \not\models_{X(F/x)}(x=y),\]
since $s(F(s)/x),\,s'(F(s')/x) \in X(F/x)$.

Since $F$ was arbitrary, we may conclude that \cref{dtoif1} holds.
\end{proof}

Next we show a translation from $\iftwo$ to $\df^3$.
\begin{lemma}\label{iftod} For any formula $\phi\in \iftwo$
 there is a formula $\phi^*\in \df^3$ such that for all
 structures  $\mA$ and  teams $X$, where $\dom{X}=\{x,y\}$, it holds that  
\[\mA\models _X\phi  \quad\text{iff}\quad  \mA\models _X\phi^*.\]
\end{lemma}
\begin{proof}
The claim follows by the following translation $\phi\mapsto \phi^*$:
For atomic and negated atomic formulas the translation is the identity, and for propositional connectives and first-order quantifiers it is defined in the obvious inductive way. The only non-trivial cases are the slashed quantifiers:
\[\begin{array}{rcl}
\exists x /\{y\}\psi  &\mapsto&  \exists z(x=z\wedge \exists x (\deppp(z,x) \wedge \psi^*)),\\
\exists x /\{x\}\psi  &\mapsto&  \exists x (\deppp(y,x) \wedge \psi^*),\\
\exists x /\{x,y\}\psi  &\mapsto&  \exists x(\deppp(x)\wedge \psi^*).\\
\end{array}\]
Again, the claim can be proved using induction on $\phi$.
We consider the case where $\phi$ is of the form $\exists x /\{y\}\psi$.
Assume $\mA\models _X \phi$. Then there is a $\{y\}$-independent function $F\colon X\rightarrow A$ such that
\begin{equation}\label{iftod0}
\mA\models_{X(F/x)} \psi. 
\end{equation}
By $\{y\}$-independence, $s(x)=s'(x)$ implies that $F(s)=F(s')$ for all $s,s'\in X$. Our goal is to show that 
\begin{equation}\label{iftod1}
\mA\models_X \exists z(x=z\wedge \exists x (\deppp(z,x) \wedge \psi^*)).
\end{equation}
Now, \cref{iftod1} holds if for $G\colon X \to A$ defined by $G(s)=s(x)$ for all $s\in X$ it holds that
\begin{equation}\label{iftod2}
\mA\models_{X(G/z)} \exists x (\deppp(z,x) \wedge \psi^*).
\end{equation}
Define $F'\colon X(G/z)\rightarrow A$ by
$F'(s)=F(s\upharpoonright\{x,y\})$. Now we claim that
\[ \mA\models_{X(G/z)(F'/x)} \deppp(z,x) \wedge \psi^*,\]
implying \cref{iftod2} and hence \cref{iftod1}.

First we show that 
\begin{equation}\label{iftod42}
\mA\models_{X(G/z)(F'/x)} \deppp(z,x).
\end{equation}
At this point it is helpful to note that every $s\in {X(G/z)(F'/x)}$ arises from an $s'\in X$ by first copying the value of $x$ to $z$ and then replacing the value of $x$ by $F(s\upharpoonright\{x,y\})$, \ie that $s(z)=s'(G(s')/z)(z)=G(s')=s'(x)$ and $s(x)=F(s')$.
Now, to show \cref{iftod42}, let $s_1,s_2\in X(G/z)(F'/x)$ with $s_1(z)=s_2(z)$ and let $s'_1,s'_2\in X$ as above, \ie $s_1$ (resp.~$s_2$) arises from $s'_1$ (resp.~$s'_2$). Then it follows that $s'_1(x)=s'_2(x)$. Hence, by $\{y\}$-independence, $F(s'_1)=F(s'_2)$, implying that $s_1(x)=F(s'_1)=F(s'_2)=s_2(x)$ which proves \cref{iftod42}.
Let us then show that  
\begin{equation}\label{iftod3}
 \mA\models_{X(G/z)(F'/x)}\psi^*.
\end{equation}
Note first that by the definition of the mapping $\phi\mapsto \phi^*$ the variable $z$ cannot appear free in $\psi^*$. By \cref{freevar}, the satisfaction of any $\df$ formula $\theta$ only depends on those variables in a team that appear free in $\theta$, therefore \cref{iftod3} holds iff
 \begin{equation}\label{iftod4}
 \mA\models_{X(G/z)(F'/x)\upharpoonright\{x,y\}}\psi^*.
\end{equation}
We have chosen  $G$ and $F'$ in such a way that 
$$X(G/z)(F'/x)\upharpoonright\{x,y\}=X(F/x),$$
hence \cref{iftod4} now follows from \cref{iftod0} and the induction hypothesis.

We omit the proof of the converse implication which is analogous.
\end{proof}

For sentences, \cref{dtoif,iftod} now imply the following.

\begin{theorem}\label{dtwo le if2}
$\dtwo\lleq \iftwo\lleq \df^3$
\end{theorem}
\begin{proof} 
The claim follows by \cref{dtoif,iftod}.
 First of all, if $\phi$ is a sentence of $\ifl$ or $\df$, 
then, by \cref{freevar}, for every model $\mA$ and team $X\neq \emptyset$ 
\begin{equation}\label{sentences}
\mA \models_X \phi\text{ iff } \mA \models_{\{\emptyset\}} \phi.
\end{equation}
It is important to note that, even if $\phi\in \dtwo$ is a sentence, it
 may happen that $\phi^*$ has free variables since variables
 in $W$ are regarded as free in subformulas of $\phi^*$ of the form
 $\exists x /W\psi$. However, this is not a problem. Let $Y$ be the set of all assigments of $\mA$ with the domain $\{x,y\}$. Now
\[\begin{array}{rcl}
 \mA\models_{\{\emptyset\}}\phi &\text{iff}&  \mA\models_Y\phi\\
 &\text{iff}& \mA\models_Y \forall x\forall y \phi \\
 &\text{iff}&  \mA\models_Y\forall x\forall y \phi^*\\
 &\text{iff}& \mA\models_{\{\emptyset\}}\forall x\forall y \phi^*,
\end{array}\]
where the first and the last equivalence hold by \cref{sentences}, the second by the semantics of the universal quantifier and the third by \cref{dtoif}. An analogous argument can be used to show that for every sentence $\phi\in \iftwo$ there is an equivalent sentence of the logic $\df^3$.
%
%
\end{proof}

\subsection{\texorpdfstring{Examples of properties definable in \dtwo}{Examples of properties definable in D\textasciicircum2}}
We now give some examples of definable classes of structures in \dtwo (and in \iftwo by \cref{dtwo le if2}).

\begin{proposition}\label{thm:dtwo expressibles}
The following properties can be expressed in \dtwo.
\begin{enumerate}
\item\label{example1} For unary relation symbols $P$ and $Q$, \dtwo can express $|P|\leq|Q|$. This shows $\dtwo \not \lleq \fo$.
\item\label{example2} If the vocabulary of $\mA$ contains a constant $c$, then \dtwo can express that $A$ is infinite.
\item\label{example3} $|A|\le k$ can be expressed already in $\df^1$.
\end{enumerate}
\end{proposition}
\begin{proof}
Let us first consider part \ref{example1}). 
Define $\phi$ by
\[\phi\dfn \forall x\exists y( \deppp(y,x)\wedge (\neg P(x)\vee Q(y))).\]
Now,  $\begin{array}[t]{r@{\;\;}c@{\;\;}l}\mA\models \phi &\text{iff}& \text{there is an injective function }F\colon A\rightarrow A\text{ with }F[P^{\mA}]\subseteq Q^{\mA}\\
                                            &\text{iff}& |P^{\mA}|\leq |Q^{\mA}|.
       \end{array}$

For part \ref{example2}), we use the same idea as above. Define $\psi$ by
\[\psi\dfn \forall x\exists y(\deppp(y,x)\wedge \neg c=y).\]
Now, $\begin{array}[t]{r@{\;\;}c@{\;\;}l}\mA\models\psi &\text{iff}& \text{there is an injective function }F\colon A\rightarrow A\text{ with }c^{\mA}\notin F[A]\\
                                                        &\text{iff}& \text{$A$ is infinite.}
      \end{array}$

Finally, we show how to express the property from part \ref{example3}). Define $\theta$ as
\[\forall x( \bigvee_{1\le i\le k}\chi_i),    \]
where $\chi_i$ is $\deppp(x)$. It is now immediate that $\mA\models \theta$ iff $|A|\le k$.
\end{proof}

It is interesting to note that part \ref{example1}) implies that \dtwo does not have a zero-one law, since the property $|P|\leq |Q|$ 
has the limit probability $\frac{1}{2}$.


\begin{proposition}\label{d less than if} 
$\dtwo \llneq \iftwo$. This holds already in the finite.
\end{proposition}
\begin{proof}
Almost all \iftwo formulas used in \cref{ifsatsection} to prove the undecidability of the (finite) satisfiability problem of \iftwo are in fact \fotwo formulas. The only exception is the formula $\phi_\mathrm{join}$ (which is part of the formula $\phi_\mathrm{grid}$, \cf \cref{def:gridlike}).
Therefore, if $\phi_\mathrm{join}$ could be expressed in \dtwo then (finite) satisfiability would be as undecidable for \dtwo as it is for \iftwo. But this is a contradiction to \cref{dtwo nexptime} and, hence, the \iftwo formula $\phi_\mathrm{join}$ is not definable in \dtwo.
\end{proof}




\section{\texorpdfstring{Satisfiability is undecidable for \iftwo}{Satisfiability is undecidable for IF\textasciicircum2}}\label{ifsatsection}
In this section we will use tiling problems, introduced by Wang \cite{Wang:1961}, to show the undecidability of \satt[\iftwo] as well as \finsat[\iftwo].
%

Here, a \emph{Wang tile} is a square in which each edge (top, right, bottom, left) is assigned a color.
We say that a set of tiles can tile the $\N\times\N$ plane if a tile can be placed on each point $(i,j)\in \N\times\N$ such that the right color of the tile in $(i,j)$ is the same as the left color of the tile in $(i+1,j)$ and the top color of the tile in $(i,j)$ is the same as the bottom color of the tile in $(i,j+1)$.
Notice that turning and flipping tiles is not allowed.

We then define some specific structures needed later.

\begin{definition}\label{def:grid}
Let $m,n \in \N\cup\{\infty\}$.
Then the model $\grid{m}{n} \dfn(G,V,H)$ where
\begin{itemize}
 \item $G=\{0,\dots,m\}\times \{0,\dots,n\}$,
 \item $H=\{((i,j),(i+1,j))\in G\times G\mid i< m,j\leq n\}$ and
 \item $V=\{((i,j),(i,j+1))\in G\times G\mid i\leq m,j< n\}$
\end{itemize}
is called the ($m\times n$)-\cim{grid}.
Instead of $\grid{\infty}{\infty}$ we sometimes just write $\grid{}{}$ and call it the \emph{infinite grid}\index{grid!infinite}.
The set $\{\grid{m}{n}\mid m,n\in \N\}$ of all \emph{finite grids}\index{grid!finite} is denoted by $\fingrid$\index{FinGrid@$\fingrid$}.
\end{definition}

\begin{definition}\label{def:tiling}
A set of \cim{colors} $C$ is defined to be an arbitrary finite set. The set of all \emph{(Wang) tiles}\index{tile} over $C$ is $C^4$, \ie a tile is an ordered list of four colors, interpreted as the colors of the four edges of the tile in the order top, right, bottom and left. A tile $t=(c_0, c_1, c_2, c_3)$ will usually be written in the form
\[\tile{c_0}{c_3}{c_1}{c_2}.\]
And we will often refer to its single colors as $t_\text{top}$, $t_\text{right}$, $t_\text{bottom}$ and $t_\text{left}$.

Let $C$ be a set of colors, $T\subseteq C^4$ a finite set of tiles and $\mA=(A,V,H)$ a first-order structure with binary relations $V$ and $H$ interpreted as vertical and horizontal successor relations. Then a $T$-\cim{tiling} of $\mA$ is a total function $t\colon A \to T$ such that for all $x,y\in A$ it holds that
\begin{enumerate}
 \item $t(x)_\text{top}=t(y)_\text{bottom}$ if $(x,y)\in V$, \ie the top color of $x$ matches the bottom color of $y$, and
 \item $t(x)_\text{right}=t(y)_\text{left}$ if $(x,y)\in H$, \ie the right color of $x$ matches the left color of $y$.
\end{enumerate}

A \emph{periodic} $T$-tiling\index{tiling!periodic} of $\mA$ (with period $m \times n$) is a $T$-tiling $t$ of $\mA$ such that there are $m,n\geq 1$ with
\[t(x)_\text{left} = t(y)_\text{left}\text{ for all }x,y\in A\text{ with }(x,y) \in H^m\]
and
\[\text{$t(x)_\text{bottom} = t(y)_\text{bottom}$ for all $x,y\in A$ with $(x,y) \in V^n$,}\]
\ie there is a rectangle of tiles of size $m\times n$ which can be concatenated arbitrarily often in both directions (if $V$ and $H$ are total relations in $\mA$).

Now additionally let $c\in C$ be a color. Then a \emph{(bordered)} $(T,c)$-\emph{tiling}\index{tiling!bordered} of $\mA$ is a $T$-tiling $t$ of $\mA$ such that for all $x\in A$ it holds that
\[\begin{array}{lcl}
t(x)_\text{left} = c &\text{if}& \text{there is no $y\in A$ with $(y,x)\in H$},\\
t(x)_\text{bottom} = c &\text{if}& \text{there is no $y\in A$ with $(y,x)\in V$},\\
t(x)_\text{right} = c &\text{if}& \text{there is no $y\in A$ with $(x,y)\in H$}\text{ and}\\
t(x)_\text{top} = c &\text{if}& \text{there is no $y\in A$ with $(x,y)\in V$},
\end{array}\]
\ie the borders of $\mA$ (so far as they exist) are all $c$-colored.
\end{definition}

Next we define tiling problems for classes of structures.
\begin{definition}
A structure $\mA=(A,V,H)$ is called (periodically) $T$-\emph{tilable} (resp.~$(T,c)$-tilable) iff there is a (periodic) $T$-tiling (resp.~bordered $(T,c)$-tiling) of $\mA$.

For any class $\calC$ of structures over the vocabulary $\{V,H\}$ we define the problems
\[\tiling[\calC] \dfn \{T\mid \text{there is an $\mA\in \calC$ such that $\mA$ is $T$-tilable}\},\]
\begin{multline*}
\ptiling[\calC] \dfn \{T\mid \text{there is an $\mA\in \calC$ such that}\\
    \text{$\mA$ is periodically $T$-tilable}\}
\end{multline*}
and
\begin{multline*}
\btiling[\calC] \dfn \{(T,c)\mid \text{there is an $\mA\in \calC$ such that}\\
\text{$\mA$ is $(T,c)$-tilable}\}.
\end{multline*}

We usually write $\tiling[\mA]$ instead of $\tiling[\{\mA\}]$ -- and analogous for the other two problems.
\end{definition}

Later we will use the following theorems to show the undecidability of \satt[\iftwo] and \finsat[\iftwo].
\begin{theorem}[\cite{be66}, \cite{Harel:1986}]\label{tiling pizeroone}
\tiling[\mG] is $\pizeroone$-complete.
\end{theorem}

\begin{theorem}[\cite{guko72}]\label{periodic tiling sigmazeroone}
\ptiling[\mG] is $\sigmazeroone$-complete.
\end{theorem}

\begin{theorem}\label{bordered tiling sigmazeroone}
$\btiling[\fingrid]$ is $\sigmazeroone$-complete.\footnote{In \cite{em96} a very similar problem is proven to be $\sigmazeroone$-hard. Their technique, however, is completely different from ours.}
\end{theorem}
\begin{namedproof}{\cref{bordered tiling sigmazeroone}}
For the upper bound note that all possible solutions are finite and can be recursively enumerated and checked.

For the lower bound we give a reduction from $\ptiling[\mG]$.
For this purpose let $C$ be a set of colors and let $T$ be an arbitrary set of tiles over $C$.
Then we define a new set of colors $C'\dfn (C\times C) \cup \{\bl,\wh,\gr\}$ where \bl, \wh and \gr are new colors not included in $C$.
And we define a new set of tiles
\[\begin{array}{rccl}
T' & \dfn &      & \left\{\tile{\gr}{\bl}{\gr}{\bl}, \tile{\wh}{\wh}{\bl}{\bl}, \tile{\bl}{\bl}{\wh}{\wh}, \tile{\bl}{\wh}{\bl}{\wh}\right\}\\
&&&  \qquad\text{\small\emph{corner tiles}}\\[2ex]
   &      & \cup & \left\{\begin{array}{@{}l@{}}\tile{(c,c)}{\gr}{\wh}{\bl}, \tile{(c,c)}{\wh}{\wh}{\bl}, \tile{\wh}{\bl}{(c,c)}{\gr}, \tile{\wh}{\bl}{(c,c)}{\wh},\\[3ex]
                                                 \left.\tile{\bl}{\wh}{\wh}{(c,c)}, \tile{\wh}{(c,c)}{\bl}{\wh}\;\;\middle|\  c\in C  \right.\end{array} \right\}\\
&&&  \qquad\text{\small\emph{edge tiles}}\\[2ex]
   &      & \cup & \set{\tile{(t_\text{top},c)}{(t_\text{left},d)}{(t_\text{right},d)}{(t_\text{bottom},c)}\ }{$t\in T$, $c,d\in C$}.\\
&&&  \qquad\text{\small\emph{inner tiles}}\\[2ex]
\end{array}\]

We will now prove that there is a periodic $T$-tiling of the infinite grid iff there is a bordered $(T',\bl)$-tiling of a finite grid. Thus, the construction of $T'$ from $T$ is a reduction from $\ptiling[\mG]$ to $\btiling[\fingrid]$ and therefore, by \cref{periodic tiling sigmazeroone}, $\btiling[\fingrid]$ is $\sigmazeroone$-hard.

Now, first let there be a periodic $T$-tiling of $\mG$ with period $m\times n$ which is composed of repetitions of the rectangle
{\small
\[\begin{array}{c@{\;\;}c@{\;\;}c@{\;\;}c}
\tile{a}{s}{t}{r} & \tile{e}{t}{v}{u} & \cdots & \tile{h}{x}{s}{w}\\[4ex]
\vdots & \vdots & \iddots &\vdots\\[2ex]
\tile{m}{k}{l}{d} & \tile{o}{l}{n}{g} & \cdots & \tile{q}{p}{k}{j}\\[4ex]
\tile{d}{b}{c}{a} & \tile{g}{c}{f}{e} & \cdots & \tile{j}{i}{b}{h}\ ,
\end{array}\]}
with $a,b,c,\dots \in C$.

Then the following is a bordered $(T',\bl)$-tiling of $\grid{(m+2)}{(n+2)}$:
{\small
\[\begin{array}{@{}c@{\;}c@{\;}c@{\,}c@{\,}c@{\;}c@{}}
\tile{\bl}{\bl}{\wh}{\wh}&\tile{\bl}{\wh}{\wh}{(a,a)} & \tile{\bl}{\wh}{\wh}{(e,e)} & \cdots & \tile{\bl}{\wh}{\wh}{(h,h)}&\tile{\bl}{\wh}{\bl}{\wh}\\[4ex]
\tile{\wh}{\bl}{(s,s)}{\wh}&\tile{(a,a)}{(s,s)}{(t,s)}{(r,a)} & \tile{(e,e)}{(t,s)}{(v,s)}{(u,e)} & \cdots & \tile{(h,h)}{(x,s)}{(s,s)}{(w,h)}&\tile{\wh}{(s,s)}{\bl}{\wh}\\[4ex]
\vdots&\vdots & \vdots & \iddots &\vdots&\vdots\\[2ex]
\tile{\wh}{\bl}{(k,k)}{\wh}&\tile{(m,a)}{(k,k)}{(l,k)}{(d,a)} & \tile{(o,e)}{(l,k)}{(n,k)}{(g,e)} & \cdots & \tile{(q,h)}{(p,k)}{(k,k)}{(j,h)}&\tile{\wh}{(k,k)}{\bl}{\wh}\\[4ex]
\tile{\wh}{\bl}{(b,b)}{\gr}&\tile{(d,a)}{(b,b)}{(c,b)}{(a,a)} & \tile{(g,e)}{(c,b)}{(f,b)}{(e,e)} & \cdots & \tile{(j,h)}{(i,b)}{(b,b)}{(h,h)}&\tile{\wh}{(b,b)}{\bl}{\wh}\\[4ex]
\tile{\gr}{\bl}{\gr}{\bl}&\tile{(a,a)}{\gr}{\wh}{\bl} & \tile{(e,e)}{\wh}{\wh}{\bl} & \cdots & \tile{(h,h)}{\wh}{\wh}{\bl}&\tile{\wh}{\wh}{\bl}{\bl}
\end{array}\]}

\smallskip
For the reverse direction let there be a bordered $(T',\bl)$-tiling of $\grid{m}{n}$.
Because the corner and edge tiles of $T'$ are the only tiles with the color $\bl$ the tiling is of the form
{\small
\[\begin{array}{c@{\;\;}c@{\;\;}c@{\;\;}c}
\tile{\bl}{\bl}{\wh}{\wh}&\tile{\bl}{\wh}{\wh}{(f,f)} & \cdots &\tile{\bl}{\wh}{\bl}{\wh}\\[4ex]
\vdots&\vdots &  \iddots &\vdots\\[2ex]
\tile{\wh}{\bl}{(b,b)}{\gr}&\tile{(d,a)}{(b,b)}{(c,b)}{(a,a)} & \cdots & \tile{\wh}{(e,e)}{\bl}{\wh}\\[4ex]
\tile{\gr}{\bl}{\gr}{\bl}&\tile{(a,a)}{\gr}{\wh}{\bl} &  \cdots & \tile{\wh}{\wh}{\bl}{\bl}\ ,
\end{array}\]}
with $a,b,c\in C$.
Since the bottom left corner tile uses the color $\gr$ instead of $\wh$ we know that $m,n\geq 3$, \ie there is at least one inner tile in the tiling.
And due to the definition of the first components of the colors of the inner tiles of $T'$ it follows that the first components in the inner tiles in the above tiling form a partial $T$-tiling of $\mG$ with size $(m-2)\times (n-2)$.
Now, because of the definition of the second components in the inner tiles of $T'$ it follows that $a=f$, $b=e$ etc., \ie the partial $T$-tiling is the defining rectangle of a periodic $T$-tiling of $\mG$ which therefore exists.
\end{namedproof}

To prove the undecidability of \satt[\iftwo] (\Cref{iftwo sat complexity}) and \finsat[\iftwo] (\Cref{iftwo finsat complexity}) we will, for every set of tiles $T$, define a formula $\phi_T$ such that $\mA$ has a $T$-tiling iff there is an expansion $\mA^*$ of $\mA$ with $\mA^*\models \phi_T$. Then we will define another formula $\phi_\mathrm{grid}$ and show that $\mA\models \phi_\mathrm{grid}$ iff $\mA$ contains a substructure isomorphic to a grid.
Therefore $\phi_T\wedge \phi_\mathrm{grid}$ is satisfiable if and only if there is a $T$-tiling of a grid.
For $\iftwo\hsat$ we will additionally ensure that this grid is the infinite grid and for $\iftwo\hfinsat$ (\Cref{iffinsatsection}) we will extend $\phi_T$ by a formula $\phi_{c,T}$ ensuring that $\mA^*$ has a bordered $(T,c)$-tiling.

\begin{definition}\label{def:tiling formula}
Let $T=\{t^0,\dots,t^k\}$ be a set of tiles and
for all $i\leq k$ let $\cmd{right}(t^i)$ (resp.~$\cmd{top}(t^i)$) be the set
\[\{t^j\in\{t^0,\dots,t^k\}\mid t^i_\text{right} = t^j_\text{left}\text{ (resp.~}t^i_\text{top} = t^j_\text{bottom})\},\]
\ie the set of tiles matching $t^i$ to the right (resp.~top).

Then we define the first-order formulas
\[\begin{array}{r@{}l}
\psi_T\dfn\forall x \forall y \bigg(\Big(&H(x,y)\rightarrow \bigwedge\limits_{i\leq k}\big(P_i(x)\rightarrow \bigvee\limits_{t^j\,\in\, \cmd{right}(t^i)} P_j(y)\big)\Big)\ \wedge\\
\Big(&V(x,y)\rightarrow \bigwedge\limits_{i\leq k}\big(P_i(x)\rightarrow \bigvee\limits_{t^j\,\in\, \cmd{top}(t^i)} P_j(y)\big)\Big)\bigg),\\[3ex]
\multicolumn{2}{l}{\theta_T\dfn\forall x \bigvee\limits_{i\leq k}\big(P_i(x)\wedge \bigwedge\limits_{\substack{j\leq k\\j\neq i}} \neg P_j(x)\big)\text{ and}}\\
\multicolumn{2}{l}{\phi_T\dfn\psi_T \wedge \theta_T}
\end{array}\]
over the vocabulary $V,H,P_0,\dots,P_k$.
Here, $\phi \to \psi$ is an abbreviation of $\dual{\phi} \vee \psi$.
\end{definition}

Note that our semantics for the implication operator leads to the same semantics as the usual first-order definition of $\phi \to \psi$ being an abbreviation for $\neg \phi \vee \psi$ because in the first-order case $\dual{\phi}$ is nothing else than the negation normal form of $\neg \phi$ (\cf \cref{dual modal logic formulas}).
However, our definition uses negation only directly in front of atomic formulas and is thus compatible with our definitions of formulas to always be in negation normal form (\cf \cref{sec:predicate logic,sec:df logic}).

With the previously defined formulas we can now express the $T$-tilability of a given structure.
\begin{lemma}\label{tiling iff tiling formula}
Let $T=\{t_0,\dots,t_k\}$ be a set of tiles and $\mA=(A,V,H)$ a structure. Then $\mA$ is $T$-tilable iff there is an expansion $\mA^*=(A,V,H,P_0,\dots,P_k)$ of $\mA$ such that $\mA^*\models\phi_T$.
\end{lemma}

Notice that $\phi_T$ is an $\fotwo$-sentence. Therefore $T$-tiling is expressible even in $\fotwo$.
The difficulty lies in expressing that a structure is (or at least contains) a grid. This is the part of the construction where \fotwo or even \dtwo formulas are no longer sufficient and the full expressivity of \iftwo is needed.

\begin{definition}\label{def:gridlike}
A structure $\mA=(A,V,H)$ is called \cim{grid-like} iff
it satisfies the conjunction $\phi_\mathrm{grid}$ of the formulas
\[\begin{array}{@{}l@{\ }c@{\ }l@{}}
\phi_{\mathrm{SWroot}}           & \dfn    &\exists x \Big(\forall y \big(\neg V(y,x)\wedge \neg H(y,x)\big)\Big),\\
\phi_{\mathrm{functional}}      & \dfn    & \forall x\forall y \big(R(x,y)\,\to\, \exists x/\{y\}\,y=x \big)\\
                                &       & \quad\text{for $R\in\{V,H\}$},\\
\phi_{\mathrm{injective}}       & \dfn    & \forall x\forall y \big(R(x,y)\, \to\, \exists y/\{x\} \,x=y \big)\\
                                &       & \quad\text{for $R\in\{V,H\}$},\\
\phi_{\mathrm{distinct}}        & \dfn    & \forall x\forall y\, \neg \big(V(x,y)\wedge H(x,y)\big),\\
\phi_{\mathrm{SWedges}}        & \dfn    & \forall x \Big(\big(\forall y \,\neg R(y,x)\big)\to\forall y \big((R'(x,y)\vee R'(y,x)) \to \forall x\,\neg R(x,y)\big)\Big)\\
                                &       & \quad\text{for $(R,R')\in\{(V,H),(H,V)\}$,}\\
\phi_{\mathrm{NEedges}}        & \dfn    & \forall x \Big(\big(\forall y \,\neg R(x,y)\big)\to\forall y \big((R'(x,y)\vee R'(y,x))\to \forall x\,\neg R(y,x)\big)\Big)\\
                                &       & \quad\text{for $(R,R')\in\{(V,H),(H,V)\}$,}\\
\phi_{\mathrm{join}}          & \dfn    & \forall x \Big(\big(\exists y V(x,y)\wedge \exists y H(x,y)\big)\to \forall y \Big(\big(V(x,y)\vee H(x,y)\big)\\
                                &       & \quad\to \exists x/\{y\}\, \big(V(y,x)\vee H(y,x)\big)\Big)\Big).
\end{array}\]
\end{definition}

The grid-likeness of a structure can alternatively be described in the following more intuitive way.
\begin{remark}\label{gridlike description}
A structure $\mA=(A,V,H)$ is grid-like iff
\begin{enumerate}
\item\label{degree} $V$ and $H$ are (graphs of) injective partial functions, \ie the in- and out-degree of every element is at most one ($\phi_{\mathrm{functional}}$ and $\phi_{\mathrm{injective}}$),
\item\label{sw existence} there exists a point $SW$ that does not have any predecessors ($\phi_{\mathrm{SWroot}}$),
\item for every element its $V$-successor is distinct from its $H$-successor ($\phi_{\mathrm{distinct}}$),
\item for every element $x$ such that $x$ does not have a $V$- (resp.~$H$-) predecessor, the $H$- (resp.~$V$-) successor and -predecessor of $x$ also do not have a $V$- (resp.~$H$-) predecessor ($\phi_{\mathrm{SWedges}}$),
\item for every element $x$ such that $x$ does not have a $V$- (resp.~$H$-) successor, the $H$- (resp.~$V$-) successor and -predecessor of $x$ also do not have a $V$- (resp.~$H$-) successor ($\phi_{\mathrm{NEedges}}$),
\item\label{join description} for every element $x$ that has a $V$-successor and an $H$-successor there is an element $y$ such that $(x,y) \in (V\circ H)\cap (H\circ V)$ or $(x,y) \in (V\circ V)\cap (H\circ H)$.
\end{enumerate}
\end{remark}
\begin{proof}
The only difficulty is the connection between property \ref{join description}) and formula $\phi_\mathrm{join}$. First note that $\phi_\mathrm{join}$ is equivalent to the first-order formula
\begin{multline*}
\forall x \Big(\big(\exists z V(x,z)\wedge \exists z H(x,z)\big)\to \\
\exists x' \forall y\Big(\big(V(x,y)\vee H(x,y)\big)\to  \big(V(y,x')\vee H(y,x')\big)\Big)\Big).
\end{multline*}
Since $\phi_{\mathrm{functional}}$ and $\phi_{\mathrm{distinct}}$ hold as well, this implies
\begin{multline*}
\forall x \Big(\big(\exists z V(x,z)\wedge \exists z H(x,z)\big)\to \exists x' \exists y_1 \exists y_2 \Big(y_1\neq y_2 \wedge V(x,y_1)\wedge H(x,y_2) \\
\wedge \big(V(y_1,x')\vee H(y_1,x')\big) \wedge \big(V(y_2,x')\vee H(y_2,x')\big)\Big)\Big).
\end{multline*}
Due to $\phi_{\mathrm{injective}}$ neither $V(y_1,x') \wedge V(y_2,x')$ nor $H(y_1,x') \wedge H(y_2,x')$ can be true if $y_1\neq y_2$. Hence, from the above it follows that
\begin{multline*}
\forall x \Big(\big(\exists z V(x,z)\wedge \exists z H(x,z)\big)\to \exists x' \exists y_1 \exists y_2 \Big(y_1\neq y_2 \\
\wedge \Big(\big(V(x,y_1)\wedge H(x,y_2) \wedge V(y_1,x') \wedge H(y_2,x')\big) \\
\vee \big(V(x,y_1)\wedge H(x,y_2) \wedge H(y_1,x') \wedge V(y_2,x')\big)\Big)\Big)\Big).
\end{multline*}
From this formula the property \ref{join description}) is immediate (with $x\dfn x$ and $y\dfn x'$). The reverse direction can be shown similarly but it will not be needed anyway.
\end{proof}

Now we will use \cref{gridlike description} to show that a grid-like structure, although it does not need to be a grid itself, must at least contain a component that is  isomorphic to a grid.
Here, by \cim{component} we mean a maximal weakly connected substructure, \ie $\mB$ is a component of $\mA=(A,V,H)$ iff $\mB$ is an induced substructure of $\mA$ such that between any two elements from  $B$ there is a path along $R\dfn V\cup H\cup V^{-1}\cup H^{-1}$ and $B$ is closed under following edges from $R$.
\begin{theorem}\label{gridlike includes grid}
Let $\mA=(A,V,H)$ be a grid-like structure. Then $\mA$ contains a component that is isomorphic to a grid.
\end{theorem}
\begin{proof}
%
Due to $\phi_\mathrm{SWroot}$ there exists a point $SW\in A$ that has a $V$- and an $H$-successor but no $V$- or $H$-predecessor.
Now let $n$ 
be the smallest natural number such that for all $x\in A$ it holds that $(SW,x)\notin V^{n+1}$.
If such a number does not exist let $n\dfn \infty$. Since, by \cref{gridlike description}\ref{degree}, $V$ is an injective partial function and $SW$ does not have a $V$-predecessor the latter case can only occur if $A$ is infinite.
Analogously define $m\in\N\cup\{\infty\}$ as the smallest number such that for all $x\in A$ it holds that $(SW,x)\notin H^{n+1}$ or as $\infty$.

We will show that $\mA$ has a component that is isomorphic to $\grid{m}{n}$.
For this purpose we first prove by induction on $1\leq k\leq n$ that $\mA$ has an in-degree complete substructure isomorphic to $\grid{m}{n}$ -- with $(0,0)$ mapped to $SW$.
Here, we call a substructure $\mH$ of $\mA$ \emph{in-degree complete} if every point in $\mH$ has the same in-degrees in $\mH$ as it has in $\mA$.

By selection of $m$ the point $SW$ has an $H^i$-successor $u_i$ for each $0 \leq i\leq m$ (whenever we write $i\leq m$ we naturally mean that the case $i=m$ is only included if $m<\infty$). Since $H$ is an injective partial function and $SW$ has no $H$-predecessor the points $u_i$ are all distinct and unique. Due to $\phi_\mathrm{SWedges}$ none of the points $u_i$ has a $V$-predecessor and therefore the $V$-successors of the points $u_i$ are not in the set $\{u_i\mid i\leq m\}$. Therefore $\mA\restricted{u_i\mid i\leq m}$ is isomorphic to $\grid{m}{1}$.
Due to $\phi_\mathrm{SWedges}$, $\phi_\mathrm{SWroot}$ and $\phi_\mathrm{injective}$ the structure $\mA\restricted{u_i\mid i\leq m}$ is in-degree complete.

For the induction step assume that there is an in-degree complete substructure $\mB$ of $\mA$ which is isomorphic to $\grid{m}{k}$ for a $k<n$ such that $(0,0)$ is mapped to $SW$. Let $h$ be the isomorphism from $\grid{m}{k}$ to $\mB$. We will now extend $h$ to $h'$ such that $h'$ is an isomorphism from $\grid{m}{(k+1)}$ to a substructure of $\mA$. Since $k+1\leq n$ there exists a point $a_0\in A$ such that $a_0$ is the $V$-successor of $h((0,k))$.
Due to $\phi_\mathrm{NEedges}$ and since $h((0,k))$ has a $V$-successor, each of the points $h((i,k))$, $i\leq m$, has a $V$-successor $a_i$. Since $V$ is a partial injective function and the points $h((i,k))$ are all distinct, the points $a_i$ are also all distinct. And since the structure $\mB$ is in-degree complete, none of the points $a_i$ nor any of their $(V\cup H)$-successors is in $B$.

Next, we will show that $a_{i+1}$ is the $H$-successor of $a_i$ for all $i<m$. For $i\leq m-2$, the point $h((i,k))$ has an $H^2$- but no $V^2$-successor in $\mB$. Hence, for all $i\leq m-2$, if the $V^2$-successor of $h((i,k))$ exists in $\mA$, it cannot be the same as the $H^2$-successor of $h((i,k))$. Also notice that each of the points $h((i,k))$, $i\leq m-2$, has a $V$- and an $H$-successor in $\mA$. Therefore due to \cref{gridlike description}\ref{join description} the $(V\circ H)$-successor and the $(H\circ V)$-successor of the point $h((i,k))$, $i\leq m-2$, are the same. Therefore the $H$-successor of $a_i$ is $a_{i+1}$ for all $i\leq m-2$.

In the case $m< \infty$ it still needs to be shown that $a_m$ is the $H$-successor of $a_{m-1}$.
For this purpose note that the point $h((m-1,k))$ has no $H^2$-successor in $\mA$ since $h((m,k))$ does not have an $H$-successor (due to $\phi_\mathrm{NEedges}$ and the selection of $m$). Hence, there cannot be a point $a$ in $\mA$ such that it is both an $H^2$- and a $V^2$-successor of $h((m-1,k))$. Now due to \cref{gridlike description}\ref{join description} and the fact that $h((m-1,k))$ has a $V$- and an $H$-successor in $\mA$, the $(H\circ V)$- and the $(V\circ H)$-successor of $h((m-1,k))$ have to be the same. Therefore $a_m$ is the $H$-successor of $a_{m-1}$.

We define $h'\dfn h\cup \{(i,k+1) \mapsto a_i \mid i\leq m\}$. Each point $a_i$ (with the exception of $a_0$ which has no $H$-predecessor at all due to $\phi_\mathrm{SWedges}$) has $a_{i-1}$ as $H$-predecessor. Hence, due to the in-degree completeness of $\mB$, injectivity of $V$ and $H$ and since each of the points $a_i$ has a $V$-predecessor in $B$, we conclude that the structure $\mA\restricted{}\rngg[h']$ is an in-degree complete substructure of $\mA$. We also notice that due to injectivity, the points $a_i$ have no reflexive loops. Due to the in-degree completeness of $\mB$ none of the $(V\cup H)$-successors of the points $a_i$, $i\leq m$, are in $B$. Hence it is straightforward to observe that $h'$ is the desired isomorphism from $\grid{m}{(k+1)}$ to an in-degree complete substructure of $\mA$.

We have now proven that $\mA$ has an in-degree complete substructure $\mB$ such that there is an isomorphism $h$ from $\grid{m}{n}$ to $\mB$ with $h((0,0))=SW$.
If $n< \infty$ then, by definition of $n$, the point $h((0,n))$ has no $V$-successor. Therefore due to $\phi_\mathrm{NEedges}$ none of the points $h((i,n))$, $i\leq m$, has a $V$-successor in $\mA$. If on the other hand $n=\infty$ then, again by definition of $n$ and $\phi_\mathrm{NEedges}$, each of the infinite many points $h((i,j))$, $i\leq m$, $j\in\N$, has a $V$-successor in $\mB$.
These facts together with functionality and injectivity of $V$ and the in-degree completenes of $\mB$ implies that in both cases $\mB$ is closed under $V$ and $V^{-1}$, \ie
\[B\cup\{b\mid \text{there is $a\in B$ with $(a,b)\in V$ or $(b,a)\in V$}\} = A.\]
Analogously it can be shown that $\mB$ is also closed under $H$ and $H^{-1}$. Hence, $\mB$ is in fact a component of $\mA$.
\end{proof}

Now we will use the previous result to define a formula that ensures containment of the infinite grid.
\begin{corollary}\label{infgrid included}
Let $\mA=(A,V,H)$ be a structure that satisfies the conjunction $\phi_\mathrm{infgrid}$ of $\phi_\mathrm{grid}$ and the formulas
\[\phi_{\mathrm{infinite}}(R)\ \dfn\ \forall x\exists y R(x,y) \quad\text{for $R\in\{V,H\}$}.\]
Then $\mA$ contains a component that is isomorphic to the infinite grid $\grid{}{}$.
\end{corollary}
\begin{proof}
By \cref{gridlike includes grid} we know that $\mA$ contains a component $\mB$ that is isomorphic to $\grid{m}{n}$ for some $m,n\in \N \cup\{\infty\}$. Now suppose that $m< \infty$ and let $h$ be the isomorphism from $\grid{m}{n}$ to $\mB$. Then $h((m,0))$ cannot have an $H$-successor in $\mA$ since $\mB$ is a component in $\mA$ (and as such is closed under $H$), $\mB$ is isomorphic to $\grid{m}{n}$ and $(m,0)$ does not have an $H$-successor in $\grid{m}{n}$. But this is a contradiction to $\phi_\mathrm{infinite}(H)$ and therefore $m=\infty$. For analogous reasons it follows that $n=\infty$ and, hence, $\mB$ is isomorphic to $\grid{}{}$.
\end{proof}

The last tool needed to prove the main theorem is the following lemma about tilability of components.
\begin{lemma}\label{tiling supergrids}
Let $C$ be a set of colors, $T$ a set of tiles over $C$, $c\in C$ a color and $\mB=(B,V,H)$ a structure. Then $\mB$ is $T$-tilable (resp.~$(T,c)$-tilable) iff there is a structure $\mA$ which is $T$-tilable (resp.~$(T,c)$-tilable) and contains a component that is isomorphic to $\mB$.
\end{lemma}
\begin{proof}
The direction from left to right is trivial since every structure has itself as a component. For the reverse direction the $T$-tilability of $\mB$ is also obvious since the matching of neighboring colors is clearly preserved under taking arbitrary substructures.

For the property of $(T,c)$-tilability on the other hand it is necessary that $\mB$ is in fact isomorphic to a \emph{component} and not only to any substructure of $\mA$:
If $t$ is a $(T,c)$-tiling of $\mA$ and $\mB$ is isomorphic to a component of $\mA$ then assume there is a $x\in B$ such that, for example, $x$ does not have a $H$-predecessor in $\mB$ but also $t(x)_\text{left} \neq c$. From this follows that $x$ does not have a $H$-predecessor in $\mA$ since $\mB$ is isomorphic to a component of $\mA$, \ie a \emph{maximal} weakly connected substructure. But this is a contradiction to $t$ being a $(T,c)$-tiling of $\mA$.
\end{proof}

The following is the main theorem of this section.
\begin{theorem}\label{iftwo sat complexity}
\satt[\iftwo] is \pizeroone-complete.
\end{theorem}
\begin{proof}
For the upper bound note that $\satt[\fo]\in\pizeroone$ by G\"odel's completeness theorem. By \cref{eso sat} it follows that $\satt[\eso]\in\pizeroone$ and by the computable translation from $\df$ to $\eso$ from \cite[Theorem~6.2]{va07} it follows that $\satt[\df^3] \in \pizeroone$. Finally, the computability of the reductions in \cref{iftod} and \cref{dtwo le if2} implies that $\satt[\iftwo]\in \pizeroone$.

The lower bound follows by \cref{tiling pizeroone} and the reduction $g$ from \linebreak
$\tiling[{\mG}]$ to $\iftwo\hsat$ defined by $g(T)\dfn \phi_{\mathrm{infgrid}} \wedge \phi_T$.
To see that $g$ indeed is such a reduction, first let $T$ be a set of tiles such that $\mG$ is $T$-tilable. Then, by \cref{tiling iff tiling formula}, it follows that there is an expansion $\mG^*$ of $\mG$ such that $\mG^*\models \phi_T$. Clearly, $\mG^*\models \phi_{\mathrm{infgrid}}$ and therefore $\mG^* \models \phi_{\mathrm{infgrid}}\wedge \phi_T$.

If, on the other hand, $\mA^*$ is a structure such that $\mA^*\models \phi_{\mathrm{infgrid}}\wedge \phi_T$, then, by \cref{infgrid included}, the $\{V,H\}$-reduct $\mA$ of $\mA^*$ contains a component that is isomorphic to $\mG$. Furthermore, by \cref{tiling iff tiling formula}, $\mA$ is $T$-tilable. Hence, by \cref{tiling supergrids}, $\mG$ is $T$-tilable.
\end{proof}

\subsection{\texorpdfstring{Finite satisfiability is undecidable for \iftwo}{Finite satisfiability is undecidable for IF\textasciicircum2}}\label{iffinsatsection}

We will now discuss the problem \finsat[\iftwo] whose undecidability proof is similar to the above. The main difference is that we will now use bordered instead of arbitrary tilings.

Therefore we first need to be able to express the containment of a finite grid.
\begin{corollary}\label{finite gridlike includes fingrid}
Let $\mA=(A,V,H)$ be a finite grid-like structure.
Then $\mA$ contains a component that is isomorphic to a grid $\grid{m}{n}$ for some $m,n\in \N$.
\end{corollary}
\begin{proof}
By \cref{gridlike includes grid} there is a component $\mB$ of $\mA$ that is isomorphic to a grid $\grid{m}{n}$ for some $m,n\in \N \cup\{\infty\}$. Thus we only have to show that $m,n< \infty$. Now suppose that $m = \infty$ and let $h$ be the isomorphism from $\grid{\infty}{n}$ to $\mB$. Then $h((0,0)), h((1,0)), h((2,0)), \dots$ is an infinite sequence of pairwise-distinct points since $h((0,0))$ does not have an $H$-predecessor and $H$ is injective. But this is a contradiction to the finiteness of $\mA$. For analogous reasons it holds that $m<\infty$.
\end{proof}

Next we define some formulas dealing with border colors.
\begin{definition}\label{def:btiling formula}
Let $C$ be a set of colors, $T=\{t^0,\dots,t^k\}$ a set of tiles over $C$ and $c\in C$ a color. Furthermore let
\[ T_{c,dir} \dfn \{t^i \mid t^i_{dir} = c\},\]
for $dir \in \{\text{top},\text{right}, \text{bottom}, \text{left}\}$,
\ie the set of tiles with $c$-colored $dir$ edge.

Then $\phi_{c,T}$ is defined as the conjunction of the \fotwo formulas
\[\begin{array}{l@{\ }c@{\ }l}
\phi_{c,T,\mathrm{bottom}}    & \dfn &    \forall x \Big(\big(\forall y \neg V(y,x)\big)\to \bigvee\limits_{t^i\in T_{c,\text{bottom}}} P_i(x)\Big),\\
\phi_{c,T,\mathrm{left}}    & \dfn &    \forall x \Big(\big(\forall y \neg H(y,x)\big)\to \bigvee\limits_{t^i\in T_{c,\text{left}}} P_i(x)\Big),\\
\phi_{c,T,\mathrm{top}}    & \dfn &    \forall x \Big(\big(\forall y \neg V(x,y)\big)\to \bigvee\limits_{t^i\in T_{c,\text{top}}} P_i(x)\Big)\text{ and}\\
\phi_{c,T,\mathrm{right}}    & \dfn &    \forall x \Big(\big(\forall y \neg H(x,y)\big)\to \bigvee\limits_{t^i\in T_{c,\text{right}}} P_i(x)\Big).\\
\end{array}\]
\end{definition}

The following obvious extension of \cref{tiling iff tiling formula} now shows how to express $(T,c)$-tilability.
\begin{lemma}\label{btiling iff btiling formula}
Let $C$ be a set of colors, $T=\{t_0,\dots,t_k\}$ a set of tiles over $C$, $c\in C$ and $\mA=(A,V,H)$ a structure. Then $\mA$ is $(T,c)$-tilable iff there is an expansion $\mA^*=(A,V,H,P_0,\dots,P_k)$ of $\mA$ such that $\mA^*\models\phi_T\wedge \phi_{c,T}$.
\end{lemma}

Finally, the following theorem is the finite analogon of \cref{iftwo sat complexity}.
\begin{theorem}\label{iftwo finsat complexity}
\finsat[\iftwo] is \sigmazeroone-complete.
\end{theorem}
\begin{proof}
For the upper bound note that since all finite structures can be recursively enumerated and the model checking problem of $\iftwo$ over finite models is clearly decidable we have $\finsat[\iftwo]\in\sigmazeroone$.

The lower bound now follows by \cref{bordered tiling sigmazeroone} and the reduction $g$ from\linebreak $\btiling[\fingrid]$ to our problem defined by
\[g((T,c))\dfn \phi_{\mathrm{grid}} \wedge \phi_T \wedge \phi_{c,T}.\]
To see that $g$ indeed is such a reduction, first let $C$ be a set of colors, $T$ a set of tiles over $C$, $c\in C$ a color and $m,n\in \N$ such that there is a bordered $(T,c)$-tiling of $\grid{m}{n}$. Then, by \cref{btiling iff btiling formula}, it follows that there is an expansion $\grid{m}{n}^*$ of $\grid{m}{n}$ such that $\grid{m}{n}^* \models \phi_T \wedge \phi_{c,T}$.
Clearly, $\grid{m}{n}^* \models \phi_{\mathrm{grid}}$ and therefore $\grid{m}{n}^* \models \phi_{\mathrm{grid}}\wedge \phi_T \wedge \phi_{c,T}$.

If, on the other hand, $\mA^*$ is a finite structure such that $\mA^*\models \phi_{\mathrm{grid}}\wedge \phi_T \wedge \phi_{c,T}$ then, by \cref{finite gridlike includes fingrid}, the $\{V,H\}$-reduct $\mA$ of $\mA^*$ contains a component that is isomorphic to $\grid{m}{n}$ for some $m,n\in \N$. Furthermore, by \cref{btiling iff btiling formula}, $\mA$ is $(T,c)$-tilable. Hence, by \cref{tiling supergrids}, $\grid{m}{n}$ is $(T,c)$-tilable.
\end{proof}

\section{\texorpdfstring{Satisfiability is $\NEXP\TIME$-complete for \dtwo}{Satisfiability is NEXPTIME-complete for D\textasciicircum2}}\label{sec:satd2}
In this section we show that \satt[\dtwo] and \finsat[\dtwo] are \NEXP-complete. Our proof uses the fact that \satt[\foctwo] and \finsat[\foctwo] are \NEXP-complete \cite{pr05}.

\begin{theorem}\label{DtoESO}
Let $\tau$ be a relational vocabulary. For every formula $\phi\in \dtwo[\tau]$ there is a sentence $\phi^*\in \eso[\tau\cup\{R\}]$ (with $\ar{R}=|\frr[\phi]|$),
\[\phi^* \dfn \exists R_1\ldots\exists R_k\psi,\]
where $R_i$ is of arity at most $2$ and $\psi\in \foctwo$,
 such that for all $\mA$ and teams $X$ with $\dom{X}=\frr(\phi)$ it holds that 
\begin{equation}\label{dtoesoequiv}
\mA\models_X \phi \quad\text{iff}\quad (\mA,\rel{X})\models \phi^*,
\end{equation}
where $(\mA,\rel{X})$ is the expansion $\mA'$ of $\mA$ into vocabulary $\tau\cup\{R\}$ defined by\linebreak \mbox{$R^{\mA'} \dfn \rel{X}$}.
\end{theorem}
\begin{proof}
Using induction on $\phi$ we will first translate $\phi$ into a sentence  $\widetilde\phi \in \eso[\tau\cup\{R\}]$ satisfying \cref{dtoesoequiv}. Then we note that $\widetilde\phi$ can be translated into an equivalent sentence $\phi^*$ that also satisfies the syntactic requirement of the theorem. The proof is a modification of the proof from \cite[Theorem~6.2]{va07}. Below we write $\phi(x,y)$ to indicate that $\frr[\phi]=\{x,y\}$. Also, the quantified relations  $S$ and $T$ below are assumed not to appear in $\widetilde{\psi}$ and $\widetilde{\theta}$. 
 \begin{enumerate}
\item Let $\phi(x,y)\in \{x=y,\neg x=y, P(x,y),\neg P(x,y)\}$. Then  $\widetilde\phi$ is defined as
\[ \forall x\forall y( R(x,y)\rightarrow \phi(x,y)).  \]
\item Let $\phi(x,y)$ be of the form $\deppp(x,y)$. Then $\widetilde\phi$ is  defined as
\[\forall x\exists ^{\le 1}yR(x,y).\]
\item Let $\phi(x,y)$ be of the form $\neg \deppp(x,y)$. Then $\widetilde\phi$ is  defined as
\[\forall x\forall y\neg R(x,y).\]
\item Let $\phi(x,y)$ be of the form $\psi(x,y)\vee \theta(y)$. Then $\widetilde\phi$ is defined as
\[
\exists S\exists T(\widetilde{\psi}(S/R)\wedge\widetilde{\theta}(T/R) \wedge \forall x\forall y (R(x,y)\rightarrow (S(x,y)\vee T(y)))).
\]
\item\label{0-ary} Let $\phi(x)$ be of the form $\psi(x)\vee \theta$. Then $\widetilde\phi$ is defined as
\[
\exists S\exists T(\widetilde{\psi}(S/R)\wedge\widetilde{\theta}(T/R) \wedge \forall x (R(x)\rightarrow (S(x)\vee T))).
\]
\item Let $\phi(x,y)$ be of the form $\psi(x)\wedge \theta(y)$. Then $\widetilde\phi$ is defined as
\[
\exists S\exists T(\widetilde{\psi}(S/R)\wedge\widetilde{\theta}(T/R)\wedge \forall x\forall y (R(x,y)\rightarrow (S(x)\wedge T(y)))).
\]
\item Let $\phi(x)$ be of the form $\exists y\psi(x,y)$.  Then $\widetilde\phi$ is defined as 
\[\exists S(\widetilde{\psi}(S/R)\wedge \forall x\exists y(R(x)\rightarrow  S(x,y))). \]      
 \item Let $\phi(x)$ be of the form $\forall y\psi(x,y)$. Then  $\widetilde\phi$ is defined as
\[\exists S(\widetilde{\psi}(S/R)\wedge \forall x \forall y(R(x)\rightarrow  S(x,y))).\]
\end{enumerate}
Note that we have not displayed all the possible cases, \eg $\phi$ of the form $\dep[x]$ or $P(x)$, for which $\widetilde\phi$ is defined analogously to the above.  
Note also that, for convenience,  we allow $0$-ary relations in the translation. The possible interpretations of a $0$-ary relation $R$ are $\emptyset$ and $\{\emptyset\}$.
For a $0$-ary $R$ we define $\mA\models R$ if and only if $R^{\mA}=\{\emptyset\}$. Clause \ref{0-ary}) exemplifies the use of $0$-ary relations in the translation. It is easy to see that $\widetilde\phi$ in \ref{0-ary}) is equivalent to 
\[
  \exists S(\widetilde{\theta}(\top/R)\vee  (\widetilde{\psi}(S/R)\wedge \forall x (R(x)\rightarrow S(x)))).
\]
Generally, the use of $0$-ary relations in the translation can be easily eliminated with no essential change.

A straightforward induction on $\phi$ shows that $\widetilde\phi$ can be transformed into $\phi^*$ of the form
\[\exists R_1\ldots\exists R_k (\forall x\forall y\psi\wedge \bigwedge_{i}\forall x \exists y\theta_i\wedge \bigwedge_{j} \forall x\exists^{\le1}y R_{m_j}(x,y)), \]
where $\psi$ and $\theta_i$ are quantifier-free.
\end{proof}

Note that if $\phi\in \dtwo$ is a sentence then the relation symbol $R$ is 0-ary and $\rel{X}$ (and $R^{\mA}$) is either $\emptyset$  or $\{\emptyset\}$.
Hence, \cref{DtoESO} implies that for an arbitrary sentence $\phi \in \dtwo[\tau]$ there is a sentence $\phi'\dfn \phi^*(\true / R) \in \eso[\tau]$ such that for all $\mA$ it holds that
\begin{equation}\label{d2eso sentences}
\begin{array}{rcl}
\mA \models \phi &\text{iff}& \mA \models_{\{\emptyset\}} \phi\\
&\text{iff}& (\mA,\{\emptyset\}) \models \phi^*\\
&\text{iff}& \mA\models \phi'.
\end{array}
\end{equation}

Note that if $\phi\in\dtwo$ does not contain any dependence atoms, \ie $\phi\in\fotwo$, then the sentence $\phi^*$ is of the form
\[\exists R_1\ldots\exists R_k (\forall x\forall y\psi\wedge \bigwedge _{i}\forall x \exists y\theta_i)\]
and the first-order part of this is in \cim{Scott normal form} (\cf \cite{sc62}). So, in \cref{DtoESO} we essentially translate formulas of $\dtwo$ into Scott normal form.

\Cref{DtoESO} now implies the main theorem of this section.
\begin{theorem}\label{dtwo nexptime}
\satt[\dtwo] and \finsat[\dtwo] are \NEXP-complete.
\end{theorem}
\begin{proof}
Let $\phi\in \dtwo[\tau]$ be a sentence. Then, by \cref{d2eso sentences}, $\phi$ is (finitely) satisfiable if and only if $\phi'$ is. Now $\phi'$ is of the form
\[\exists R_1\ldots\exists R_k\psi,\]
where $\psi\in \foctwo$. Clearly, $\phi'$ is (finitely) satisfiable iff
$\psi$ is (finitely) satisfiable as a $\foctwo[\tau\cup\{R_1,\dots,R_k\}]$ sentence. Now since the mapping $\phi\mapsto \phi'$ is computable in polynomial time and (finite) satisfiability of $\psi$ can be checked in \NEXP \cite{pr05}, we get that $\satt[\dtwo], \finsat[\dtwo]\in \NEXP$.

On the other hand, since $\fotwo$ is a sublogic of $\dtwo $ and \satt[\fotwo], \finsat[\fotwo] are \NEXP-hard \cite{grkova97}, it follows that \satt[\dtwo] and \finsat[\dtwo] are $\NEXP$-hard as well.
\end{proof}

\section{Conclusion}

We have studied the complexity of the two-variable fragments of dependence logic and independence-friendly logic.
We have shown that both the satisfiablity and finite satisfiability problems for \dtwo are decidable and, in fact, even \NEXP-complete.
Also we have proved that both problems are undecidable for \iftwo; the satisfiability and finite satisfiabity problems for \iftwo are \pizeroone-complete and \sigmazeroone-complete, respectively.
\Cref{table:results} gives an overview of our complexity results as well as previously-known results.

\begin{table}[!t]
\begin{center}
\begin{tabular}{@{}c|c|c@{}}
\bfseries{Logic}            & \bfseries{Complexity of \problem{Sat} / \problem{FinSat}}   & \bfseries{Reference}\\\hline
\fo, $\fo^3$            & \pizeroone ~/ \sigmazeroone           & \cite{ch36,tu36}\\
\eso, \df, \ifl         & \pizeroone ~/ \sigmazeroone           & \Cref{eso sat}, \cite{ch36,tu36}\\
\fotwo                  & \NEXP                             & \cite{grkova97}\\
\foctwo                 & \NEXP                             & \cite{pr05}\\
\dtwo                   & \NEXP                             & \Cref{dtwo nexptime}\\
\iftwo                  & \pizeroone ~/ \sigmazeroone           & \Cref{iftwo sat complexity,iftwo finsat complexity}\\
\end{tabular}
\caption{Complexity for extensions / restrictions of first-order logic.}
{\small The results are completeness results for the full relational vocabulary.}
\label{table:results}
\end{center}
\end{table}

While the full logics \df and \ifl are expressively equivalent over sentences, we have shown that the finite variable variants \dtwo and \iftwo are not, the latter being more expressive (\Cref{dtwo le if2,d less than if}).
This was obtained as a by-product of the deeper result concerning the decidability barrier between these two logics.

An interesting open question concerns the complexity of the \emph{validity} (or \emph{tautology}) problem for \dtwo and \iftwo, \ie the problem in which a formula over a vocabulary $\tau$ is given and the problem is to decide whether \emph{all} $\tau$-structures satisfy the formula. For plain first-order logic the complexity of the validity problem is the dual of the complexity of the satisfiability problem because for all $\phi\in\fo$ it holds that
\[\phi\in \fo\hsat \quad\text{iff}\quad \neg\phi \notin \fo\htaut.\]
This, however, does not hold for \df or \ifl since for these logics the law of excluded middle does not hold (\cf \Cref{no excluded middle}). Hence, it is not even clear whether $\dtwo\htaut$ is decidable.



  \cleardoublepage
  
  \addcontentsline{toc}{chapter}{Bibliography}

\begin{thebibliography}{BMM{\etalchar{+}}11}

\bibitem[AB09]{arba09}
S.~Arora and B.~Barak, \emph{Computational complexity: A modern approach},
  Cambridge University Press, 2009.

\bibitem[AV09]{abva08}
S.~Abramsky and J.~V{\"a}{\"a}n{\"a}nen, \emph{From {IF} to {BI}}, Synthese
  \textbf{167} (2009), no.~2, 207--230.

\bibitem[BBC{\etalchar{+}}10]{babocrrescvo10}
M.~Bauland, E.~B{\"o}hler, N.~Creignou, S.~Reith, H.~Schnoor, and H.~Vollmer,
  \emph{The complexity of problems for quantified constraints}, Theory of
  Computing Systems \textbf{47} (2010), 454--490, 10.1007/s00224-009-9194-6.

\bibitem[BdRV02]{blrive02}
P.~Blackburn, M.~de~Rijke, and Y.~Venema, \emph{Modal logic}, Cambridge Tracts
  in Theoretical Computer Scie, vol.~53, Cambridge University Press, Cambridge,
  2002.

\bibitem[Ber66]{be66}
R.~Berger, \emph{The undecidability of the domino problem}, Memoirs of the
  American Mathematical Society, no.~66, American Mathematical Society, 1966.

\bibitem[BMM{\etalchar{+}}11]{bememuscthvo11}
O.~Beyersdorff, A.~Meier, M.~Mundhenk, T.~Schneider, M.~Thomas, and H.~Vollmer,
  \emph{Model checking {CTL} is almost always inherently sequential}, Logical
  Methods in Computer Science (2011).

\bibitem[CDJ09]{cadeja09}
X.~Caicedo, F.~Dechesne, and T.~M.~V. Janssen, \emph{Equivalence and quantifier
  rules for logic with imperfect information}, Logic Journal of the IGPL
  \textbf{17} (2009), no.~1, 91--129.

\bibitem[CES86]{clemsi86}
E.~M. Clarke, E.~A. Emerson, and A.~P. Sistla, \emph{Automatic verification of
  finite-state concurrent systems using temporal logic specifications}, ACM
  Trans. Program. Lang. Syst. \textbf{8} (1986), no.~2, 244--263.

\bibitem[Chu36]{ch36}
A.~Church, \emph{A note on the {E}ntscheidungsproblem}, Journal of Symbolic
  Logic \textbf{1} (1936), no.~1, 40--41.

\bibitem[CKS81]{chkost81}
A.~K. Chandra, D.~C. Kozen, and L.~J. Stockmeyer, \emph{Alternation}, J. ACM
  \textbf{28} (1981), no.~1, 114--133.

\bibitem[Coo71]{co71}
S.~A. Cook, \emph{The complexity of theorem-proving procedures}, STOC '71:
  Proceedings of the third annual ACM symposium on Theory of computing (New
  York, NY, USA), ACM, 1971, pp.~151--158.

\bibitem[DLN{\etalchar{+}}92]{dolenahonuma92}
F.~M. Donini, M.~Lenzerini, D.~Nardi, B.~Hollunder, W.~Nutt, and
  A.~Marchetti-Spaccamela, \emph{The complexity of existential quantification
  in concept languages}, Artif. Intell. \textbf{53} (1992), no.~2-3, 309--327.

\bibitem[EFT94]{ebflth94}
H.~Ebbinghaus, J.~Flum, and W.~Thomas, \emph{Mathematical logic}, Undergraduate
  texts in mathematics, Springer-Verlag, 1994.

\bibitem[EL12]{eblo12}
J.~Ebbing and P.~Lohmann, \emph{Complexity of model checking for modal
  dependence logic}, Proceedings SOFSEM 2012: Theory and Practice of Computer
  Science, Lecture Notes in Computer Science, vol. 7147, Springer Berlin /
  Heidelberg, 2012, pp.~226--237.

\bibitem[End70]{en70}
H.~B. Enderton, \emph{Finite partially-ordered quantifiers}, Mathematical Logic
  Quarterly \textbf{16} (1970), no.~8, 393--397.

\bibitem[EVW02]{etvawi02}
K.~Etessami, M.~Y. Vardi, and T.~Wilke, \emph{First-order logic with two
  variables and unary temporal logic}, Inf. Comput. \textbf{179} (2002), no.~2,
  279--295.

\bibitem[GK72]{guko72}
Y.~S. Gurevich and I.~O. Koryakov, \emph{Remarks on {Berger}'s paper on the
  domino problem}, Siberian Mathematical Journal \textbf{13} (1972), 319--321.

\bibitem[GKV97]{grkova97}
E.~Gr{\"a}del, P.~G. Kolaitis, and M.~Y. Vardi, \emph{On the decision problem
  for two-variable first-order logic}, The Bulletin of Symbolic Logic
  \textbf{3} (1997), no.~1, 53--69.

\bibitem[GO99]{grot99}
E.~Gr{\"a}del and M.~Otto, \emph{On logics with two variables}, Theor. Comput.
  Sci. \textbf{224} (1999), 73--113.

\bibitem[GOR97a]{grotro97a}
E.~Gr{\"a}del, M.~Otto, and E.~Rosen, \emph{Two-variable logic with counting is
  decidable}, Logic in Computer Science, 1997. LICS '97. Proceedings., 12th
  Annual IEEE Symposium on, jun. 1997, pp.~306 --317.

\bibitem[GOR97b]{grotro97}
E.~Gr{\"a}del, M.~Otto, and E.~Rosen, \emph{Undecidability results on
  two-variable logics}, STACS '97: Proceedings of the 14th Annual Symposium on
  Theoretical Aspects of Computer Science (London, UK), Springer-Verlag, 1997,
  pp.~249--260.

\bibitem[Har86]{Harel:1986}
D.~Harel, \emph{Effective transformations on infinite trees, with applications
  to high undecidability, dominoes, and fairness}, Journal of the ACM
  \textbf{33} (1986), no.~1, 224--248.

\bibitem[Hem01]{he01}
E.~Hemaspaandra, \emph{The complexity of poor man's logic}, Journal of Logic
  and Computation \textbf{11} (2001), no.~4, 609--622, Corrected
  version:~\cite{he05}.

\bibitem[Hem05]{he05}
E.~Hemaspaandra, \emph{The complexity of poor man's logic}, CoRR
  \textbf{cs.LO/9911014v2} (2005).

\bibitem[Hen61]{he61}
L.~Henkin, \emph{Some remarks on infinitely long formulas}, Infinitistic
  Methods (Warsaw), Proceedings Symposium Foundations of Mathematics, Pergamon,
  1961, pp.~167--183.

\bibitem[Hen67]{he67}
L.~Henkin, \emph{Logical systems containing only a finite number of symbols},
  Presses De l'Universit\'e De Montr\'e{}al, Montreal, 1967.

\bibitem[Hin96]{hi96}
J.~Hintikka, \emph{The principles of mathematics revisited}, Cambridge
  University Press, 1996.

\bibitem[Hod97a]{ho97}
W.~Hodges, \emph{Compositional semantics for a language of imperfect
  information}, Logic Journal of the IGPL \textbf{5} (1997), 539--563.

\bibitem[Hod97b]{ho97b}
W.~Hodges, \emph{Some strange quantifiers}, Structures in Logic and Computer
  Science: A Selection of Essays in Honor of A. Ehrenfeucht (J.~Mycielski,
  G.~Rozenberg, and A.~Salomaa, eds.), Lecture Notes in Computer Science, vol.
  1261, London: Springer, 1997, pp.~51--65.

\bibitem[HS89]{hisa89}
J.~Hintikka and G.~Sandu, \emph{Informational independence as a semantical
  phenomenon}, Logic, Methodology and Philosophy of Science VIII (J.~E.
  Fenstad, I.~T. Frolov, and R.~Hilpinen, eds.), vol. 126, Elsevier, Amsterdam,
  1989, pp.~571--589.

\bibitem[HSS10]{hescsc10}
E.~Hemaspaandra, H.~Schnoor, and I.~Schnoor, \emph{Generalized modal
  satisfiability}, J. Comput. Syst. Sci. \textbf{76} (2010), no.~7, 561--578.

\bibitem[KKLV11]{kokulovi11}
J.~Kontinen, A.~Kuusisto, P.~Lohmann, and J.~Virtema, \emph{Complexity of
  two-variable dependence logic and {IF}-logic}, Proceedings LICS, 2011,
  pp.~289--298.

\bibitem[KO05]{kiot05}
E.~Kieronski and M.~Otto, \emph{Small substructures and decidability issues for
  two-variable first-order logic}, Proceedings of 20th {IEEE} Symposium on
  Logic in Computer Science {LICS}'05, 2005, pp.~448--457.

\bibitem[Lad77]{la77}
R.~E. Ladner, \emph{The computational complexity of provability in systems of
  modal propositional logic}, Siam Journal on Computing \textbf{6} (1977),
  no.~3, 467--480.

\bibitem[Lew79]{le79}
H.~Lewis, \emph{Satisfiability problems for propositional calculi},
  Mathematical Systems Theory \textbf{13} (1979), 45--53.

\bibitem[LV10]{lovo10}
P.~Lohmann and H.~Vollmer, \emph{Complexity results for modal dependence
  logic}, Proceedings 19th Conference on Computer Science Logic, Lecture Notes
  in Computer Science, vol. 6247, Springer Berlin / Heidelberg, 2010,
  pp.~411--425.

\bibitem[Mei11]{me11}
A.~Meier, \emph{On the complexity of modal logic variants and their fragments},
  Ph.D. thesis, Institut f\"ur Theoretische Informatik, Leibniz Universit\"at
  Hannover, 2011, Cuvillier.

\bibitem[MMTV09]{memuthvo09}
A.~Meier, M.~Mundhenk, M.~Thomas, and H.~Vollmer, \emph{The complexity of
  satisfiability for fragments of {CTL} and {CTL$^\star$}}, International
  Journal of Foundations of Computer Science \textbf{20} (2009), no.~5,
  901--918.

\bibitem[Mor75]{mo75}
M.~Mortimer, \emph{On languages with two variables}, Mathematical Logic
  Quarterly \textbf{21} (1975), no.~1, 135--140.

\bibitem[MSS11]{masase11}
A.~L. Mann, G.~Sandu, and M.~Sevenster, \emph{Independence-friendly logic: A
  game-theoretic approach}, London Mathematical Society Lecture Note Series,
  Cambridge University Press, 2011.

\bibitem[Ott97]{ot97}
M.~Otto, \emph{Bounded variable logics and counting -- {A} study in finite
  models}, vol.~9, Springer-Verlag, 1997, IX+183 pages.

\bibitem[Pap94]{pa94}
C.~M. Papadimitriou, \emph{Computational complexity}, Addison-Wesley, Reading,
  Massachusetts, 1994.

\bibitem[PH05]{pr05}
I.~Pratt-Hartmann, \emph{Complexity of the two-variable fragment with counting
  quantifiers}, J. of Logic, Lang. and Inf. \textbf{14} (2005), 369--395.

\bibitem[PRA01]{pereaz01}
G.~Peterson, J.~Reif, and S.~Azhar, \emph{Lower bounds for multiplayer
  noncooperative games of incomplete information}, Computers \& Mathematics
  with Applications \textbf{41} (2001), no.~7-8, 957 -- 992.

\bibitem[PST97]{paszte97}
L.~Pacholski, W.~Szwast, and L.~Tendera, \emph{Complexity of two-variable logic
  with counting}, LICS '97. Proceedings., 12th Annual IEEE Symposium on Logic
  in Computer Science., 1997, pp.~318 --327.

\bibitem[Sco62]{sc62}
D.~Scott, \emph{A decision method for validity of sentences in two variables},
  J. Symbolic Logic \textbf{27} (1962), 377.

\bibitem[Sev09]{se09}
M.~Sevenster, \emph{Model-theoretic and computational properties of modal
  dependence logic}, Journal of Logic and Computation \textbf{19} (2009),
  no.~6, 1157--1173.

\bibitem[Tur36]{tu36}
A.~Turing, \emph{On computable numbers, with an application to the
  {E}ntscheidungsproblem}, Proceedings of the London Mathematical Society,
  Series 2 \textbf{42} (1936), 230--265.

\bibitem[V{\"a}{\"a}07]{va07}
J.~V{\"a}{\"a}n{\"a}nen, \emph{Dependence logic: A new approach to independence
  friendly logic}, London Mathematical Society student texts, no.~70, Cambridge
  University Press, 2007.

\bibitem[V{\"a}{\"a}08]{va08}
J.~V{\"a}{\"a}n{\"a}nen, \emph{Modal dependence logic}, New Perspectives on
  Games and Interaction (K.~R. Apt and R.~van Rooij, eds.), Texts in Logic and
  Games, vol.~4, Amsterdam University Press, 2008, pp.~237--254.

\bibitem[vEB96]{em96}
P.~van Emde~Boas, \emph{The convenience of tilings}, Tech. Report CT-1996-01,
  Institute for Logic, Language and Computation, 1996.

\bibitem[Wal70]{wa70}
J.~Walkoe, Wilbur~John, \emph{Finite partially-ordered quantification}, The
  Journal of Symbolic Logic \textbf{35} (1970), no.~4, pp. 535--555 (English).

\bibitem[Wan61]{Wang:1961}
H.~Wang, \emph{Proving theorems by pattern recognition {II}}, Bell System
  Technical Journal \textbf{40} (1961), 1--41.

\bibitem[Wra77]{wr77}
C.~Wrathall, \emph{Complete sets and the polynomial-time hierarchy},
  Theoretical Computer Science \textbf{3} (1977), no.~1, 23 -- 33.

\bibitem[Yan11]{ya11}
F.~Yang, \emph{Expressing second-order sentences in intuitionistic dependence
  logic}, Proceedings of ESSLLI workshop on Dependence and Independence in
  Logic, 2011, pp.~118--132.

\end{thebibliography}
\newcommand{\etalchar}[1]{$^{#1}$}
\providecommand{\bysame}{\leavevmode\hbox to3em{\hrulefill}\thinspace}
\providecommand{\MR}{\relax\ifhmode\unskip\space\fi MR }
\providecommand{\MRhref}[2]{%
  \href{http://www.ams.org/mathscinet-getitem?mr=#1}{#2}
}
\providecommand{\href}[2]{#2}

  \cleardoublepage
  
\begin{theindex}
{\small
{\bf Symbols}
  \item $R(T)$\dotfill \hyperpage{16}
  \item $X(A/x)$\dotfill \hyperpage{12}
  \item $X(F/x)$\dotfill \hyperpage{12}
  \item $\LL[M]$\dotfill \hyperpage{27}
  \item $\PiP{k}$\dotfill \hyperpage{17}
  \item $\SigmaP{k}$\dotfill \hyperpage{17}
  \item $\dep$\dotfill \hyperpage{2}, \textbf{23}
  \item $\dual{\cdot}$\dotfill \hyperpage{15}
  \item $\equivpm$\dotfill \hyperpage{18}
  \item $\exist$\dotfill \hyperpage{17}
  \item $\existk$\dotfill \hyperpage{8}, \textbf{13}
  \item $\exists x/W$\dotfill \hyperpage{23}
  \item $\feqv$\dotfill \hyperpage{20}
  \item $\fimp$\dotfill \hyperpage{20}
  \item $\leqpm$\dotfill \hyperpage{18}
  \item $\leqv$\dotfill \hyperpage{20}
  \item $\lleq$\dotfill \hyperpage{20}
  \item $\nor$\dotfill \hyperpage{6}, \textbf{23}
  \item $\phi(\theta / \psi)$\dotfill \hyperpage{12}
  \item $\pizeroone$\dotfill \hyperpage{17}
  \item $\restricted{}$\dotfill \hyperpage{25}
  \item $\sigmazeroone$\dotfill \hyperpage{17}
  \item $\sucteams[R]{T}$\dotfill \hyperpage{16}
  \item $t^{\mA}\langle s\rangle$\dotfill \hyperpage{11}

  \indexspace
{\bf A}
  \item accessibility relation\dotfill \hyperpage{16}
  \item alternating Turing machine\dotfill \hyperpage{17}
  \item \AP\dotfill \hyperpage{17}
  \item arity
    \subitem bounded\dotfill \hyperpage{6}, \textbf{27}
    \subitem dependence atom\dotfill \hyperpage{23}
  \item assignment\dotfill \hyperpage{11}
  \item atomic proposition\dotfill \hyperpage{14}, \hyperpage{16}

  \indexspace
{\bf C}
  \item $\CL$\dotfill \hyperpage{14}
  \item classical disjunction\dotfill \hyperpage{6}, \textbf{24}
  \item classical logic\dotfill \hyperpage{14}
  \item $\co\calC$\dotfill \hyperpage{17}
  \item colors\dotfill \hyperpage{94}
  \item completeness\dotfill \hyperpage{18}
  \item complexity operator\dotfill \hyperpage{17}
  \item component\dotfill \hyperpage{101}
  \item computational complexity\dotfill \hyperpage{4}
  \item $\co\NP$\dotfill \hyperpage{17}
  \item conservative extension\dotfill \hyperpage{3}
  \item counting quantifier\dotfill \hyperpage{8}, \textbf{13}

  \indexspace
{\bf D}
  \item $\df$\dotfill \hyperpage{2}, \textbf{23}
  \item $\df^k$\dotfill \hyperpage{26}
  \item decidable\dotfill \hyperpage{17}
  \item dependence\dotfill \hyperpage{1}
    \subitem atom\dotfill \hyperpage{23}
    \subitem disjunction\dotfill \hyperpage{6}, \textbf{24}
  \item dependence logic\dotfill \hyperpage{2}, \textbf{23}
    \subitem first-order\dotfill \hyperpage{23}
    \subitem intuitionistic\dotfill 
		\see{intuitionistic dependence logic}{7}
    \subitem modal\dotfill \hyperpage{5}, \hyperpage{26}
      \subsubitem bounded arity\dotfill \hyperpage{6}, \textbf{27}
    \subitem propositional\dotfill \hyperpage{27}
    \subitem two-variable\dotfill \hyperpage{26}
  \item determination\dotfill \hyperpage{1}
  \item disjunction
    \subitem classical\dotfill \hyperpage{6}, \textbf{24}
    \subitem dependence\dotfill \hyperpage{6}, \textbf{24}
  \item $\dom{\cdot}$\dotfill \hyperpage{11}
  \item downward closure property\dotfill \hyperpage{25}
    \subitem $\MDL$\dotfill \hyperpage{28}
  \item dual formula\dotfill \hyperpage{15}

  \indexspace
{\bf E}
  \item equivalence
    \subitem formulas\dotfill \hyperpage{20}
    \subitem logics\dotfill \hyperpage{20}
    \subitem polynomial-time\dotfill \hyperpage{18}
  \item $\eso$\dotfill \hyperpage{3}, \textbf{13}
  \item \ensuremath {\mathsf  {EXP}}\xspace  \dotfill \hyperpage{16}
  \item expressiveness\dotfill \hyperpage{20}

  \indexspace
{\bf F}
  \item $\fingrid$\dotfill \hyperpage{94}
  \item $\finsatt$\dotfill \hyperpage{19}
  \item first-order logic\dotfill \hyperpage{11}
    \subitem $k$-variable\dotfill \hyperpage{7}, \textbf{13}
      \subsubitem with counting\dotfill \hyperpage{8}, \textbf{14}
  \item flat\dotfill \hyperpage{26}
    \subitem $\MDL$\dotfill \hyperpage{28}
  \item $\fo$\dotfill \hyperpage{2}, \textbf{11}
  \item $\foc^k$\dotfill \hyperpage{8}, \textbf{14}
  \item $\fo^k$\dotfill \hyperpage{7}, \textbf{13}
  \item frame\dotfill \hyperpage{15}
  \item $\fr{\cdot}$\dotfill \hyperpage{12}

  \indexspace
{\bf G}
  \item grid\dotfill \hyperpage{94}
    \subitem finite\dotfill \hyperpage{94}
    \subitem infinite\dotfill \hyperpage{94}
  \item grid-like\dotfill \hyperpage{100}

  \indexspace
{\bf H}
  \item hardness\dotfill \hyperpage{18}

  \indexspace
{\bf I}
  \item $\IDL$\dotfill \hyperpage{7}, \textbf{29}
  \item $\ifl$\dotfill \hyperpage{2}, \textbf{23}
  \item $\iftwo$\dotfill \hyperpage{26}
  \item implication\dotfill \hyperpage{6}
    \subitem intuitionistic\dotfill \hyperpage{7}, \textbf{29}
    \subitem linear\dotfill \hyperpage{7}
  \item independence\dotfill \hyperpage{2}
  \item independence-friendly logic\dotfill \hyperpage{2}, \textbf{23}
    \subitem two-variable\dotfill \hyperpage{26}
  \item intuitionistic dependence logic\dotfill \hyperpage{7}, 
		\textbf{29}
    \subitem modal\dotfill \hyperpage{7}, \textbf{29}
    \subitem propositional\dotfill \hyperpage{7}, \textbf{30}

  \indexspace
{\bf K}
  \item Kripke structure\dotfill \hyperpage{15}

  \indexspace
{\bf L}
  \item $\LL[M]$\dotfill \hyperpage{27}
  \item labeling function\dotfill \hyperpage{16}
  \item law of excluded middle\dotfill \hyperpage{25}

  \indexspace
{\bf M}
  \item $\problem{MC}$\dotfill \hyperpage{19}
  \item $\MDL$\dotfill \hyperpage{5}, \hyperpage{26}
  \item $\MDLk$\dotfill \hyperpage{6}, \textbf{27}
  \item $\MDL[M]$\dotfill \hyperpage{27}
  \item $\MIDL$\dotfill \hyperpage{7}, \textbf{29}
  \item $\ML$\dotfill \hyperpage{14}
  \item modal logic\dotfill \hyperpage{14}
  \item model\dotfill \hyperpage{11}
  \item model checking\dotfill \hyperpage{6}, \textbf{19}

  \indexspace
{\bf N}
  \item negation normal form\dotfill \hyperpage{11}, \hyperpage{14}
  \item \NEXP\dotfill \hyperpage{16}
  \item \NP\dotfill \hyperpage{16}

  \indexspace
{\bf O}
  \item oracle Turing machine\dotfill \hyperpage{17}

  \indexspace
{\bf P}
  \item \ensuremath {\mathsf  {P}}\xspace  \dotfill \hyperpage{16}
  \item partially-ordered quantifier\dotfill \hyperpage{2}
  \item $\PDL$\dotfill \hyperpage{27}
  \item \ensuremath {\mathsf  {PH}}\xspace  \dotfill \hyperpage{17}
  \item $\PIDL$\dotfill \hyperpage{7}, \textbf{30}
  \item polynomial hierarchy\dotfill \hyperpage{17}
  \item Poor Man's Logic\dotfill \hyperpage{8}, \textbf{35}
  \item problem\dotfill \hyperpage{4}
    \subitem decision\dotfill \hyperpage{16}
  \item $\ap$\dotfill \hyperpage{14}, \hyperpage{16}
  \item $\ensuremath {\mathsf  {P}}\xspace  \ensuremath {\mathsf  {SPACE}}\xspace  $\dotfill 
		\hyperpage{16}

  \indexspace
{\bf R}
  \item $R(T)$\dotfill \hyperpage{16}
  \item reduction
    \subitem function\dotfill \hyperpage{18}
    \subitem logics\dotfill \hyperpage{20}
    \subitem polynomial-time\dotfill \hyperpage{18}
  \item $\rel{\cdot}$\dotfill \hyperpage{12}

  \indexspace
{\bf S}
  \item $\sat$\dotfill \hyperpage{18}
  \item satisfaction relation\dotfill \hyperpage{12}
  \item satisfiability\dotfill \hyperpage{6}, \textbf{18}
    \subitem finite\dotfill \hyperpage{19}
  \item Scott normal form\dotfill \hyperpage{108}
  \item second-order logic\dotfill \hyperpage{13}
    \subitem existential\dotfill \hyperpage{3}, \textbf{13}
  \item semantics
    \subitem team\dotfill \hyperpage{12}
  \item slashed quantifier\dotfill \hyperpage{2}, \textbf{23}
  \item $\so$\dotfill \hyperpage{13}
  \item space\dotfill \hyperpage{4}
  \item state\dotfill \hyperpage{16}
  \item structure
    \subitem first-order\dotfill \hyperpage{11}
    \subitem Kripke\dotfill \hyperpage{15}
  \item substitution\dotfill \hyperpage{12}
  \item successor\dotfill \hyperpage{16}
    \subitem team\dotfill \hyperpage{16}

  \indexspace
{\bf T}
  \item team\dotfill \hyperpage{3}, \textbf{11}, \hyperpage{15}
    \subitem semantics\dotfill \hyperpage{12}
  \item tile\dotfill \hyperpage{94}
  \item tiling\dotfill \hyperpage{94}
    \subitem bordered\dotfill \hyperpage{95}
    \subitem periodic\dotfill \hyperpage{95}
  \item time\dotfill \hyperpage{4}
  \item two-variable logic\dotfill \hyperpage{7}, \textbf{13}

  \indexspace
{\bf W}
  \item world\dotfill \hyperpage{16}
    \subitem successor\dotfill \hyperpage{16}}

\end{theindex}

  \cleardoublepage
  
\end{document}